\documentclass[prd,preprintnumbers,aps,preprint,amsmath,amssymb,superscriptaddress,floatfix,nofootinbib,unsortedaddress]{revtex4}
\pdfoutput=1

\usepackage{amssymb, bm, amsmath, subfigure,physymb}
\usepackage[final]{graphicx}
\usepackage[colorlinks=true,linktocpage=true,linkcolor=blue,citecolor=blue]{hyperref}
\usepackage[usenames,dvipsnames]{color}
\usepackage[T1]{fontenc}
\usepackage{subfigure}
\usepackage{slashed}
\usepackage{multirow,array}
\usepackage[utf8]{inputenc}
\usepackage{array}
\usepackage{enumerate}
\usepackage{amsfonts}
\usepackage{dsfont}
\usepackage{mathtools}
\usepackage{mathrsfs}
\usepackage{slashed}
\usepackage{latexsym}
\usepackage{hyperref}
\usepackage{epstopdf} 
\usepackage{bbm}
\usepackage[english]{babel}
\usepackage{booktabs}
\usepackage{empheq}
\usepackage{tensor}
\usepackage[toc,header]{appendix}
\usepackage{physics}
\usepackage{comment}

\pdfoutput=1
\usepackage{amsfonts}
\usepackage{soul}
\usepackage[usenames,dvipsnames]{color}

\graphicspath{{FiguresGreensFunctionsAndICCING/}}

\newcommand{\ptau}{\sqrt{p_T^2 + \pqty{p_\eta/\tau}^{\!2}}}
\newcommand{\E}[2]{\tensor*{E}{*^{\hspace{1pt} #1}_{\hspace{1.5pt} #2}}}
\newcommand{\N}[2]{\tensor*{N}{_{\!\! a}*^{\hspace{1.8pt} #1}_{\hspace{1pt} #2}}}
\newcommand{\dE}[2]{\tensor*{\delta\! E}{*^{\hspace{1pt} #1}_{\hspace{1.5pt} #2, \hspace{0.8pt}\kt}}}
\newcommand{\dN}[2]{\tensor*{\delta\! N}{_{\!\! a}*^{\hspace{1.8pt} #1}_{\hspace{1pt} #2, \hspace{0.8pt}\kt}}}
\newcommand{\G}[2]{\tensor*{\tilde{G}}{*^{#1}_{#2}}}
\newcommand{\F}[2]{\tensor*{\tilde{F}}{*^{#1}_{#2}}}
\newcommand{\GCS}[2]{\tensor*{G}{*^{#1}_{#2}}}
\newcommand{\FCS}[2]{\tensor*{F}{*^{#1}_{#2}}}
\newcommand{\tens}[3]{\tensor*{#1}{*^{#2}_{#3}}}
\newcommand{\Li}{\text{Li}}
\newcommand{\Ylm}{\tens{Y}{m}{l}}
\newcommand{\xt}{\mathbf{x}}
\newcommand{\pt}{\mathbf{p}}
\newcommand{\kt}{\mathbf{k}}
\newcommand{\rt}{\mathbf{r}}
\newcommand{\vkt}{\vqty{\mathbf{k}}}
\newcommand{\vrt}{\vqty{\mathbf{r}}}
\newcommand{\del}{\partial}
\newcommand{\imsr}{\int \frac{\dd{p_\eta}}{(2\pi)} \int \frac{\dd[2]{\pt}}{(2\pi)^2}}
\newcommand{\thydro}{\tau_\text{hydro}}
\DeclareMathOperator\artanh{arctanh}
\newcommand{\q}[1]{\grqq {#1}\grqq{}}
\newcommand{\teq}{\text{eq}}
\newcommand{\kompost}{K{\o}MP{\o}ST }
\newcommand{\BG}{\text{BG}}

\begin{document}

\title{Pre-Equilibrium Evolution of Conserved Charges with ICCING Initial Conditions}

\author{P. Carzon}
\email[Email: ]{pcarzon2@franciscan.edu}
\affiliation{Illinois Center for Advanced Studies of the Universe \&  Department of Physics, 
University of Illinois at Urbana-Champaign, Urbana, IL 61801, USA}
\affiliation{Department of Physics, 
Franciscan University, Steubenville, OH 43952, USA}
\author{M. Martinez}
\affiliation{Department of Physics, North Carolina State University, Raleigh, NC 27695, USA
}
\author{J. Noronha-Hostler}
\affiliation{Illinois Center for Advanced Studies of the Universe \&  Department of Physics, 
University of Illinois at Urbana-Champaign, Urbana, IL 61801, USA}

\author{P. Plaschke}
\email[Email: ]{pplaschke@physik.uni-bielefeld.de}
\affiliation{Fakultät für Physik, Universität Bielefeld, D-33615 Bielefeld, Germany}
\author{S. Schlichting}
\affiliation{Fakultät für Physik, Universität Bielefeld, D-33615 Bielefeld, Germany}
\author{M. Sievert}
\affiliation{Department of Physics, New Mexico State University, Las Cruces, NM 88003, USA}

\date{\today}
\begin{abstract}
Heavy-ion collisions can be well described through relativistic viscous hydrodynamics, but questions still remain when hydrodynamics is applicable because the initial state may begin very far-from-equilibrium.
Thus, a pre-equilibrium evolution phase is used to bridge the gap between the initial state and hydrodynamics. \kompost is one such pre-equilibrium model that propagates the energy-momentum tensor by decomposing it into the background and fluctuations around that background, whose evolution is captured by Green's functions. 
We extend this formalism to include conserved charges and calculate the corresponding non-equilibrium Green's functions in the relaxation time approximation. The ICCING algorithm initializes conserved charges in the initial state by sampling $g \rightarrow q\bar{q}$ splitting probabilities and is, thus, perfectly positioned to implement Green's functions for charge propagation. We show that this method alters the initial state charge geometries and is applicable in central to mid-central collisions. 

\end{abstract}

\maketitle

\tableofcontents

\section{Introduction}

Ultra-relativistic heavy-ion collisions provide an opportunity to study the extreme limits of deconfined quarks and gluons.  In the very early stages of the collisions, the energy is predominantly composed of saturated gluons emerging from the low-$x$ wave functions of the colliding nuclei \cite{H1:2009pze}.  This initial state is followed by pre-equilibrium dynamics, leading to the formation of a quark-gluon plasma (QGP) characterized by deconfined quarks and gluons acting as a nearly perfect fluid~\cite{Schlichting:2019abc,Berges:2020fwq}.  The measured distributions of final-state hadrons resulting from the freeze-out of this fluid thus encode a complex superposition of the features of the initial-state geometry, pre-equilibrium dynamics, and hydrodynamic evolution.  Thus, simulations of all stages of heavy-ion collisions are crucial to reconstruct the early stages of the collision and interpret experimental data (see \cite{Heinz:2013th} and citations within).  

To simulate heavy-ion collisions, one starts with an initial state characterization of the energy-momentum tensor $T^{\mu\nu}$ and currents $J^\mu$ of conserved charges which is far from thermodynamic equilibrium. 
The initial state at proper time $\tau_0$ and any pre-equilibrium dynamics which follow determine the initial conditions at proper time $\tau_{hydro} > \tau_0$ of the hydrodynamic equations of motion.  This hydrodynamic evolution continues until the system has frozen out into baryons and mesons. Apples-to-apples comparisons to experiments are possible after a further simulation of the hadronic gas phase, where the system is described in terms of hadrons and their interactions \cite{Petersen:2018jag, Bass:1998ca, Bleicher:1999xi}.  

Because the QGP behaves as a nearly perfect liquid \cite{Heinz:2013th}, the geometric structure from the initial-state $T^{\mu\nu}$ and $J^\mu$ leaves an observable imprint on the final state hadron distributions \cite{Teaney:2010vd,Gardim:2011xv,Niemi:2012aj,Teaney:2012ke,Qiu:2011iv,Gardim:2014tya,Betz:2016ayq,Hippert:2020kde}. This both makes models of the initial conditions particularly important in the prediction of experimental observables and also allows us to constrain these models by direct confrontation with data.  Specifically, observables which are less sensitive to the hydrodynamic phase, such as the cumulant ratio $v_n\left\{4\right\}/v_n\left\{2\right\}$ in central collisions, can provide a direct window into initial state effects \cite{Giacalone:2017uqx,Sievert:2019zjr,Rao:2019vgy}. 

Until recently, initial-state models have primarily focused on descriptions of the energy density $\epsilon = T^{00}$.  Recent progress has systematically included more initial state variables including initial flow $T^{0i}$ and initial shear $T^{ij}$ \cite{Gardim:2011qn, Gardim:2012yp, Gale:2012rq, Schenke:2019pmk, Liu:2015nwa, Kurkela:2018wud,Plumberg:2021bme,Chiu:2021muk}.  Initial conditions of conserved charge densities $\rho = J^0$ \cite{Werner:1993uh, Itakura:2003jp, Shen:2017bsr, Akamatsu:2018olk, Mohs:2019iee} have also been developed, though primarily concerned with baryon density $\rho_B$ due to its role in the search for the QCD critical point.  Recently, an open-source Monte Carlo event generator, known as ICCING (Initial Conserved Charges in Nuclear Geometry), was developed which is capable of initializing all three conserved charge densities 
\cite{Carzon:2019qja,Martinez:2019jbu}: baryon density, strangeness density, and electric charge density (BSQ). ICCING is a model-agnostic algorithm which constructs the initial conditions for the BSQ charge densities for a given energy density, by treating the energy density as being composed of gluons and stochastically sampling their probability to split into quark-antiquark ($q\bar{q}$) pairs.  The $g \rightarrow q \bar{q}$ splitting probabilities can be specified according to any desired microscopic model, among many other parts of the code. The ICCING algorithm is constructed in a modular fashion to account for a variety of different physical inputs relevant throughout the code. The sampling process is done by seeding a random point from the input energy distribution and selecting a fraction of energy from a circle centered on that point. The radius of this `gluon', the probability distribution from which the fraction of energy is sampled, and the minimum amount of energy allowed for a gluon are external inputs set by the user for this step of the process. This extends to the rest of the algorithm which is explained in full in Refs.~\cite{Carzon:2019qja,Martinez:2019jbu}.

To constrain the parameters used in ICCING initial conditions from experimental data, a necessary next step would be to evolve these initial conditions using a 2+1D viscous hydrodynamics code that simultaneously solves all of the hydrodynamic equations of motion, including all of the conserved currents as well as the energy-momentum tensor.  This task is technically challenging, not only because of the challenges associated with evolving the new conserved currents themselves, but also because of the need for a fully four-dimensional equation of state. These issues have started to be addressed and are expected to appear soon \cite{Plumberg:inPrep}. There is a further challenge in the connection of the initial state to the hydrodynamic evolution since the former is far from equilibrium. Recent work \cite{Plumberg:2021bme} has shown that moving directly from initial conditions to hydro produces a large fraction of fluid cells that violate nonlinear causality constraints \cite{Bemfica:2020xym}, about 30\%, while there is uncertainty about the causal status of the remaining cells. This study found that including a pre-equilibrium evolution stage reduces the number of acausal cells but does not fully eliminate them.  Generally, a proper description of the pre-equilibrium dynamics is not only theorectically desirable, but can also affect flow observables in small and large collision systems~\cite{Ambrus:2021fej,Kurkela:2018qeb,Ambrus:2022qya}.
Based on a microscopic description in QCD kinetic theory, the authors of~\cite{Kurkela:2018wud,Kurkela:2018vqr} developed the non-equilibrium linear response formalism \kompost which allows to propagate the energy-momentum tensor $T^{\mu\nu}$ from early times up to the point where a fluid dynamical description becomes applicable. The \kompost code was used in \cite{Plumberg:2021bme} as the pre-equilibrium stage. Coupling the ICCING code to a pre-equilibrium stage would further extend its usefulness. So far the pre-equilibrium description in \kompost itself does not contain what is needed to evolve the conserved charge densities from ICCING, but the methods that are used, namely non-equilibrium Green's functions, could be applied to our case.

The main idea of \kompost is to extract the energy-momentum tensor $T^{\mu\nu}(x)$ at time $\tau=\tau_{hydro}$ from an initial state model. For this the system is evolved by using effective kinetic theory from an initial time $\tau_0$ to $\tau_{hydro}$. Fluctuations $\delta T^{\mu\nu}(x)$ of the energy-momentum tensor around the background energy-momentum tensor $T_\text{BG}^{\mu\nu}(x)$ are considered within this framework. In practise the perturbations are assumed to be small and therefore can be linearised. This gives rise to linear response theory, where the complete energy-momentum tensor $T^{\mu\nu}(x)$ can be obtained by a sum of $T_\text{BG}^{\mu\nu}(x)$ and a term involving non-equilibrium Green's functions, which capture the evolution of the perturbations. This provides a powerful method to compute $T^{\mu\nu}(x)$ as numerical simulations only need to be done once to obtain the background evolution and the Green's functions. In the past few years different groups have begun to incorporate \kompost within their fluid dynamics simulations, making direct connections to experimental data \cite{NunesdaSilva:2020bfs,Gale:2021emg,Borghini:2022iym}.

While \kompost is based on QCD kinetic theory, more recently studies have been made in which the same Green's functions are computed in simpler models, such as the Boltzmann equation in Relaxation Time Approximation (RTA) \cite{Kamata:2020mka,Ke:2022tqf}. The assumption of relaxation time approximation drastically simplifies the theoretical description and allows for an efficient way to compute the non-equilibrium Green's functions. In the relaxation time approximation the general formalism of \kompost can also be expanded in a much simpler way to include conserved charges and compute Green's functions for the corresponding current. These Green's functions for charge and energy propagation can be included in ICCING by some careful reworking and provide a meaningful pre-equilibrium evolution for the conserved charge densities.

To test the effect of these new pre-equilibrium charge evolution equations, we will look at the event averaged 2-particle eccentricities (See App.~\ref{app:Eccentricities}) which describe the geometry of the initial state and have been shown to be good predictors of the final state flow harmonics \cite{Niemi:2012aj} except in peripheral collisions where non-linear corrects become significant \cite{Noronha-Hostler:2015dbi, Sievert:2019zjr, Rao:2019vgy}. Because of this linear mapping, it is possible to cancel out many of the medium effects by taking the ratio of 4-particle to 2-particle cumulants, which also are a measure of the fluctuations of a certain type of initial state geometry. These are well understood for the energy density and have only started to be studied for BSQ charge densities \cite{Carzon:2019qja}.

In this paper we couple the Green's functions coming from relaxation time approximation to the ICCING algorithm by treating energy and charge differences after gluon splittings as small perturbation around the background. For that we first introduce the basics about our Green's functions calculation in Sec. \ref{sec:Theory}. Sec. \ref{sec:Applications} is dedicated to the applications of these response functions in the ICCING algorithm, while we present our results in detail in Sec. \ref{sec:Results}. Conclusions are found in Sec. \ref{sec:Conclusion}.

Besides this we provide additional calculations in the appendix about the background evolution (App. \ref{app:BackgroundEvolution}), the perturbations around this background (App. \ref{app:Perturbations}) and about the Green's functions (App. \ref{app:NonEqGreensFunctions}). In App. \ref{app:IdentitiesSphericalHarmonics} we mention technical aspects regarding spherical harmonics and in App. \ref{app:Eccentricities} the definition is given for the initial state eccenctricities and cumulants.

\section{Green's Functions from Kinetic Theory \label{sec:Theory}}

In order to introduce the non-equilibrium Green's functions we follow the same idea as \cite{Kurkela:2018vqr} by dividing the space-time dynamic into a background evolution and perturbations around this background. On the technical side we follow \cite{Kamata:2020mka}, where the authors solved the equations of motions in term of moments of the distribution functions. This formalism will be extended to include conserved charges.

Although we are not primarily interested in the background evolution in this study, we need to address it briefly since its dynamics enters the time evolution of the energy and number density. Therefore Sec. \ref{sec:Theory_BoltzmannEquationInRTA} and Sec. \ref{sec:Theory_BackgroundEvolution} are dedicated to introduce the background evolution. Further results on our study regarding the background evolution can be found in App. \ref{app:BackgroundEvolution}. Afterwards we will consider the dynamics of small space-time perturbations in Sec. \ref{sec:Theory_PerturbationsAroundBjorkenFlow} (see App. \ref{app:Perturbations} for further details). The evolution of these perturbations will be captured in terms of non-equilibrium Green's functions, which are introduced in Sec. \ref{sec:Theory_NonEquilibriumGF}. For completeness, the Green's functions that are not relevant for our study are presented in App. \ref{app:NonEqGreensFunctions}.

\subsection{Boltzmann Equation in Relaxation Time Approximation} \label{sec:Theory_BoltzmannEquationInRTA}

Starting point of our analysis is the Boltzmann equation in relaxation time approximation (\textsl{RTA})
\begin{subequations}\label{eq:BEqsInRTA}
\begin{alignat}{2}
\label{e:Boltzmann1}
p^\mu\partial_\mu f &= C[f] &&= -\frac{p_\mu u^\mu(x)}{\tau_R}\bqty{f-f_\text{eq}\pqty{p_\mu\beta^\mu(x),\mu(x)}}\, , \\
\label{e:Boltzmann2}
p^\mu\partial_\mu f_a &= C[f_a] &&= -\frac{p_\mu u^\mu(x)}{\tau_R}\bqty{f_a-f_{a,\text{eq}}\pqty{p_\mu\beta^\mu(x),\mu_a(x)}}\, ,
\end{alignat}
\end{subequations}
where $x=\pqty{x^0,\xt,x^3}$ describes a four-dimensional vector in Minkowski space, $\beta(x)=u^\mu(x)/T(x)$ with $u^\mu(x)$ being the local-rest frame velocity obtained by Landau matching, $T(x)$ being the effective temperature and $a$ standing for the quarks up, down and strange. By $f$ and $f_{a}$ we denote the singlet and valence distribution functions 
\begin{subequations}
\begin{align}
f&= \nu_g f_g + \nu_q \sum\nolimits _a \bqty{f_{q_a} + \overline{f}_{q_a}}\, , \\
f_a &= \nu_q\bqty{f_{q_a}-\overline{f}_{q_a}}\, ,
\end{align}
\end{subequations}
with \textsl{g} standing for \textsl{gluon}, \textsl{q} standing for \textsl{quark}, $a=u,d,s$, such that $N_f=3$, and $\nu_g = 16, \nu_q = 6$ the spin-color degeneracy factor. The corresponding equilibrium distribution functions are the Bose–Einstein distribution function for gluons and the Fermi-Dirac distribution function for quarks and antiquarks. Each quark flavor has its own chemical potential $\mu_a (x)$ which governs the evolution of the valence charge distribution Eq.~\eqref{e:Boltzmann2}, while the flavor singlet distribution \eqref{e:Boltzmann1} evolves according to the effective chemical potential $\mu(x)$ for all flavors.
We also emphasize that we assume the same relaxation time $\tau_R$ for every species; while this is a fairly restrictive assumption, we use it as a first step to explore the geometrical impact of charge diffusion. As we consider the evolution in a conformal system, $\tau_R$ is proportional to the inverse temperature such that \cite{Denicol:2018wdp}
\begin{align}\label{eq:RelaxationTime}
\tau_R T(\tau) = 5\frac{\tilde{\eta}T}{e+P}=\textsl{const}\, ,
\end{align}
where $\tilde{\eta}$ is the shear viscosity, $e$  the energy density, $P$ the pressure and $T$ the effective temperature of the system.
The velocity $u^\mu(x)$ is the local rest-frame velocity which is determined via the Landau-matching conditions
\begin{subequations}\label{eq:LandauMatchingT}
\begin{align}
\tens{T}{\mu\nu}{}(x)u_\nu(x) &= e(x)u^\mu(x)\, ,
\end{align}
\end{subequations}
which ensures energy-momentum conservation~\cite{Rocha:2021zcw}. 
In addition to this, we also have the matching conditions for the conserved charges $n_{a}(x)$,
\begin{subequations}\label{eq:LandauMatchingN}
\begin{align}
\tens{N}{\mu}{\! a}(x) u_\mu(x) &= n_a(x)\, ,
\end{align}
\end{subequations}
such that the effective temperature $T(x)$ and effective chemical potential $\mu_{a}(x)$ can be determined from\footnote{Here the reader should think of $T(x)$ and $\mu_{a}(x)$ as effective quantities, which are defined in such a way, that they correspond to the temperature and chemical potential in thermal equilibrium. The subscript \textsl{eff} is dropped for better readability.} 
\begin{subequations}\label{eq:LandauMatchingTMU}
\begin{align}
e(x) &= e_\teq(T(x),\mu(x))\, , \\
n_a(x) &= n_{a,\teq}(T(x),\mu(x))\, .
\end{align}
\end{subequations}
We emphasize that the above-mentioned quantities $\tens{T}{\mu\nu}{}(x)$ respectively $\tens{N}{\mu}{\! a}(x)$ are defined to have contributions from all species of particles such that
\begin{subequations}
\begin{align}
\tens{T}{\mu\nu}{} &= \tens{T}{\mu\nu}{g} + \sum\nolimits_a \bqty{\tens{T}{\mu\nu}{a} + \tens{\overline{T}}{\mu\nu}{\! a}}\, , \\
\tens{N}{\mu}{\! a} &= \tens{N}{\mu}{q_a} - \tens{\overline{N}}{\mu}{\! q_a}\, .
\end{align}
\end{subequations}
As we are interested in longitudinally boost-invariant expanding systems it is convenient to work in Milne coordinates
\begin{align}
\tau = \sqrt{\pqty{x^0}^2-\pqty{x^3}^2} \quad , \quad \eta = \artanh\!\left(\frac{x^3}{x^0}\right)\, ,
\end{align}
such that $\tens{g}{}{\mu\nu}=\text{diag}\pqty{+1,-1,-1,-\tau^2}$ and $\sqrt{-g(x)}=\tau$. Furthermore we consider the quarks and gluons to be massless, such that their momentum can be parameterized as
\begin{align}
p^\mu=\pqty{p_T\cosh(y),\pt,p_T\sinh(y)}
\end{align}
with $y=\artanh\!\pqty{p^3/p^0}$ being the momentum space rapidity and $p_T\equiv\vqty{\pt}$.
In the Milne coordinates the Boltzmann equation takes the following form
\begin{subequations}\label{eq:BEinMilneCoordinates}
\begin{align}
\bqty{p^\tau\partial_\tau + p^{i}\partial_i + p^\eta\partial_\eta} f\pqty{x,p} &= -\frac{p_\mu u^\mu(x)}{\tau_R}\bqty{f\pqty{x,p} - f_\text{eq}\pqty{p_\mu\beta^\mu(x),\mu(x)}}\, , \\
\bqty{p^\tau\partial_\tau + p^{i}\partial_i + p^\eta\partial_\eta} f_a\pqty{x,p} &= -\frac{p_\mu u^\mu(x)}{\tau_R}\bqty{f_a\pqty{x,p}-f_{a,\text{eq}}\pqty{p_\mu\beta^\mu(x),\mu_a(x)}}\, ,
\end{align}
\end{subequations}
where
\begin{align}
p^\tau = p_T\cosh(y-\eta) \quad , \quad p^\eta = \frac{1}{\tau} p_T\sinh(y-\eta) \, ,
\end{align}
and $i=x,y$. It turns out that, when analyzing the dynamics of a boost invariant medium, it is more convenient to work with the (dimensionless) longitudinal momentum variable
\begin{align}
p_\eta = -\tau p_T\sinh(y-\eta) \, .
\end{align}
With respect to this coordinate we arrive at the following form of the Boltzmann equation
\begin{subequations}
\begin{align}
\bqty{p^\tau\partial_\tau + p^{i}\partial_i - \frac{p_\eta}{\tau^2}\partial_\eta} f\pqty{x,p} &= -\frac{p_\mu u^\mu(x)}{\tau_R}\bqty{f\pqty{x,p} - f_\text{eq}\pqty{p_\mu\beta^\mu(x),\mu(x)}}\, , \\
\bqty{p^\tau\partial_\tau + p^{i}\partial_i - \frac{p_\eta}{\tau^2}\partial_\eta} f_a\pqty{x,p} &= -\frac{p_\mu u^\mu(x)}{\tau_R}\bqty{f_a\pqty{x,p}-f_{a,\text{eq}}\pqty{p_\mu\beta^\mu(x),\mu_a(x)}}\, ,
\end{align}
\end{subequations}
where $p^\tau=\ptau$ represents the massless on-shell condition.

\subsection{Background Evolution} \label{sec:Theory_BackgroundEvolution}
In order to investigate the dynamics of the system in the pre-equilibrium phases, we make the assumption that the system can be divided into a background and  perturbations around this background. 
In the pre-equilibrium stage the plasma experiences a rapid longitudinal expansion. However, in the transverse plane the plasma is initially at rest and the expansion only builds up on timescales that are comparable to the systems size. Therefore, we can neglect the transverse expansion at early times and consider the idealized situation of Bjorken flow. Accordingly, the background is assumed to be longitudinally boost-invariant, parity invariant under spatial reflections along the longitudinal axis as well as azimuthally symmetric and translationally invariant in the transverse plane. The aforementioned symmetries constrain the distribution functions of the background to the following form
\begin{subequations}
\begin{align}
f\pqty{x,p} &= f_{\BG}\pqty{\tau,p_T,\vqty{p_\eta}}\, , \\
f_a\pqty{x,p} &= f_{a,\BG}\pqty{\tau,p_T,\vqty{p_\eta}} \, .
\end{align}
\end{subequations}
For such a system the energy-momentum tensor is diagonal in Milne coordinates with entries
\begin{align}
\tens{T}{\mu\nu}{\BG} = \text{diag}\pqty{e,P_T,P_T,P_L/\tau^2}\, ,
\end{align}
where $e$ is the energy density, $P_T$ the transverse and $P_L$ the longitudinal pressure. As the energy-momentum tensor is diagonal in Milne coordinates, the Landau-matching Eq. (\ref{eq:LandauMatchingT}) is solved trivially with
\begin{subequations}
\begin{align}
u^\mu &= \pqty{u^\tau, \mathbf{u}, u^\eta} = \pqty{1,0,0,0}\, , \\
e &= e_\teq\, .
\end{align}
\end{subequations}
Regarding the conserved charges and their respective currents one finds
\begin{align}
\tens{N}{\mu}{\! a} = \pqty{\tens{N}{\tau}{\! a},\mathbf{\tens{N}{}{\! a}},\tens{N}{\eta}{\! a}} = \pqty{n_a, 0,0,0}\, ,
\end{align}
where $n_a\equiv n_{q_{a}} - \overline{n}_{q_{a}}$.

Finally, the Boltzmann equation takes the familiar form
\begin{subequations}\label{eq:BEBackgroundRTA}
\begin{align}
\tau\partial_\tau f_{\BG}\pqty{\tau,p_T,\vqty{p_\eta}} &= -\frac{\tau}{\tau_R}\bqty{f_{\BG}\pqty{\tau,p_T,\vqty{p_\eta}} - f_\teq\pqty{\frac{p^\tau}{T(\tau)},\mu(\tau)}}\, , \\
\tau\partial_\tau f_{a,\BG}\pqty{\tau,p_T,\vqty{p_\eta}} &= -\frac{\tau}{\tau_R}\bqty{f_{a,\BG}\pqty{\tau,p_T,\vqty{p_\eta}}-f_{a,\text{eq}}\pqty{\frac{p^\tau}{T(\tau)},\mu_a(\tau)}}\, .
\end{align}
\end{subequations}
Our strategy to solve these equations follows \cite{Kamata:2020mka,Grad} and consists of expanding the distribution functions in terms of spherical harmonics $\Ylm(\phi,\theta)$ according to
\begin{subequations}\label{eq:EandNMoments}
\begin{align}
\E{m}{l}(\tau) &= \tau^{1/3} \imsr p^\tau \Ylm\pqty{\phi_\pt,\theta_\pt} f_{\BG}\pqty{\tau,p_T,\vqty{p_\eta}}\, ,\label{eq:EMoments} \\
\N{m}{l}(\tau) &= \imsr \tens{Y}{m}{l} \pqty{\phi_\pt,\theta_\pt} f_{a,\BG}\pqty{\tau, p_T, \vqty{p_\eta}}\, ,\label{eq:NMoments}
\end{align}
\end{subequations}
and solve the equations of motions for these moments. Since the evolution of the background is not the main focus of this work, we simply note that a detailed analysis for vanishing charge densities can be found in \cite{Kamata:2020mka}. While in this work, we will only consider energy-momentum and charge perturbations on top of a charge neutral background, we provide additional discussion on the background evolution in the presence of non-vanishing density in App. \ref{app:BackgroundEvolution}.

As we are interested in the evolution around vanishing background charge density, we should mention that we can extract the background energy density at any given time as a function of the initial energy density according to the following method\footnote{For the sake of readability we drop the subscript $BG$ for the energy density for a moment.}.
By following \cite{Kamata:2020mka,Giacalone:2019ldn,Du:2020zqg}, we can compute the  energy density at late times $\pqty{\tau^{4/3}e(\tau)}_\infty$ as a function of the initial energy density as 
\begin{align}
\pqty{\tau^{4/3}e(\tau)}_\infty = C_\infty \pqty{\frac{4\pi\tilde{\eta}/s}{T(\tau_0)\tau_0^{1/4}}}^{4/9} (e\tau)_0 \, ,
\end{align}
where the constant $C_{\infty}\approx 0.9$~\cite{Kamata:2020mka,Giacalone:2019ldn} quantifies how efficiently the initial energy is converted into thermal energy\footnote{See App. \ref{app:BackgroundEvolutionInConformalSystems} for further details on how to determine the constant $C_\infty$.}, $\tilde{\eta}/s$ is the shear-viscosity to entropy density ratio\footnote{Note that in Eq. (\ref{eq:RelaxationTime}) we had $\Tilde{\eta}T/(e+P)$ instead of $\Tilde{\eta}/s$. However, since we now look at a vanishing background charge density, $\Tilde{\eta}T/(e+P)$ actually simplifies to $\Tilde{\eta}/s$.}, which is constant for a conformal system with vanishing net charge density. In our calculations we choose $\Tilde{\eta}/s=\frac{1}{4\pi}$, although the actual value can be scaled out of the equations of motion. By $(e\tau)_0$ we denote the initial energy density per unit area and rapidity
\begin{align}
(e\tau)_0 \equiv
\frac{\dd{E_{0}}}{\dd{\eta}\dd[2]{\xt}} &= \frac{\dd{E_{0,g}}}{\dd{\eta}\dd[2]{\xt}} + \sum_a\pqty{\frac{\dd{E_{0,q_a}}}{\dd{\eta}\dd[2]{\xt}} + \frac{\dd{E_{0,\overline{q}_a}}}{\dd{\eta}\dd[2]{\xt}}} 
\label{eq:BackgroundInitialEnergyE}
\end{align}
which becomes constant in the limit $\tau \to 0$, in kinetic theory.

Inverting the Landau matching condition Eq. (\ref{eq:LandauMatchingT}) for vanishing chemical potential gives
\begin{align} \label{eq:EffectiveTemperature}
T(\tau)=\bqty{\frac{30}{\pi^2\nu_{\text{eff}}}e(\tau)}^{1/4}\,
\end{align}
with $\nu_{\text{eff}}=\nu_g+\frac{7}{8} 2N_f \nu_q$ the overall effective degeneracy factor of all partons, such that $\pqty{\tau^{4/3}e(\tau)}_\infty$ can be expressed as
\begin{align}
\pqty{\tau^{4/3}e(\tau)}_\infty = C_\infty \pqty{4\pi\tilde{\eta}/s}^{4/9} \pqty{\frac{\pi^2\nu_\text{eff}}{30}}^{1/9} (e\tau)_0^{8/9}\, .
\end{align}
In the next step we introduce an attractor curve $\mathcal{E}(\tilde{w})$, which depends on the dimensionless time-variable \cite{Giacalone:2019ldn}
\begin{align}\label{eq:wTilde}
\tilde{w} = \frac{T(\tau)\tau}{4\pi\tilde{\eta}/s}\, .
\end{align}
This attractor curve smoothly interpolates between free-streaming at early times and viscous hydrodynamics at late times
\begin{subequations}
\begin{align}
\mathcal{E}(\tilde{w}\ll1) &= C_\infty^{-1} \tilde{w}^{4/9} \, , \\
\mathcal{E}(\tilde{w}\gg 1) &= 1-\frac{2}{3\pi\tilde{w}^{4/9}}\, .
\end{align}
\end{subequations}
The attractor curve connects the asymptotic value $\pqty{\tau^{4/3}e(\tau)}_\infty$ to its counterpart at any given time $\pqty{\tau^{4/3}e(\tau)}$ according to
\begin{align}   \label{e:attractor1}
\mathcal{E}(\tilde{w}) = \frac{\pqty{\tau^{4/3}e(\tau)}}{\pqty{\tau^{4/3}e(\tau)}_\infty} \,,
\end{align}
and has been calculated in \cite{Kamata:2020mka,Giacalone:2019ldn} for the Boltzmann equaton in RTA and in \cite{Giacalone:2019ldn,Du:2020zqg} for Yang-Mills and QCD kinetic theory.
Based on this attractor curve, we can therefore relate the initial energy density to the energy density at a later time via
\begin{align}   \label{e:attractor2}
e_\text{\BG}\pqty{e(\tau_0)} = C_\infty \pqty{4\pi\tilde{\eta}/s}^{4/9} \pqty{\frac{\pi^2\nu_\text{eff}}{30}}^{1/9} \frac{(e\tau)_0^{8/9}}{\tau^{4/3}} \mathcal{E}(\tilde{w})   \: ,
\end{align}
assuming that the system can locally be described by conformal Bjorken flow up to this time scale.

\subsection{Perturbations around Bjorken flow} \label{sec:Theory_PerturbationsAroundBjorkenFlow}

So far we addressed the evolution of a homogeneous, boost invariant background. Now we will consider linearized perturbations around this background, caused by small space-time dependent variations of the initial energy or charge densities. We will linearize the kinetic equations, such that we can derive an evolution equation for the perturbations of the distribution functions $\delta f$ and $\delta f_a$
\begin{subequations}\label{eq:GeneralEoMForPerturbation}
\begin{align}
\begin{aligned}
&\bqty{p^\tau \del_\tau + p^{i}\del_i - \frac{p_\eta}{\tau^2}\del_\eta} \delta f(x,p) \\
&= - \frac{p^\tau}{\tau_R} \delta f\pqty{x,p} + \frac{p_\mu\delta u^\mu(x)}{\tau_R} \bqty{\pqty{f_{\text{eq}} - f\pqty{x,p}} + \frac{p^\tau}{T(\tau)} f^{\scriptscriptstyle (1,0)}_{\text{eq}}} \\
&\phantom{=} -\frac{p^\tau}{\tau_R} \frac{\delta T(x)}{T(\tau)} \bqty{\frac{T(\tau)}{\tau_R} \pdv{\tau_R}{T} \pqty{f_{\text{eq}} - f\pqty{x,p}} + \frac{p^\tau}{T(\tau)} f^{\scriptscriptstyle (1,0)}_{\text{eq}}} +\frac{p^\tau}{\tau_R} \sum_a \delta \mu_{a}(x)  \bqty{f^{\scriptscriptstyle (0,1)}_{q_a,\text{eq}} + \overline{f}^{\scriptscriptstyle (0,1)}_{q_a,\text{eq}}}
\end{aligned}
\end{align}
and
\begin{align}
\begin{aligned}
&\bqty{p^\tau \del_\tau + p^{i}\del_i - \frac{p_\eta}{\tau^2}\del_\eta} \delta f_{a}(x,p) \\
&= - \frac{p^\tau}{\tau_R} \delta f_{a}\pqty{x,p} + \frac{p_\mu\delta u^\mu(x)}{\tau_R} \bqty{\pqty{f_{a,\text{eq}} - f_{a}\pqty{x,p}} + \frac{p^\tau}{T(\tau)} f^{\scriptscriptstyle (1,0)}_{a,\text{eq}}} \\
&\phantom{=} -\frac{p^\tau}{\tau_R} \frac{\delta T(x)}{T(\tau)} \bqty{\frac{T(\tau)}{\tau_R} \pdv{\tau_R}{T} \pqty{f_{a,\text{eq}} - f_{a}\pqty{x,p}} + \frac{p^\tau}{T(\tau)} f^{\scriptscriptstyle (1,0)}_{a,\text{eq}}} + \frac{p^\tau}{\tau_R}\delta \mu_a(x)  f^{\scriptscriptstyle (0,1)}_{a,\text{eq}}\, ,
\end{aligned}
\end{align}
\end{subequations}
where
\begin{subequations}
\begin{align}
\delta f\pqty{x,p} &= \nu_g \delta f_g\pqty{x,p} + \nu_q\sum_a \bqty{\delta f_{q_a}\pqty{x,p} + \delta \overline{f}_{q_a}\pqty{x,p}}\, , \\
\delta f_a\pqty{x,p} &= \nu_q\bqty{\delta f_{q_a}\pqty{x,p} - \delta \overline{f}_{q_a}\pqty{x,p}}\, .
\end{align}
\end{subequations}
Here we used a shorter notation for the derivatives, namely
\begin{align}
f^{\scriptscriptstyle (n,m)} (x,y) \equiv \pdv[n]{x} \pdv[m]{y} f(x,y) \, .
\end{align}
As
\begin{align}
f_\teq &= f_\teq \pqty{\frac{p^\tau}{T(\tau)},\mu(\tau)}\, , \\
f_{a,\teq} &= f_{a,\teq}\pqty{\frac{p^\tau}{T(\tau)},\mu(\tau)}\, ,
\end{align}
the derivatives are with respect to $p^\tau/T(\tau)$ for the $(1,0)$-derivative and with respect to $\mu(\tau)$ for the $(0,1)$-derivative.

The perturbations of the rest-frame velocity, $\delta u^\mu(x)$, the temperature, $\delta T(x)$, and the chemical potential, $\delta \mu_{a}(x)$, are determined by the linearised Landau-matching conditions. Details can be found in App. \ref{app:LinEoM}. \\

\subsubsection{Evolution equations for perturbations in the transverse plane}
From now on we will concentrate on perturbations in the transverse plane, i.e. only for the transverse coordinates $\xt$. To solve the equations of motion we will expand the perturbations in a Fourier basis such that
\begin{align}
\delta f_i\pqty{\tau,\mathbf{x}, \pt, \vqty{p_\eta}} &= \int \frac{\dd[2]{\mathbf{k}}}{(2\pi)^2} \delta f_{i,\hspace{0.8pt} \mathbf{k}}\pqty{\tau,\pt, \vqty{p_\eta}} e^{i\mathbf{k}\cdot\mathbf{x}}
\end{align}
for $i\in\Bqty{g,q_a,\overline{q}_a}$ and where $\delta f_{i,\hspace{0.8pt} \mathbf{k}}\pqty{\tau,\pt, \vqty{p_\eta}}\equiv\delta f_{i}\pqty{\tau,\mathbf{k},\pt, \vqty{p_\eta}}$. The definition of $\delta f_{\mathbf{k}}\pqty{\tau,\pt, \vqty{p_\eta}}$ and $\delta f_{a,\hspace{0.8pt} \mathbf{k}}\pqty{\tau,\pt, \vqty{p_\eta}}$ is analogous to the cases before. Decomposing the velocity perturbation in the transverse plane into components parallel, $\delta u_\mathbf{k}^\parallel(\tau)$, and transverse, $\delta u_\mathbf{k}^\perp(\tau)$, to the wave vector in the transverse plane $\kt$ we therefore find
\begin{subequations}\label{eq:PerturbationFourierExpansion}
\begin{align}
\delta T\pqty{\tau,\mathbf{x}} &= \int \frac{\dd[2]{\kt}}{(2\pi)^2} \delta T_\mathbf{k}\pqty{\tau} e^{i\kt\cdot\mathbf{x}} \, , \\
\delta \mu_a\pqty{\tau,\mathbf{x}} &= \int \frac{\dd[2]{\kt}}{(2\pi)^2} \delta \mu_{a,\hspace{0.8pt} \mathbf{k}}\pqty{\tau} e^{i\kt\cdot\mathbf{x}} \, , \\
\delta u^{i}\pqty{\tau,\mathbf{x}} &= \int \frac{\dd[2]{\kt}}{(2\pi)^2} \bqty{\delta u_\mathbf{k}^\parallel(\tau)\delta^{ji} + \delta u_\mathbf{k}^\perp(\tau)\epsilon^{ji}} \frac{\kt^{j}}{\abs{\kt}} e^{i\kt\cdot\mathbf{x}} \, , \\
\delta &u^\tau\pqty{\tau,\mathbf{x}} = 0 \quad , \quad \delta u^\eta(\tau,\xt) = 0 \, .
\end{align}
\end{subequations}
Note that $\delta u^\eta(\tau,\xt) = 0$ vanishes identically due to the assumption that boost invariance along the beam axis is not broken for the perturbations.  This assumption could be relaxed in future work, but is important here for coupling to a 2+1D geometry as in ICCING. \\
Denoting
\begin{subequations}\label{eq:ScalarAndCrossProductKP}
\begin{align}
\frac{\kt\cdot\pt}{\abs{\kt}p^\tau} &= \delta^{ij} \frac{\kt^{i}\pt^{j}}{\abs{\kt}p^\tau} = \cos(\phi_{\pt\kt})\sin(\theta_\pt)\, , \\
\frac{\kt\times\pt}{\abs{\kt}p^\tau} &= \epsilon^{ij} \frac{\kt^{i}\pt^{j}}{\abs{\kt}p^\tau} = \sin(\phi_{\pt\kt})\sin(\theta_\pt)\, ,
\end{align}
\end{subequations}
where $\phi_{\pt\kt}\equiv\phi_\pt-\phi_\kt$ is the angle between $\pt$ and $\kt$ in the transverse plane and $\sin(\theta_\pt)=p_T/p^\tau$ and inserting the Fourier integrals above into Eq. (\ref{eq:GeneralEoMForPerturbation}) allows us to find an evolution equation for $\delta f_{\mathbf{k}}$ and $\delta f_{a,\hspace{0.8pt} \mathbf{k}}$, such that we have
\begin{subequations}\label{eq:EoMForPerturbation}
\begin{align}\label{eq:EoMForPerturbedf}
\begin{aligned}
\tau &\del_\tau \delta f_{\mathbf{k}}\pqty{\tau,\pt,\vqty{p_\eta}} \\
&= -\bqty{i\tau \abs{\kt} \frac{\kt\cdot\pt}{\abs{\kt} p^\tau} + \frac{\tau}{\tau_R} } \delta f_{\mathbf{k}}\pqty{\tau,\pt,\vqty{p_\eta}} \\
&\phantom{=} -\frac{\tau}{\tau_R} \pqty{ \delta u_\mathbf{k}^\parallel(\tau)\frac{\kt\cdot\pt}{\abs{\kt} p^\tau} +\delta u_\mathbf{k}^\perp(\tau)\frac{\kt\times\pt}{\abs{\kt} p^\tau} } \bqty{\pqty{f_{\text{eq}} - f\pqty{x,p}} + \frac{p^\tau}{T(\tau)} f^{\scriptscriptstyle (1,0)}_{\text{eq}}} \\
&\phantom{=} -\frac{\tau}{\tau_R} \frac{\delta T_\mathbf{k}(\tau)}{T(\tau)} \bqty{\frac{T(\tau)}{\tau_R} \pdv{\tau_R}{T} \pqty{f_{\text{eq}} - f\pqty{x,p}} + \frac{p^\tau}{T(\tau)} f^{\scriptscriptstyle (1,0)}_{\text{eq}}} +\frac{\tau}{\tau_R}\sum_a\delta \mu_{a,\hspace{0.8pt} \kt}(x) \bqty{f^{\scriptscriptstyle (0,1)}_{q_a,\text{eq}} + \overline{f}^{\scriptscriptstyle (0,1)}_{q_a,\text{eq}}}
\end{aligned}
\end{align}
and
\begin{align}\label{eq:EoMForPerturbedfa}
\begin{aligned}
\tau &\del_\tau \delta f_{a,\hspace{0.8pt} \kt}\pqty{\tau,\pt,\vqty{p_\eta}} \\
&= -\bqty{i\tau \abs{\kt} \frac{\kt\cdot\pt}{\abs{\kt} p^\tau} + \frac{\tau}{\tau_R} } \delta f_{a,\hspace{0.8pt} \kt}\pqty{\tau,\pt,\vqty{p_\eta}} \\
&\phantom{=} -\frac{\tau}{\tau_R} \pqty{ \delta u_\mathbf{k}^\parallel(\tau)\frac{\kt\cdot\pt}{\abs{\kt} p^\tau} +\delta u_\mathbf{k}^\perp(\tau)\frac{\kt\times\pt}{\abs{\kt} p^\tau} } \bqty{\pqty{f_{a,\text{eq}} - f_{a}\pqty{x,p}} + \frac{p^\tau}{T(\tau)} f^{\scriptscriptstyle (1,0)}_{a,\text{eq}}} \\
&\phantom{=} -\frac{\tau}{\tau_R} \frac{\delta T_\kt(\tau)}{T(\tau)} \bqty{\frac{T(\tau)}{\tau_R} \pdv{\tau_R}{T} \pqty{f_{a,\text{eq}} - f_{a}\pqty{x,p}} + \frac{p^\tau}{T(\tau)} f^{\scriptscriptstyle (1,0)}_{a,\text{eq}}}  +\frac{\tau}{\tau_R}\delta \mu_{a,\hspace{0.8pt} \kt}(x) f^{\scriptscriptstyle (0,1)}_{a,\text{eq}}\, .
\end{aligned}
\end{align}
\end{subequations}
In Eqs. (\ref{eq:EoMForPerturbation}) we sorted the terms by the several perturbations. The first term on the right hand side corresponds to free-streaming, while the second term describes the relaxation of the perturbations. In the following lines one sees that the perturbations of the velocity, temperature and chemical potential cause a change of the equilibrium distribution, while the velocity and temperature perturbations also affect the relaxation of the out-of-equilibrium background.

\subsubsection{Solving the equations of motion in the transverse plane}
In order to solve Eq. (\ref{eq:EoMForPerturbation}) we follow the same strategy as for the background. Therefore, we define the perturbed moments according to
\begin{subequations} \label{eq:PerturbedEandNMoments}
\begin{align}
\dE{m}{l}(\tau) &= \tau^{1/3} \imsr p^\tau \Ylm\pqty{\phi_{\pt\kt},\theta_\pt} \delta f_\kt\pqty{\tau,\pt,\vqty{p_\eta}}\, , \label{eq:EPerturbedMoments} \\
\dN{m}{l}(\tau) &= \imsr \Ylm\pqty{\phi_{\pt\kt},\theta_\pt} \delta f_{a,\kt}\pqty{\tau,\pt,\vqty{p_\eta}}\, . \label{eq:NPerturbedMoments}
\end{align}
\end{subequations}
As the derivation of the equations of motion for the moments are not of primary interest here, we will shift the explicit calculation into App. \ref{app:Perturbations}. \\
Similar to the background, we are able to obtain the components of $\tens{\delta T}{\mu\nu}{\kt}$ and $\tens{\delta N}{\mu}{\! a,\hspace{0.8pt} \kt}$ as combinations of low order moments. A full list can be found in App. \ref{app:EoMPerturbedMoments}. Furthermore we can also relate the perturbations of the intensive quantities to the perturbation of the extensive quantities for $n_a=0$ according to Eq. (\ref{eq:IntensivePerturbationsZeroDensity}) by
\begin{subequations}
\begin{align}
\frac{\delta T_\kt}{T} &= \frac{\delta e_\kt}{4e}\, , \\
\delta \mu_{a,\hspace{0.8pt} \kt} &= \frac{6}{\nu_q} \frac{\delta n_{a.\hspace{0.8pt} \kt}}{T^2}\, .
\end{align}
\end{subequations}
Therefore we can replace $\delta T_\kt$ and $\delta\mu_{a,\hspace{0.8pt} \kt}$ with $\delta e_\kt$ and $\delta n_{a,\hspace{0.8pt} \kt}$, which is useful since we can express these quantities by low order moments again
\begin{subequations}\label{eq:PerturbationsWrittenAsMoments}
\begin{align}
\tau^{4/3} \delta e_\kt(\tau) &= \sqrt{4\pi} \dE{0}{0}(\tau) \, , \\
\tau^{4/3} (e+P_T) \delta u^\parallel_\kt(\tau) &= -\sqrt{\frac{2\pi}{3}} \qty\Big(\dE{+1}{1}(\tau) - \dE{-1}{1}(\tau)) \, , \\
\tau^{4/3} (e+P_T) \delta u^\perp_\kt(\tau) &= i\sqrt{\frac{2\pi}{3}} \qty\Big(\dE{+1}{1}(\tau) + \dE{-1}{1}(\tau)) \, , \\
\tau \delta n_{a,\hspace{0.8pt} \kt} &= \sqrt{4\pi}\dN{0}{0}\, ,
\end{align}
\end{subequations}
This results in a closed set of equations since all appearing perturbations can be written as linear combinations of moments.

The equations of motions will be solved numerically. For this we truncate the evolution at $l_\text{max}$=512. In order to find a reasonable value we compared our results to the analytical free-streaming equations. This comparison shows that convergence is reached much faster for $\dE{m}{l}$ than for $\dN{m}{l}$. More details can be found in Fig. \ref{fig:ConvergencePlots} in App. \ref{app:Numerics}.

We note that, in addition to \cite{Kamata:2020mka} we also need to invert the matching conditions Eq. (\ref{eq:LandauMatchingTMU}). This we do numerically at each time step. More details on this can be found in App. \ref{app:ConnectingIntensiveAndExtensiveQuantities}.

\subsection{Non-equilibrium Green's Functions} \label{sec:Theory_NonEquilibriumGF}
In principle, we can obtain all information about the evolution of the system from the moments. However, we find it more convenient to consider Green's functions of the energy-momentum tensor and the charge current. Therefore, we will consider linear response functions $\G{\mu\nu}{\alpha\beta}$ for the energy-momentum tensor respectively and $(\tilde{F}_{a b})_\alpha^\mu$ for the charge current. Here $\G{}{}$ and $\F{}{}$ describe the Green's functions in momentum space. As we will show below, the Green's functions can be related to macroscopic quantities (Eq. (\ref{eq:GssInMomentumSpace}), Eq. (\ref{eq:FssInMomentumSpace})), which are however related to low order moments according to Eqs. (\ref{eq:PerturbationsWrittenAsMoments}). This is another powerful property of our formalism as it is easy to quantify the systems response to perturbations once one have solved the equations of motions. 

For our study, only $\G{\tau\tau}{\tau\tau}$ and $(\tilde{F}_{a b})_\tau^\tau$ are relevant, such that we only present the result for these two here. The other Green's functions can be found in App. \ref{app:NonEqGreensFunctions}.

\subsubsection{Non-equilibrium Green's Functions of the Energy-Momentum Tensor}\label{sec:GreensFunctionsT}
We will follow the construction of the response functions according to \cite{Kurkela:2018vqr,Kurkela:2018wud} and express $\tens{\delta T}{\mu\nu}{\kt}(\tau)$ as
\begin{align}\label{eq:ResponseFunctionEquationForT}
\frac{\tens{\delta T}{\mu\nu}{\kt}(\tau)}{e(\tau)} = \frac{1}{2} \G{\mu\nu}{\alpha\beta}\pqty{\kt,\tau,\tau_0} \frac{\tens{\delta T}{\alpha\beta}{\kt}(\tau_0)}{e(\tau_0)}\, .
\end{align}
In the following we will omit the explicit dependence on $\tau_0$ for better readability since we are mainly interested in the limit $\tau_0/\tau_R\rightarrow 0$, where the kinetic framework describes the equilibration process from directly after the collision until the onset of the hydrodynamic regime. Besides this, we also introduce the propagation phase $\kappa$ by
\begin{align}
\kappa = \vkt (\tau-\tau_0)\, .
\end{align}
When expressing the evolution equations for the moments in terms of $\kappa$, this change of variable introduces additional terms in the time derivative \cite{Kamata:2020mka}, which were taken into account. 

Similarly to \cite{Kurkela:2018vqr}, we will decompose the response functions into a basis of Lorentz scalars (\textsl{s}), vectors (\textsl{v}) and tensors (\textsl{t}). For $\G{\tau\tau}{\tau\tau}$ this means
\begin{align}
\G{\tau\tau}{\tau\tau}\pqty{\kt,\tau} &= \G{s}{s}\pqty{\kappa,x}\, .
\end{align}
\noindent Since the normalization of the linearized perturbation is 
arbitrary, we adopt the convention
\begin{align}
\frac{\delta e(\tau_0)}{e(\tau_0)} = 1
\end{align}
such that we can express the decomposed response function in terms of $\tens{\delta T}{\mu\nu}{\kt}(\tau)$ (see \cite{Kurkela:2018vqr}) according to
\begin{align}\label{eq:GssInMomentumSpace}
\G{s}{s}\pqty{\kappa,x} &= \frac{\tens{\delta T}{\tau\tau}{\kt}(x)}{e(x)} = \frac{\delta e_\kappa(x)}{e(x)}\, .
\end{align}

\subsubsection{Non-equilibrium Green's Functions of the Current of Conserved Charges}\label{sec:GreensFunctionsN}
For the Green's functions corresponding to the conserved charges, we follow the same strategy. Before we compute them, we first recall that in Bjorken flow
\begin{align}
\tau n(\tau) = \text{\textsl{const}}\, .
\end{align}
In particular we have $\tau n(\tau)=\tau_0 n(\tau_0)$, i.e. we need a slightly different definition for the charge Green's functions
\begin{align}
\tau\tens{\delta N}{\mu}{\! a,\hspace{0.8pt} \kt}(\tau) = (\tilde{F}_{a b})_\alpha^\mu \pqty{\kt,\tau,\tau_0} \, \tau_0\tens{\delta N}{\alpha}{\! b,\hspace{0.8pt} \kt}(\tau_0)\, .
\end{align}
Note that in general different flavours can couple to each other via the response function. However, for vanishing densities, there is no coupling between the response for different quark flavors and all flavors will have the same response functions, such that the response matrix is proportional to the identity in flavor space, i.e. $(\tilde{F}_{a b})_\alpha^\mu=\F{\mu}{\alpha}\tens{\delta}{}{ab}$.

We will decompose the charge Green's functions also in a scalar-vector-tensor basis such that we have
\begin{align}
\F{\tau}{\tau}\pqty{\kt,\tau} &= \F{s}{s}\pqty{\kappa,x}\, .
\end{align}
It is possible to express the response functions in terms of $\tens{\delta N}{\mu}{\! a,\hspace{0.8pt} \kt}$. Adapting the normalization
\begin{align}
\tau_0\tens{\delta N}{\tau}{\kt}(\tau_0) = 1
\end{align}
we find
\begin{align}\label{eq:FssInMomentumSpace}
\F{s}{s}\pqty{\kappa,x} &= \tau\tens{\delta N}{\tau}{\kappa}(x)=\tau\delta n_\kappa(x)\, .
\end{align}
Note that we also drop the index \textsl{a} on the components $\tens{\delta N}{i}{\kt}$ as they will be the same for all species for vanishing background number charge densities.

\subsubsection{Numerical Results for the non-equilibrium Green's Functions}\label{sec:GreensFunctionsResults}
\begin{figure}[t!]
\begin{center}
\includegraphics[width=0.49\textwidth]{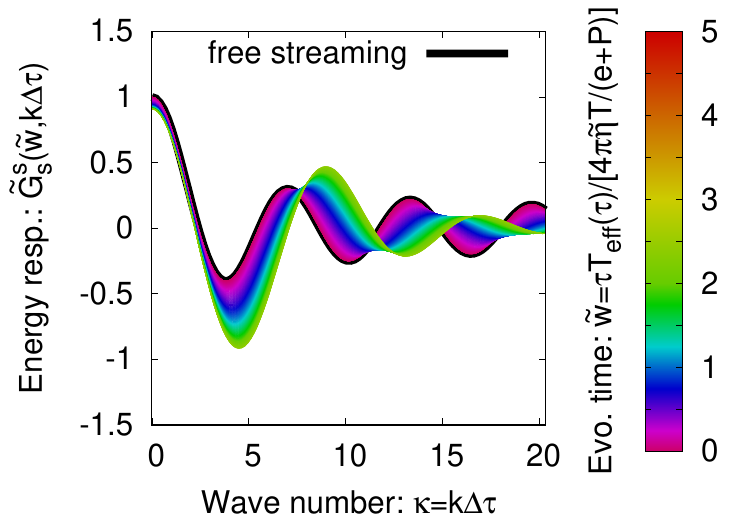}
\includegraphics[width=0.49\textwidth]{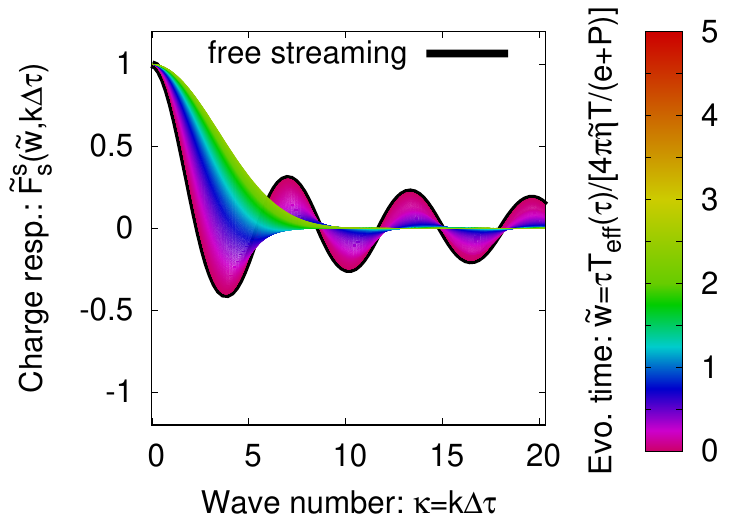}
\caption{\textsl{Left:} Evolution of the energy-momentum Green's function $\G{s}{s}$ in response to initial energy perturbations. \textsl{Right:} Evolution of the charge Green's function $\F{s}{s}$ in response to initial charge perturbations. The different curves in each panel correspond to different times $\tilde{w}$. \label{fig:ResponseFunctionsToInitialEnergyAndChargePerturbationsKappaMode}}
\end{center}
\end{figure}
We present the results for the response functions $\G{s}{s}$ and $\F{s}{s}$ in Fig. \ref{fig:ResponseFunctionsToInitialEnergyAndChargePerturbationsKappaMode}, where we plotted the different response functions in dependence of the propagation phase $\kappa$ and time $\tilde{w}$ (Eq. (\ref{eq:wTilde})). The different panels correspond to the different response functions and are labelled by the components of the energy-momentum tensor and the charge current that they affect, e.g. $\G{s}{s}$ is labelled by \q{energy response} as this function describes the response of $\tens{\delta T}{\tau\tau}{\kt}(\tau)$, which corresponds to the energy. Each curve in each panel corresponds to the response function at a different time as it is indicated by the colour code. Besides this we also plotted the free-streaming behavior for each response function, which corresponds to the black line at $\tilde{w}=0$. 

Due to the fact that we consider the perturbations for the charge density and chemical potential around vanishing background densities, the evolution of the response function $\G{s}{s}$ does not change in comparison to the results in \cite{Kamata:2020mka}. This can already be seen at the level of the equation of motion in Eq. (\ref{eq:EoMPerturbedEinKappaMode}) for vanishing densities. For early times ($\tilde{w}\ll 1$) one observes the free-streaming behavior, characterized by wave-like modes with both peaks of excess density and troughs of depleted density (the diffusion wake). Towards later times larger $\kappa$-modes become damped, which can be explained by viscous effects of the medium. At the onset of the hydrodynamic regime ($\tilde{w}\sim 1$) only long wave-length modes  survive, which indicates that the free-streaming initial conditions are getting washed out during the evolution of the system. The shift of the peak for later times towards larger values of the propagation phase can be understood by noting that at early times, shortly after the collision, the system is highly anisotropic and expands in the transverse plane with a phase-velocity close to the speed of light. As the system evolves in time, it will become more and more isotropic and the phase-velocity will approach the speed of sound resulting in the shift of the peak. 

For the charge density response $\F{s}{s}$ we see that for early times ($\tilde{w}\ll 1$) the behavior is similar to the one of $\G{s}{s}$. However for later times the damping of the modes sets in earlier than for $\G{s}{s}$, such that for $\kappa\gtrsim 10$ there are already no visible deviations from zero any more. Due to the damping of these functions we see that at $\tilde{w}\sim 1$, when the hydrodynamic regime sets in, again only long wave-length modes will survive. Moreover, the absence of oscillations in the spectrum signals the transition from propagating to diffusive behavior of the charge perturbations, as will become evident in coordinate space. 

We note that these results are obtained for perturbations around zero density. Therefore further studies are necessary to clarify the impact of perturbations around non-vanishing densities. In particular, considering non-vanishing densities will remove the degeneracy between the flavours leading to interesting phenomena like cross-diffusion \cite{Fotakis:2022usk, Fotakis:2021diq, Greif:2017byw}.

\subsection{Green's Functions in Coordinate Space}\label{sec:GreensFunctionsInCoordinateSpace}
So far we computed the Green's functions in Fourier space, which provides useful insight into the underlying physics and dynamics of such far-from-equilibrium systems. However, the Green's functions in position space will provide additional and useful information to understand the system's evolution. 
\begin{figure}[t!]
\begin{center}
\includegraphics[width=0.49\textwidth]{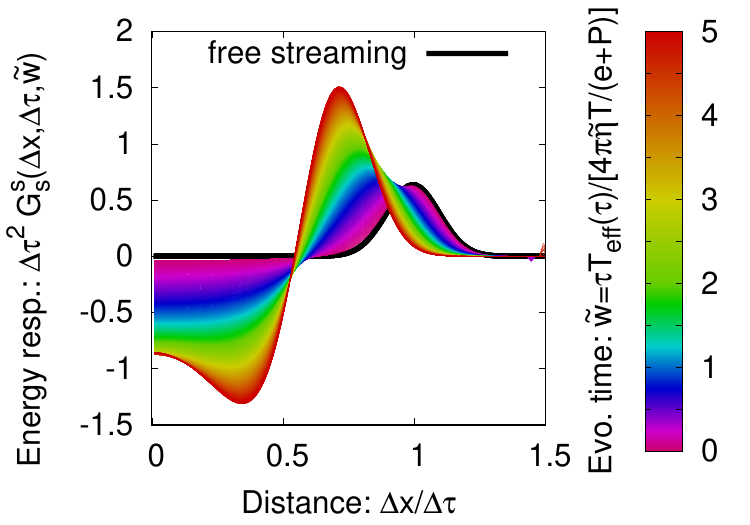}
\includegraphics[width=0.49\textwidth]{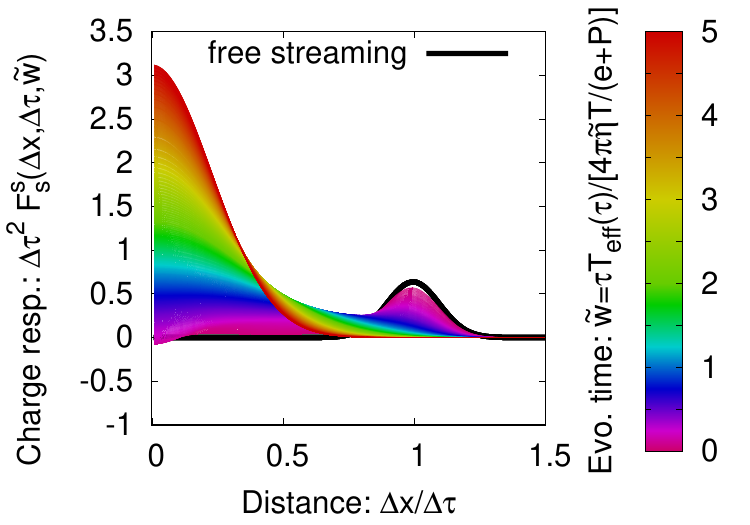}
\caption{\textsl{Left:} Evolution of the energy Green's function $\G{s}{s}$ in response to initial energy perturbations in coordinate space. \textsl{Right:} Evolution of the charge Green's function $\F{s}{s}$ in response to initial charge perturbations in coordinate space. The different curves in each panel correspond to different times $\tilde{w}$. \label{fig:ResponseFunctionsToInitialEnergyAndChargePerturbationsCoordinateSpace}}
\end{center}
\end{figure}
Similar to the decomposition in Fourier space, we can decompose the Green's functions in coordinate space as well into a basis of scalars, vectors, and tensors, such that for the two Green's functions we find
\begin{subequations}
\begin{align}
\GCS{\tau\tau}{\tau\tau}\pqty{\rt,\tau} &= \GCS{s}{s}\pqty{\vrt,\tau}\, ,\\
\FCS{\tau}{\tau}\pqty{\rt,\tau} &= \FCS{s}{s}\pqty{\vrt,\tau}\, .
\end{align}
\end{subequations}
for the energy-momentum tensor and charge current, respectively.
The relation to their counterparts in Fourier space is given by the following Fourier-Hankel transforms
\begin{subequations}
\begin{align}
\GCS{s}{s}\pqty{\vrt,\tau} &= \frac{1}{2\pi} \int \dd{\vkt} \vkt J_0\pqty{ \vkt \vrt } \G{s}{s} \pqty{\vkt,\tau} \, , \\
\FCS{s}{s}\pqty{\vrt,\tau} &= \frac{1}{2\pi} \int \dd{\vkt} \vkt J_0\pqty{ \vkt \vrt } \F{s}{s} \pqty{\vkt,\tau} \, ,
\end{align}
\end{subequations}
where $J_\nu$ are the Bessel functions of the first kind.

The results for the Green's functions in coordinate space are presented in Fig. \ref{fig:ResponseFunctionsToInitialEnergyAndChargePerturbationsCoordinateSpace}. The corresponding Green's function is plotted as a function of $\Delta x/\Delta\tau$, i.e. the propagation distance in units of the elapsed time, while the color coding indicates the evolution time $\tilde{w}$. \\
In the evolution of $\GCS{s}{s}$ the propagation of sound waves is clearly visible. In the free streaming evolution and also still at early times the waves propagate with (almost) the speed of light. Towards later times, the peak shifts to smaller values of $\Delta x/\Delta\tau$ approaching the speed of sound $c_{s}=\sqrt{1/3}$ and exhibits a negative contribution at small $\Delta x/\Delta\tau$
which corresponds to the diffusion wake.  \\
Since at early times the net charge density is carried by free-streaming particles, the charge response $\FCS{s}{s}$ has the same free-streaming behavior. One observes, that this behavior of the charge Green's function $\FCS{s}{s}$ persists up to $\tilde{w}\sim 0.5$. Subsequently, the Green's function transitions to a different behavior, where one can clearly see the diffusion of charges that results in a pronounced peak centered around $\Delta x/\Delta\tau=0$, and no longer the free propagation of charges.

\section{Applications \label{sec:Applications}}
Now that we have obtained both the energy-momentum and charge dependent Green's functions, we can couple them to ICCING initial conditions. The upgraded version of ICCING 2.0 will be available on \hyperlink{https://github.com/pcarzon/ICCING}{GitHub}\footnote{https://github.com/pcarzon/ICCING} upon publication. 
We will study the impacts of combining linearized pre-equilibrium Green's functions with initial geometries produced by ICCING.  As we will show, the assumption of linear response 
leads to nontrivial effects on the resulting energy and charge perturbations. 

\subsection{Using Green's Functions in ICCING}
The starting point of the construction of an initial state profile of the energy-momentum tensor and the conserved charges, is the generation of an initial energy density profile based on the initial-state model TRENTO.\footnote{In practice, the output of TRENTO is a ``reduced thickness function'' which is taken to be proportional to the entropy density.  The coefficient of proportionality is fixed by comparison to experimental multiplicities in central collisions, and the properly-normlaized entropy density is converted to an energy density using the lattice QCD based equation of state from Refs.~\cite{Borsanyi:2013bia, Alba:2017hhe}.} Then, this energy density at $\tau_0$ is used to compute the background energy density at any proper times $\tau > \tau_0$. To describe the fluctuations of conserved charges about this background, we use ICCING \cite{Carzon:2019qja,Martinez:2019jbu}.  The ICCING algorithm is a Monte Carlo event generator run subsequent to TRENTO which simulates the fluctuations due to $g\rightarrow q\Bar{q}$ pair production. In particular, ICCING generates $2+1D$ distributions of the fluctuating BSQ charge densities: baryon number, strangeness, and electric charge.  By incorporating the Green's functions into the energy and charge redistribution algorithm in ICCING, we can study the impact of perturbations in both energy and charge on the evolution.

A single gluon splitting produces two types of perturbations relative to the background.  The first is a negative energy perturbation (``hole'') due to removing a gluon from the background.  The second is a positive energy perturbation, displaced relative to the gluon, which deposits the energy density corresponding to the quark/anti-quark pair.  All three (quark, antiquark, and gluon) can be treated as perturbation around the background.  The propagation of the perturbations generated by ICCING are treated via the Green's function $\GCS{s}{s}$ until some time $\tau_\text{hydro}$ when hydrodynamics becomes applicable. 

In addition to the perturbations in the energy densities, the charge densities of quarks created by the gluon splitting process can also be evolved by the same formalism. Since TRENTO does not provide any charge information, we can treat the quark charges as perturbations around a vanishing background charge density, which is exactly how the charge Green's functions are constructed.
First we evolve the background according to the energy attractor $\mathcal{E}(\tilde{w})$ using Eq.~\eqref{e:attractor2}, then we can use the Green's functions $\GCS{s}{s}$ and $\FCS{s}{s}$ in order to describe the propagation of energy and respectively charge perturbations that occur whenever a splitting happens. \\
We can compactly express this as
\begin{subequations}\label{eq:FullPropagationViaGF}
\begin{align}
e\pqty{\thydro, \xt} &= e_\BG\pqty{e_\text{Trento}(\tau_0),\xt} \Big[ 1 + \int_\odot \frac{\dd[2]{x_0}}{\pqty{\Delta\tau}^2} \pqty{\Delta\tau}^2 \GCS{s}{s}\pqty{\frac{\abs{\xt-\xt_0}}{\Delta\tau}, \tilde{w}} \frac{1}{e_\text{Trento}(\tau_0,\xt_0)} \nonumber \\
&\qquad \times \pqty{\delta e_q(\tau_0,\xt_0) + \delta e_{\overline{q}}(\tau_0,\xt_0) - \delta e_g(\tau_0,\xt_0)}\Big] \, , \\
n_i\pqty{\thydro, \xt} &= \int_\odot \frac{\dd[2]{x_0}}{\pqty{\Delta\tau}^2} \pqty{\Delta\tau}^2 \FCS{s}{s}\pqty{\frac{\abs{\xt-\xt_0}}{\Delta\tau}, \tilde{w}} \frac{\tau_0}{\tau} \bqty{\delta n_q(\tau_0,\xt_0) - \delta n_{\overline{q}}(\tau_0,\xt_0)} \, .
\end{align}
\end{subequations}
where we integrate the Green's function evolution over all $\xt_0$ in the past causal light cone $|\xt_0 - \xt| < \Delta \tau$. 

Furthermore $\xt$ is the point of interest and $\Delta\tau\equiv\thydro-\tau_0$. We have implemented the pre-equilibrium evolution given in Eq.~\eqref{eq:FullPropagationViaGF} in a new C++ class GreensFunctions.h which interacts with the Event class Event.h in nontrivial ways and can be found at the GitHub link above upon publication.

%
%
\begin{figure}
    \includegraphics[width=0.85 \textwidth]{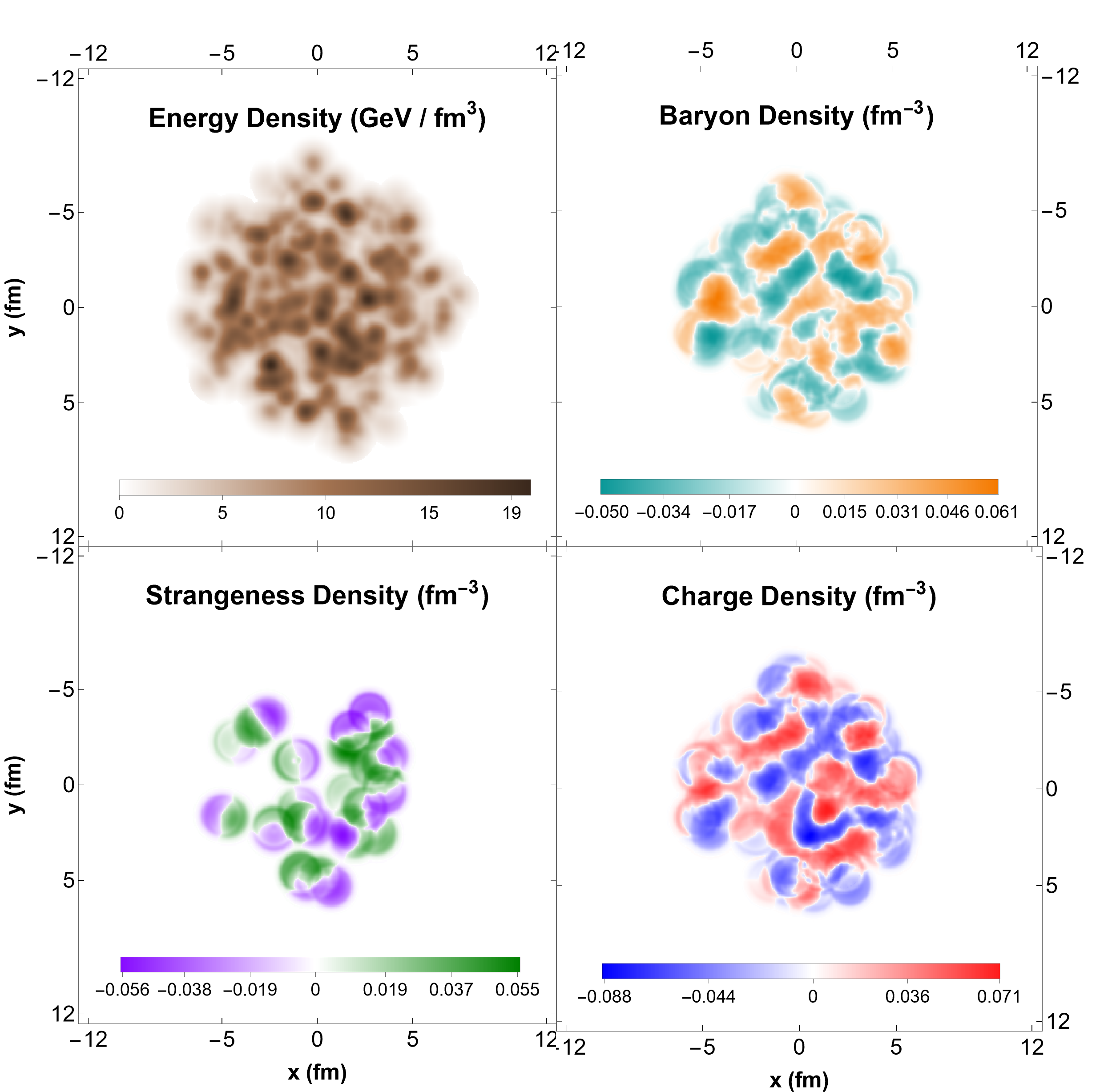}
	\caption{Density distributions for an ICCING event with Green's function evolution of energy and charge perturbations from $g \to q\bar{q}$ splittings after 1 fm/c of evolution.}
	\label{f:FullEventDensities}
\end{figure}
%
%

The energy and BSQ charge densities of a single, peripheral ICCING PbPb event at $\sqrt{s_{NN}} = 5.02 \, \mathrm{TeV}$ evolved for $1$ fm/c using the Green's functions are shown in Fig.~\ref{f:FullEventDensities} for our default parameter set (see Table 2 in \cite{Carzon:2020xwp}, with the exception of $\tau_0$ which here equals 0.1 fm).

The behavior of the baryon and electric charge distributions are similar to default ICCING \cite{Carzon:2020xwp} and follow the bulk geometry while the strangeness distribution is more rarefied due to the larger mass threshold required to produce strange quarks. A significant difference is seen in the size of the charge fluctuations, whose radius now  depends on the evolution time.  In order to understand the effects of applying the Green's functions, we can look at individual quark splittings, the background evolution, and radius dependence separately.

%
%
\begin{figure}
    \begin{centering}
	\includegraphics[width=0.85 \textwidth]{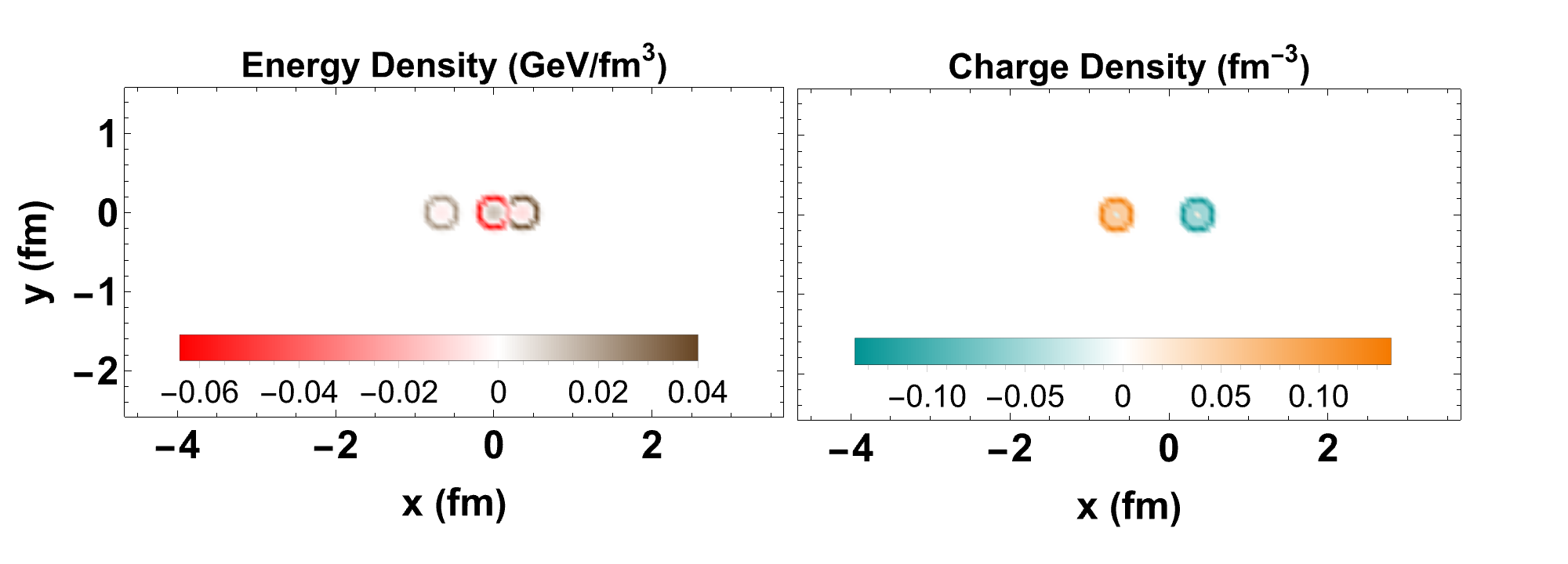} \,
	\includegraphics[width=0.85 \textwidth]{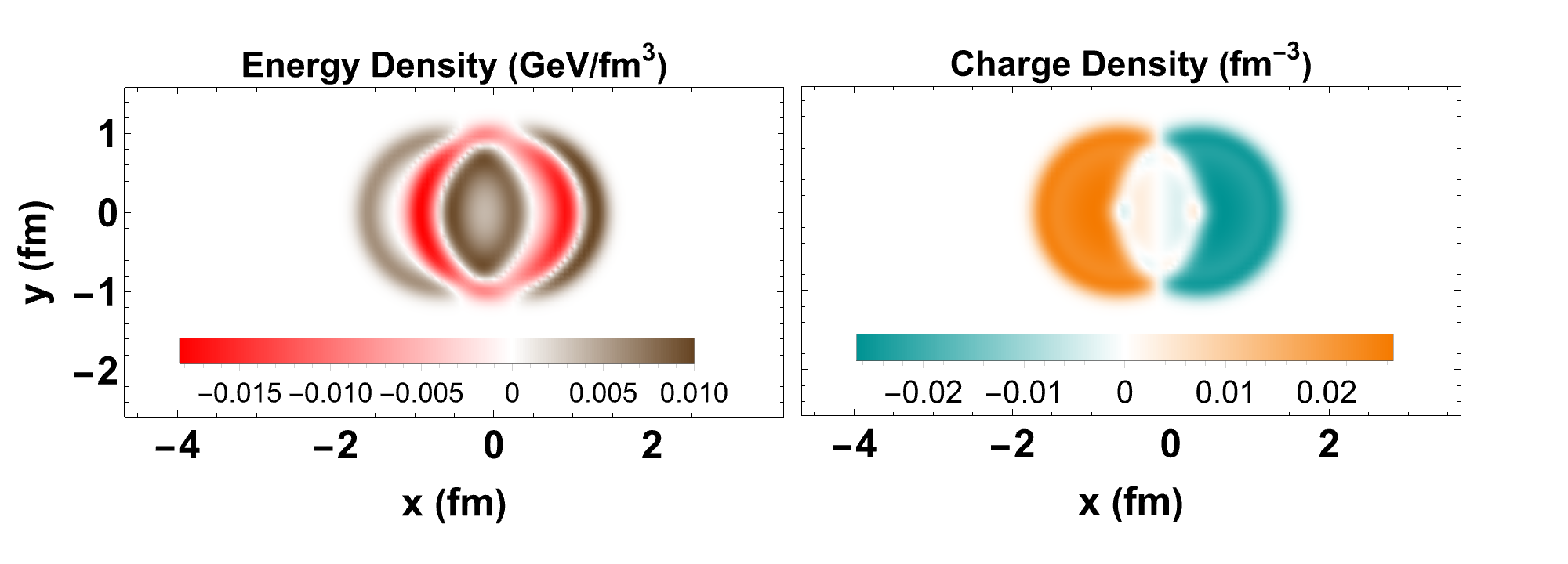}
	\end{centering}
	\caption{Density distributions for a single strange quark splitting compared for evolution times of $\tau_{hydro} = 0.2$ fm, on the left, and $\tau_{hydro} = 1$ fm, on the right. }
	\label{f:SingleQuarkEvolution}
\end{figure}
%
%

To start, it is important to have a grasp on the full effect the evolution has on a single quark/anti-quark pair. The energy density and charge density for a quark splitting evolved for $0.2  \: \mathrm{fm/c}$ and $1 \: \mathrm{fm/c}$ are shown in Fig.~\ref{f:SingleQuarkEvolution}. The top panel of Fig.~\ref{f:SingleQuarkEvolution} occurs soon after the quark splits and the bottom panel is at the end of the evolution. For small evolution times, we clearly see three different types of perturbations in the top row of Fig.~\ref{f:SingleQuarkEvolution}: (mostly) positive-energy perturbations corresponding to the deposition of the $q \bar q$ pair, and a (mostly) negative-energy perturbation corresponding to the subtraction of the parent gluon from the background. The quarks also come with associated positive/negative charge densities being deposited, whereas the gluon subtraction has no impact on the charge densities.

The dominant effect seen in Fig.~\ref{f:SingleQuarkEvolution} is that the energy and charge perturbations grow in size over time and have a wave-like structure leading to non-trivial interference. Note that the central positions of the quarks do not change due to the evolution prescribed by the Green's functions; rather, they are determined from the $g \rightarrow q \bar q$ splitting function used by ICCING. The Green's functions do not interact with any part of the quark sampling algorithm in ICCING and only determine how the energy and charge densities of the perturbations are distributed.

%
%
\begin{figure}
    \begin{centering}
	\includegraphics[width=0.85 \textwidth]{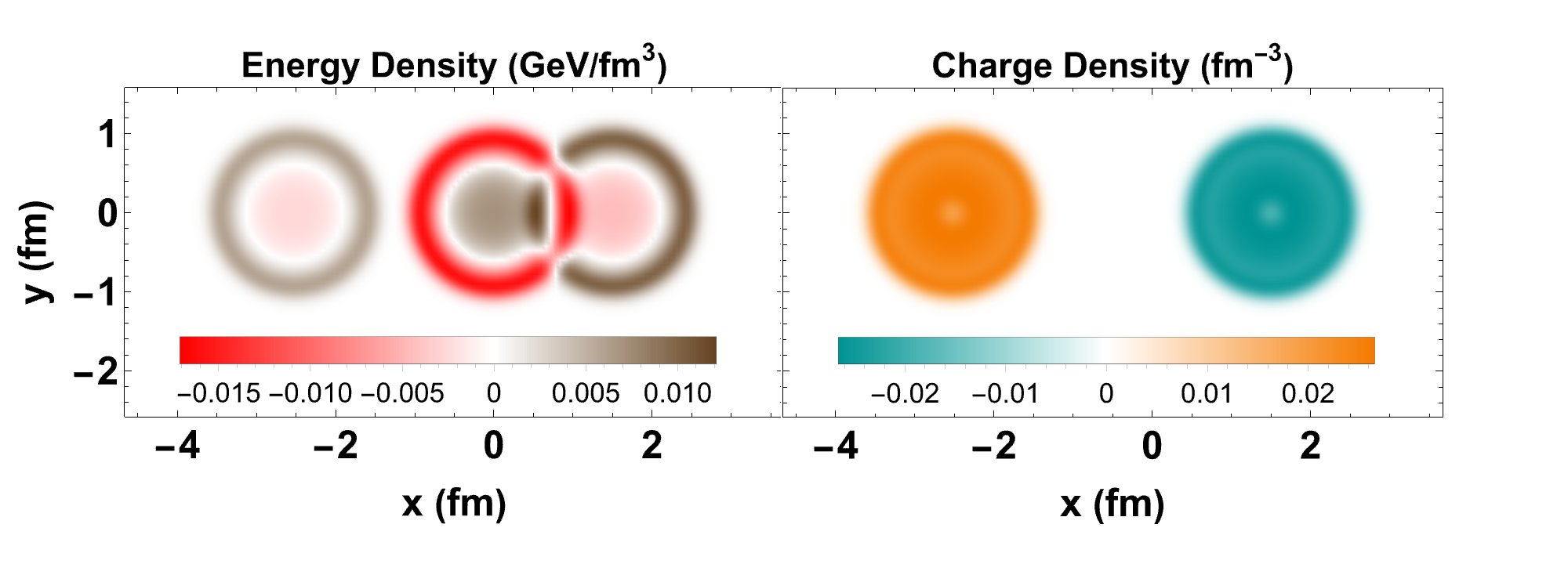} \,
	\includegraphics[width=0.85 \textwidth]{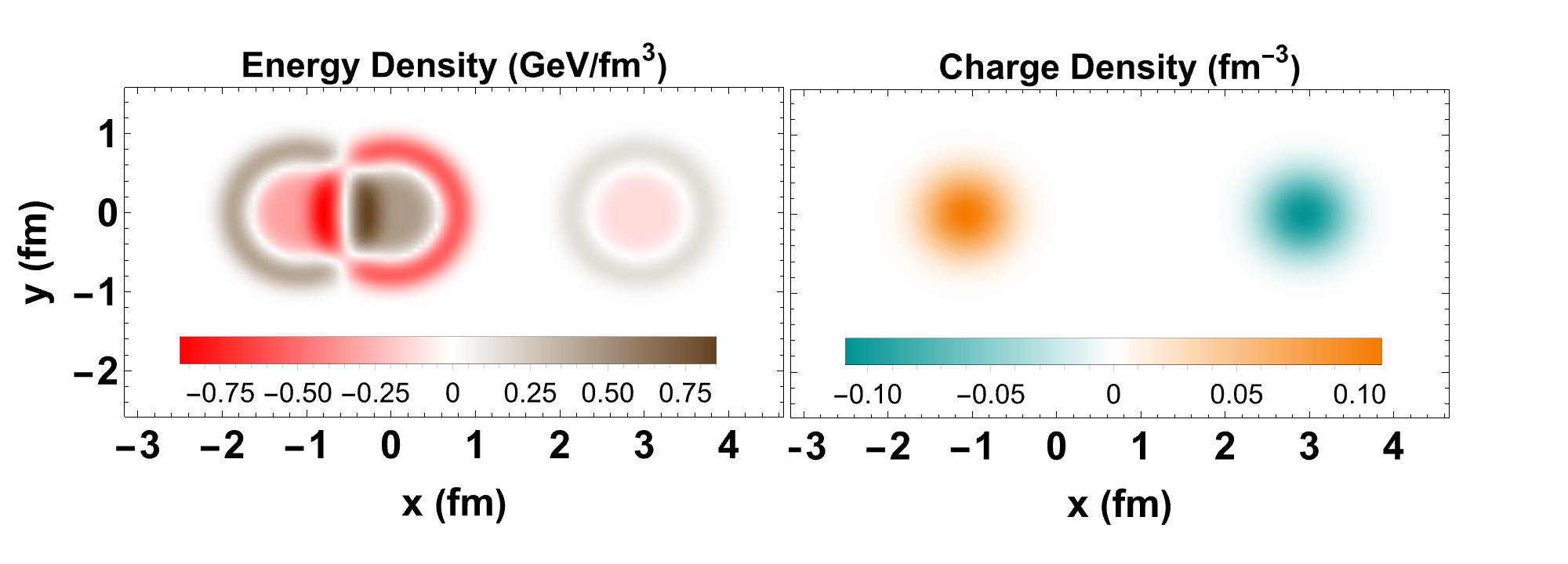}
	\end{centering}
	\caption{Density distributions for a single strange quark splitting from different areas of the event: a cold region on the top, and a hot region on the bottom. Here the separation of the two quarks is artificially increased to better illustrate the behavior of the Green's functions.  }
	\label{f:TempQuarks}
\end{figure}
%
%

Next, we can look at how the Green's functions distribute the energy and charge for the quarks.  To illustrate the spatial profiles of energy and charge density produced by the Green's functions, Fig.~\ref{f:TempQuarks} shows the results of a single $g \rightarrow q \bar q$ splitting in a low-temperature region (top) and a high-temperature region (bottom).  The Green's functions have different behavior depending on the local value of the initial energy density $e(\tau_0)$. This dependence arises because the natural unit of time $\tilde{w}$ depends on the effective temperature $T$ (see Eq.~\eqref{eq:wTilde}).  As a result, splittings which occur in hot spots transition more quickly from propagating behavior for small $\tilde{w}$ to diffusive behavior for large $\tilde{w}$.  This is clearly seen in the charge density plots of Fig.~\ref{f:TempQuarks}.  The top panel, where the splitting occurs at low temperatures, retains significant spatial structure of the charge distribution associated with the propagating modes of the Green's function.  But if the splitting occurs at higher temperatures (bottom panel), the charge density is distributed according to the diffusive modes from Fig.~\ref{fig:ResponseFunctionsToInitialEnergyAndChargePerturbationsKappaMode}.  This results in charge distributions which wash out the spatial structure of the Green's functions as seen in Fig.~\ref{f:TempQuarks}.

There is also a connection here to the Knudsen number ($Kn$) since $Kn=\tau_R/\tau\propto (\tau T)^{-1}$ so $\tilde{\omega}\propto Kn^{-1}$ \cite{Kamata:2020mka} such that at late times one expects a smaller $Kn$ number. This provides physical intuition for the dependence of the charge fluctuation on the location of splitting, hotter spots in the medium will have larger $Kn$ and thus produce more Gaussian charge densities while cold spots will have smaller $Kn$ and produce charge densities in a shock wave form. The implications of this difference in behavior based on the location of splitting may become more important when analyzing events across systems of different energy.  

While this result is interesting and an effect of the physics included in the Green's functions, it could be worrying since one of the core assumptions of the ICCING algorithm is that all charge must be correlated with some energy.  The problem with a linearized treatment of the Green's functions is that large negative corrections to the energy density could overturn the background energy density, resulting in grid points with net negative energy.  This problem would be nonphysical and require some sort of remedy.  Moreover, even if the net energy density is not driven negative by a large perturbation, one may still be unable to match a fluid cell with very low energy density but very high charge density to a reasonable equation of state.

These potential problems arising from large perturbations could be solved by going back to the linearized approximation made in Sec.~\ref{sec:Theory_PerturbationsAroundBjorkenFlow}, which is broken when the local redistributed energy or charge density is greater than or close to the background. There are several ways in which to solve this problem, the first of which would be by artificially damping the magnitude of the perturbations relative to the background in order ensure that the linearization remains valid. This would introduce a new problem though, since the artificial damping may not affect a $q \bar q$ pair equally.  If the quark is deposited in the periphery of the event, but the antiquark is deposited closer to the center, then the positive and negative charge densities will be damped in different amounts, leading to a violation of charge conservation.  To correct this, one could suppress the quark and anti-quark in the same way mirroring the effect. 

Another possible solution -- the one we pursue here -- is to veto any quark splittings that would create energy density perturbations that are large with respect to the background. This is a very simple solution but adds a new complication because it effectively eliminates quark production at the edge of the event, and reducing the "cold quark" Green's functions contribution.  Another unintended effect of this solution would be that a quark/anti-quark pair produced near the periphery with a smaller radius, for example from evolving for only $0.5 \: \mathrm{fm/c}$, would survive the veto, but a pair evolved for longer time would be rejected. Evidently, this problem could be cured by also including the transverse expansion of the background energy density in the pre-equilibrium stage, but that is beyond the scope of this paper and is left for future work. Despite its shortcomings, the solution of vetoing quark splittings if the energy perturbation is not small compared to the background has been chosen here both for its simplicity and its flexibility.

%
%
\begin{figure}
    \begin{centering}
	\includegraphics[width=0.45 \textwidth]{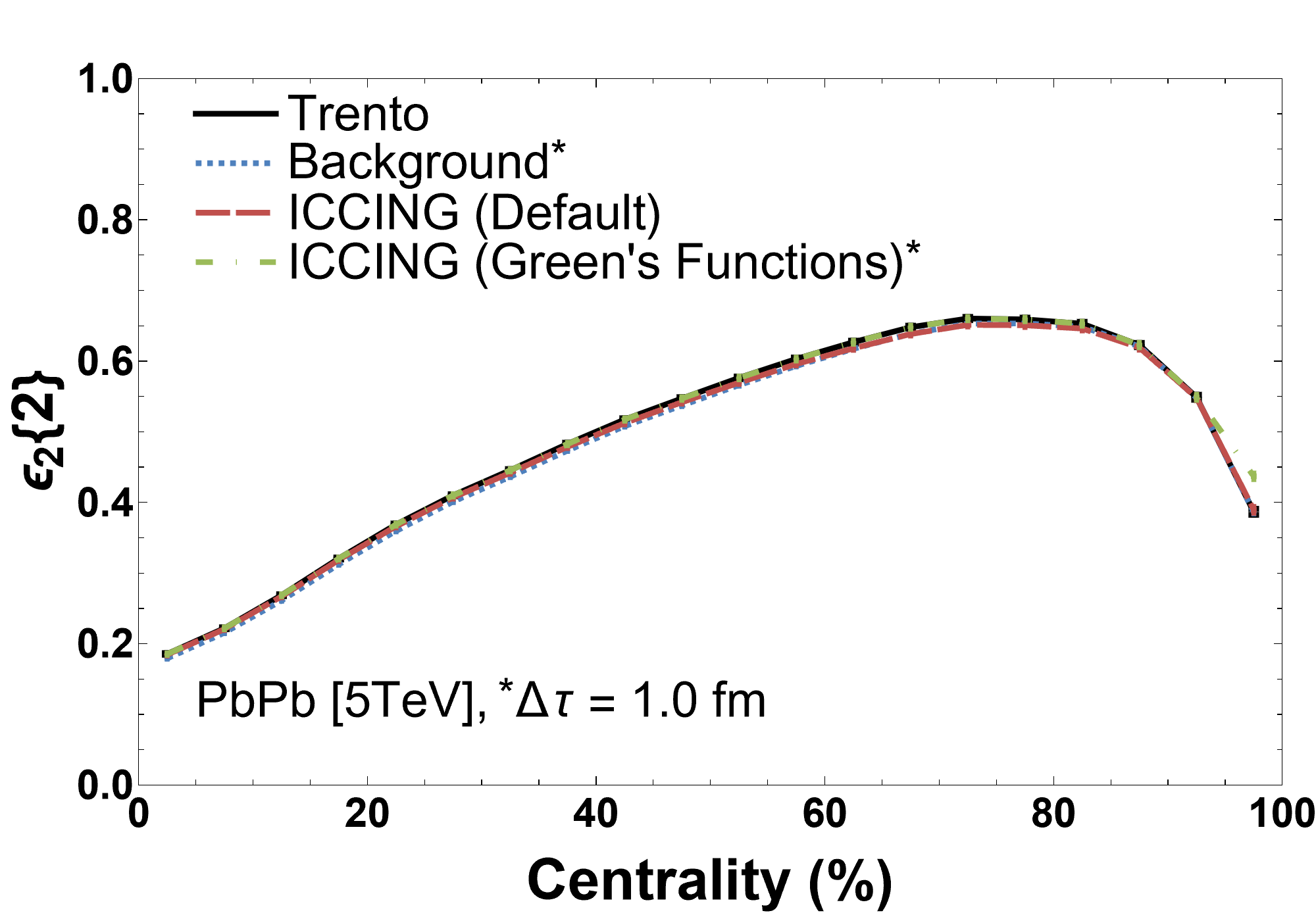} \,
	\includegraphics[width=0.45 \textwidth]{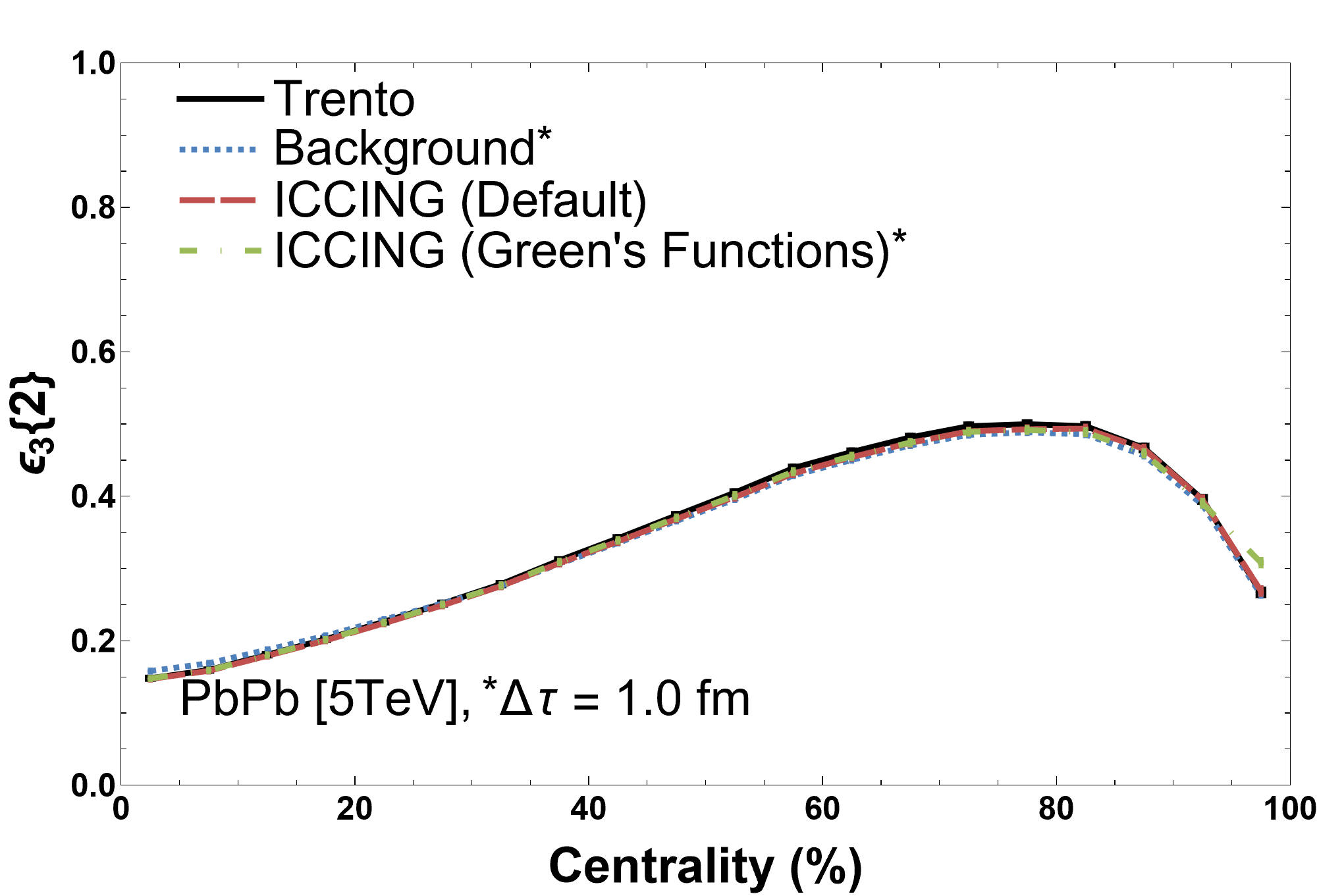}
	\end{centering}
	\caption{Comparison of $\varepsilon_n \{2\}$ across energy and BSQ distributions for different Green's function evolution times.}
	\label{f:BackgroundGeometry}
\end{figure}
%
%

Because our procedure should be only a small effect on the total energy density, we do not expect large changes to the energy density eccentricities, which are defined in App.~\ref{app:Eccentricities}. Thus, before looking at the eccentricities of the charge densities, we should look at the effect that the different processes have on the energy density with the hope that any affect is minimal. Since the energy density eccentricities are a good predictor of the final state and these initial conditions have been used extensively in comparisons to experimental data, the hope is for a minimal effect. In Fig.~\ref{f:BackgroundGeometry}, the energy ellipticity and triangularity is plotted for the original trento event, the locally evolved background used for the Green's function evolution, the trento event after default ICCING, and the full ICCING coupled to Green's functions simulation. When comparing the evolved background with the trento profile, we observe small changes at the percent level which can be attributed to the phenomenon of inhomogenous longitudinal cooling~\cite{Ambrus:2021fej,Ambrus:2022qya}. In essence, thermalization proceeds more quickly in more highly energetic regions, leading to a slightly faster decrease of the energy density of the hotter regions of the QGP as compared to the colder regions of the QGP. However, as we see in Fig.~\ref{f:BackgroundGeometry}, in practice this effect is rather small. Similarly, we see that for default ICCING, there is a slight modification in peripheral events and the most central events that should not have a significant effect on the agreement with experimental data. 
Adding both the modifications from the ICCING sampling and the Green's functions evolution, has no significant effect on the energy geometry beyond the background evolution. This indicates that any small changes in energy density distribution generated by ICCING are quickly washed out and won't make it to the final state. Additionally, previous comparisons to experimental data for all charge particles should still be valid.

\section{Impact on Eccentricities}\label{sec:Results}

Now let us look at the contributions, that different parts of the Green's function evolution have on the event averaged eccentricities. Because we will be dealing with time dependent quantities, we will define the time evolution for applying the Green's function:
\begin{equation}
    \Delta \tau\equiv \tau_{hydro}-\tau_0
\end{equation}
where $\tau_0$ is our initial time when we begin the Green's function evolution and $\tau_{hydro}$ is where we stop the evolution and switch to hydrodynamics.  We will start with studying the consequences of our perturbative cutoff effect on the eccentricities, then we will compare the Green's function expansion to a trivial Gaussian smearing to determine any non-trivial effects.  After these two effects are studied we will explore the time dependence of the Green's function on various eccentricities, which are the main results for this work. 

Eccentricities of charge are defined the same as for energy, see App.~\ref{app:Eccentricities},  except that the center of mass is taken to be that of the energy density and the observable is calculated for the positive and negative charge densities separately, since otherwise the observable would be zero. When there is no charge density from quark/anti-quark splittings, the eccentricity is defined as zero. While adequate, the definition of the eccentricities for energy are not the best possible estimators for the the charge density and more development can be made in this direction \cite{Carzon:2020xwp}. 

\begin{figure}
    \begin{centering}
	\includegraphics[width=0.45 \textwidth]{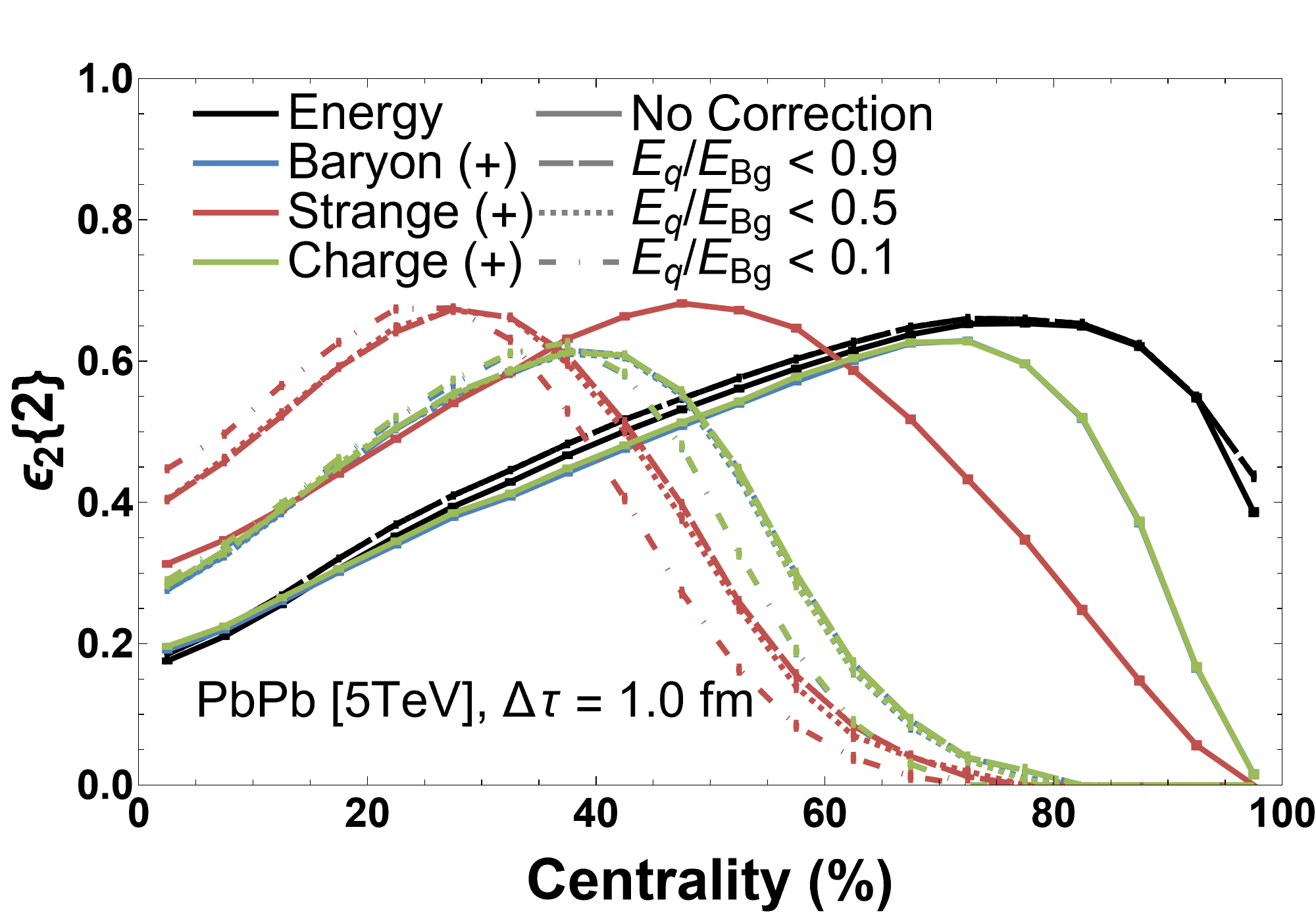} \,
	\includegraphics[width=0.45 \textwidth]{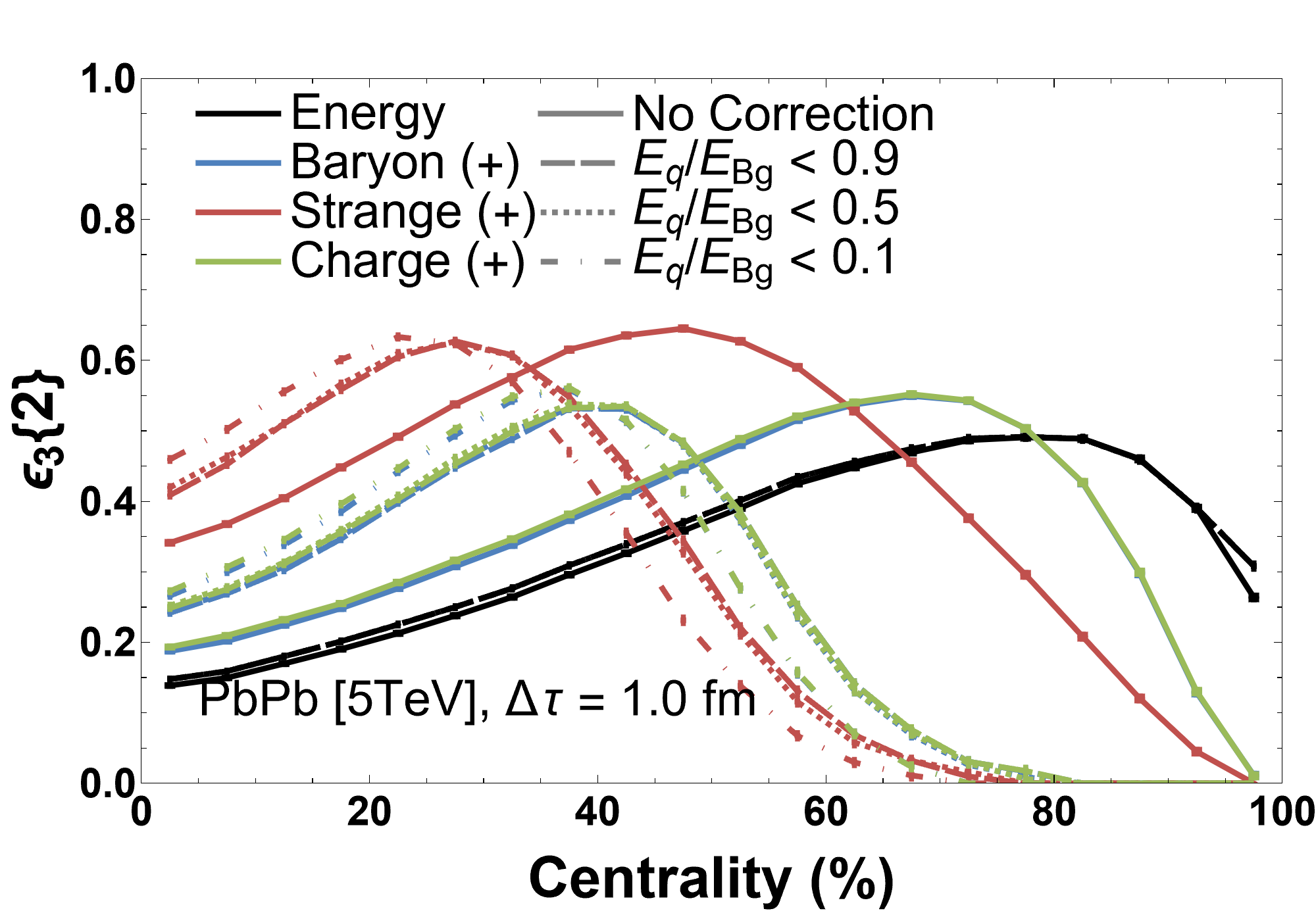}
	\end{centering}
	\caption{Comparison of $\varepsilon_n \{2\}$ for different perturbation cutoff values with a Green's function evolution of $\Delta \tau = 1.0$ fm/c. The solid line is with a Green's function but no cutoff, default ICCING is not shown. }
	\label{f:PerturbativeCutoff}
\end{figure}

First, we study the effect of the suppression of gluon splitting to ensure positive energy densities. In order to avoid problematic regions with negative energy density, we restrict quarks from splitting if their redistributed energy densities approach a certain threshold compared to the background. The selection criteria is examined for each point in the quark densities and is determined by
\begin{subequations}
\begin{align}
E_q/E_{bg} < P,
\end{align}
\end{subequations}
where $P$ is the perturbative cutoff. In Fig.~\ref{f:PerturbativeCutoff}, several values were selected for an evolution time of $\Delta\tau=1$ fm/c  to illustrate the effect this cutoff has on the charge densities. Both $\varepsilon_2\left\{2\right\}$ and $\varepsilon_3\left\{2\right\}$ are shown. The effect of applying the perturbative cutoff vs no cutoff at all is clearly the dominate effect. This signifies that there are many quark/anti-quark pairs above 40\% Centrality that produce negative energy and thus the mismatch between the locally evolved background and non-local quark perturbations is quite significant. For $P=0.9$ and evolution of $1$ fm/c, the percentage of events that produce no quarks at all in the $65-75\%$ Centrality class is $98.9\%$. The peak of the charge eccentricities thus signifies that number effects dominate the charge geometry. With such an extreme response to this perturbation parameter it is reasonable to assume the model, as it is currently formulated, breaks down at this point and should not be used beyond an evolution time of $1$ fm/c. However, we do not find a significant difference between the most generous value of a $0.9$ cutoff  vs the $0.5$ mark with only small change when $P$ goes to $0.1$.  In the remaining results we will explore only the $0.9$ cutoff since we do not anticipate a strong dependence on the cutoff for other observables as well.

%
%
%

\begin{figure}
    \includegraphics[width=0.85 \textwidth]{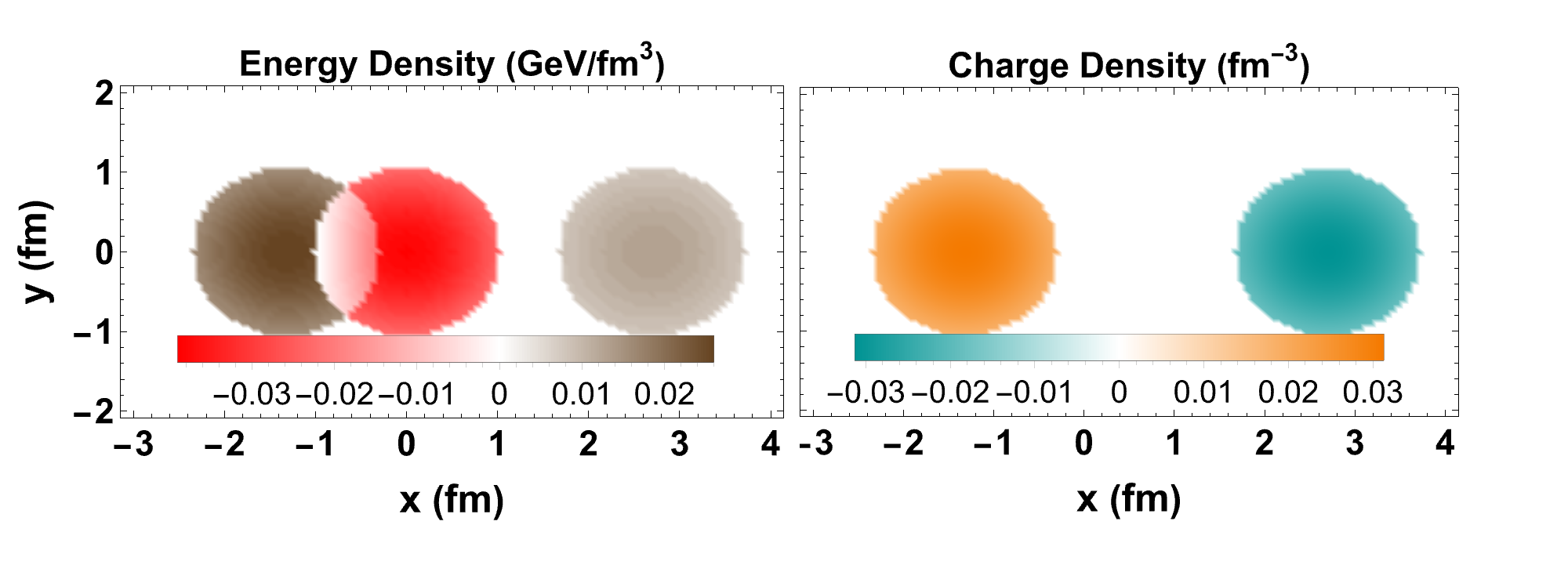}
	\caption{Illustrative density profiles of the Gaussian smearing option at $\Delta \tau = 1.0$ fm/c which separates structure introduced by the Green's functions from the radial dependence.}
	\label{f:GaussianSmearing}
\end{figure}

\begin{figure}
	\includegraphics[width=0.85 \textwidth]{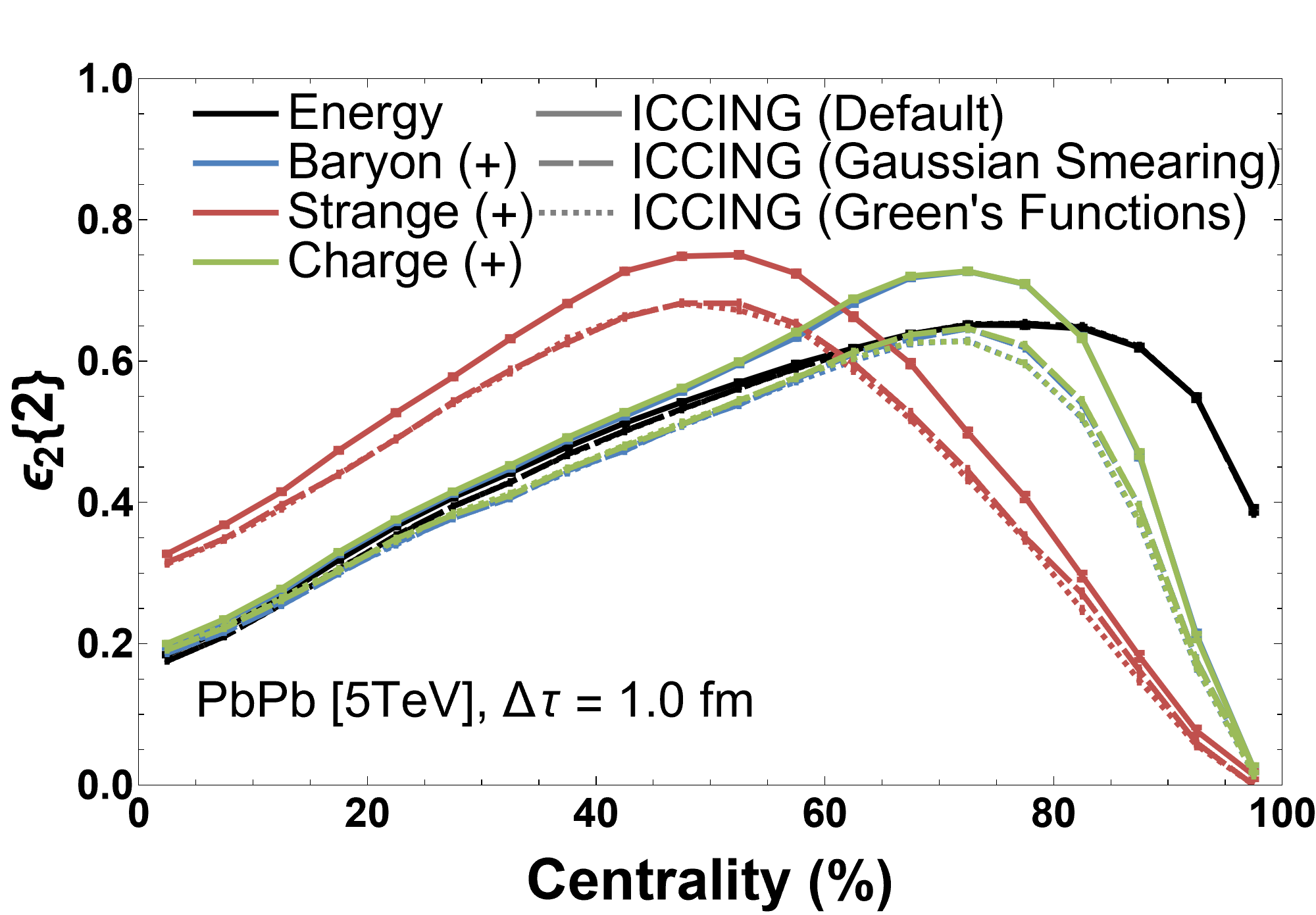}
	\caption{Comparison of $\varepsilon_2 \{2\}$ between default ICCING, Gaussian Smearing, and Green's Functions with an evolution of $\Delta \tau = 1.0$ fm/c. The perturbation cutoff is not used here. }
	\label{f:GaussianSmearingEccs}
\end{figure}

Next, we will try to disentangle the effect of the expanding radius from the structure introduced to the quark densities based on the background energy. In our Green's function approach, the overall size of the quarks expand over time and that may be the dominate (albeit trivial) effect of applying the Green's function. Thus, to determine any non-trivial consequences of the Green's function, we apply a simple Gaussian smearing to the quarks, as illustrated in Fig.~\ref{f:GaussianSmearing}, and compare the Gaussian smearing, defined as:
\begin{equation}
    \GCS{s}{s}(r,t)=\FCS{s}{s}(r,t)=\frac{\exp(-r^2/R(t)^2)}{\pi R^2 (t)},
\end{equation} 
to the Green's function method. In Fig.~\ref{f:GaussianSmearingEccs}, we see that there is a negligible difference between the Green's function and a simple Gaussian smearing, implying that the dominant effect, when looking at event averaged geometry, is the size of the density perturbations and not the structure introduced by the Green's Functions.

%
%
%
%

%
%
%

\begin{figure}
    \includegraphics[width=0.85 \textwidth]{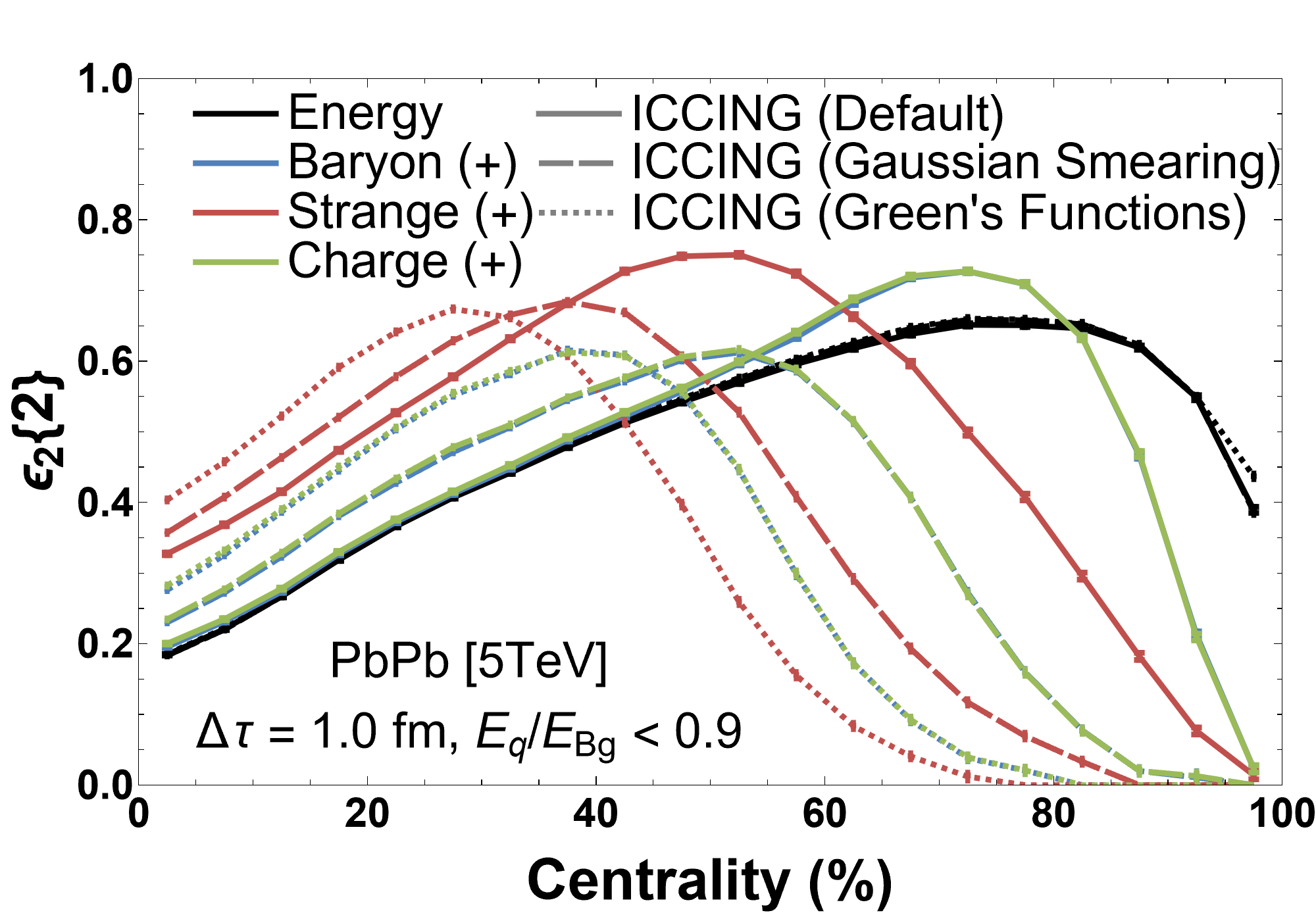}
	\caption{Including perturbation cutoff of $0.9$ for comparison between Default ICCING, Gaussian smearing, and Green's functions for $\Delta \tau = 1.0$ fm/c.}
	\label{f:RadialPerturbative}
\end{figure}

In Fig.~\ref{f:RadialPerturbative}, we show $\varepsilon_2 \{2\}$ adding back in the perturbation cutoff and evolving for $\Delta\tau=1.0$ fm/c. The solid curves are from default ICCING without any pre-evolution, the dashed curves add in evolution but only allow the radius of the quark and gluon density perturbations to change while holding the density profile fixed, and the dotted curves add in the full Green's functions. Several things are happening in Fig.~\ref{f:RadialPerturbative} that need to be disentangled. First, the Gaussian smearing with the peturbative cut-off (compared to default ICCING with no time evolution) has the general effect of shifting the peak in $\varepsilon_2\left\{2\right\}$ to lower centralities and also leading to a larger $\varepsilon_2\left\{2\right\}$ in central collisions. The shift from the Gaussian smearing compared to default ICCING occurs both because as the quarks grow in size the positive and negative densities will cancel out more and wash out the geometry in regions with low densities (i.e. peripheral collisions) and because quark/anti-quark splitting is suppressed due to the perturbation parameter.

%
%
%
%

Now that we know the effect of just a trivial Gaussian smearing, the next question is what effect does the non-trivial Green's function have? 
In Fig.~\ref{f:RadialPerturbative} we can see that the Green's function shifts the peak of the eccentricities even further to lower centralities for all BSQ densities.
To understand this effect, let us break down the fundamental differences between a trivial Gaussian smearing and the Green's functions.
There are two differences between the Gaussian smearing and Green's functions density perturbations one of which is that the density profile of the Gaussian smearing is smooth and mostly uniform with sharp edges and only negative values of energy coming from the gluon hole, as shown in Fig.~\ref{f:GaussianSmearing}. The Green's functions, on the other hand, have a density profile that has a wave structure with the largest energy density values coming from the ring at the edge of the quarks and gluons, as previously shown in Fig.~\ref{f:SingleQuarkEvolution}. 
The Green's function density profiles also contain negative energy at the center of the quarks and a large amount around the edge of the gluon. Applying the perturbative cutoff removes any net negative energy from the final output in these two methods, which strongly affects Green's function method because of the concentration of the energy density around the edge of the quarks. While the Gaussian smearing also breaks perturbative assumptions the effect is much smaller than for the Green's function. The structure of the Green's function density perturbations is relatively 'microscopic' and so when compared against the Gaussian smearing, without the perturbation cutoff in Fig.~\ref{f:GaussianSmearingEccs}, there is no difference. Since the perturbation cutoff is defined here as 'microscopic', then a difference is seen in Fig.~\ref{f:RadialPerturbative} when including the more complicated structure of the Green's functions. The sensitivity to 'microscopic' differences in the density perturbations may disappear with different choices of the perturbation cutoff method. However, the unique structure of the Green's function density perturbations will still be important when coupling to hydrodynamics since there would be a non-trivial change to gradients.

\begin{figure}
    \begin{centering}
	\includegraphics[width=0.45 \textwidth]{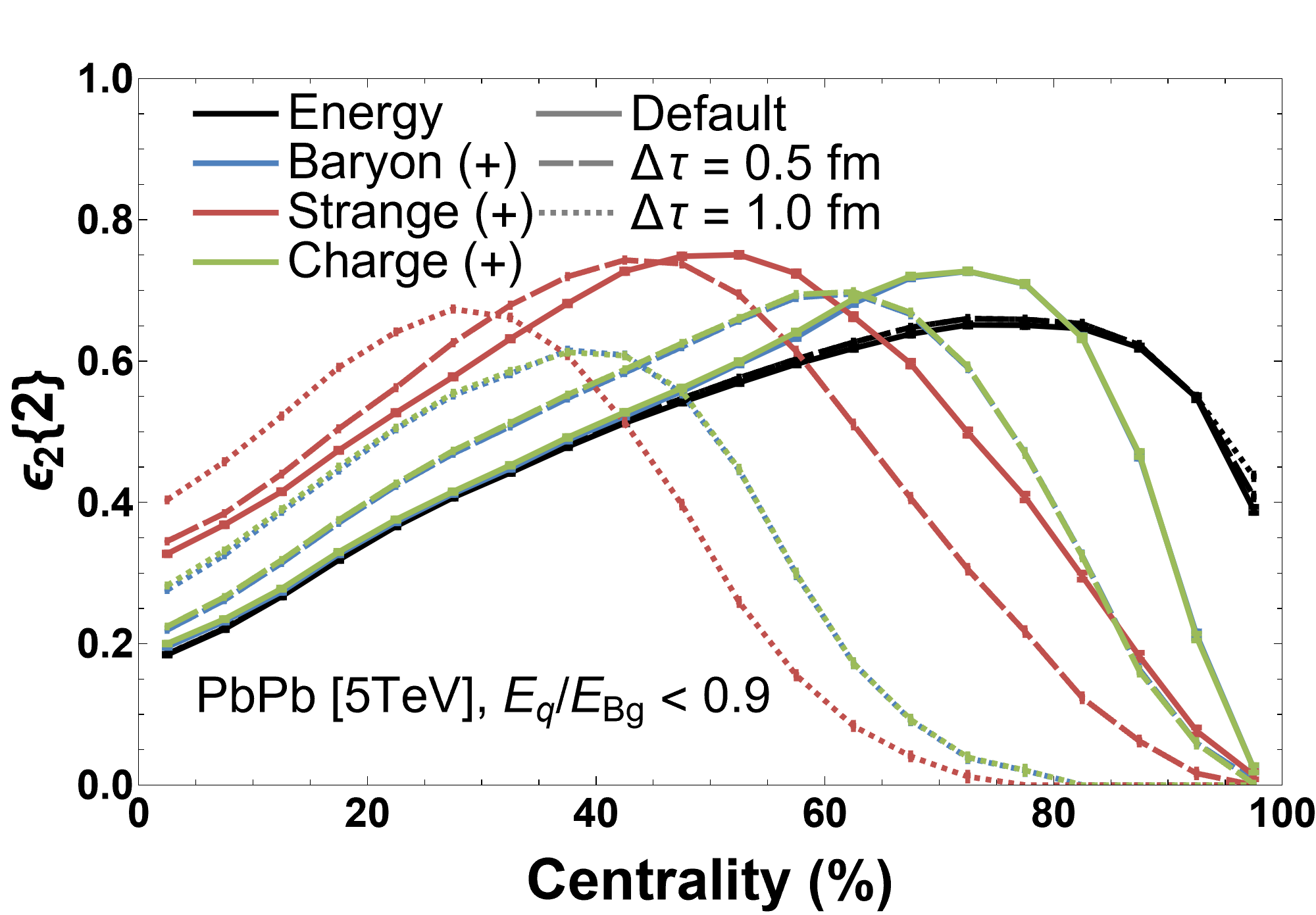} \,
	\includegraphics[width=0.45 \textwidth]{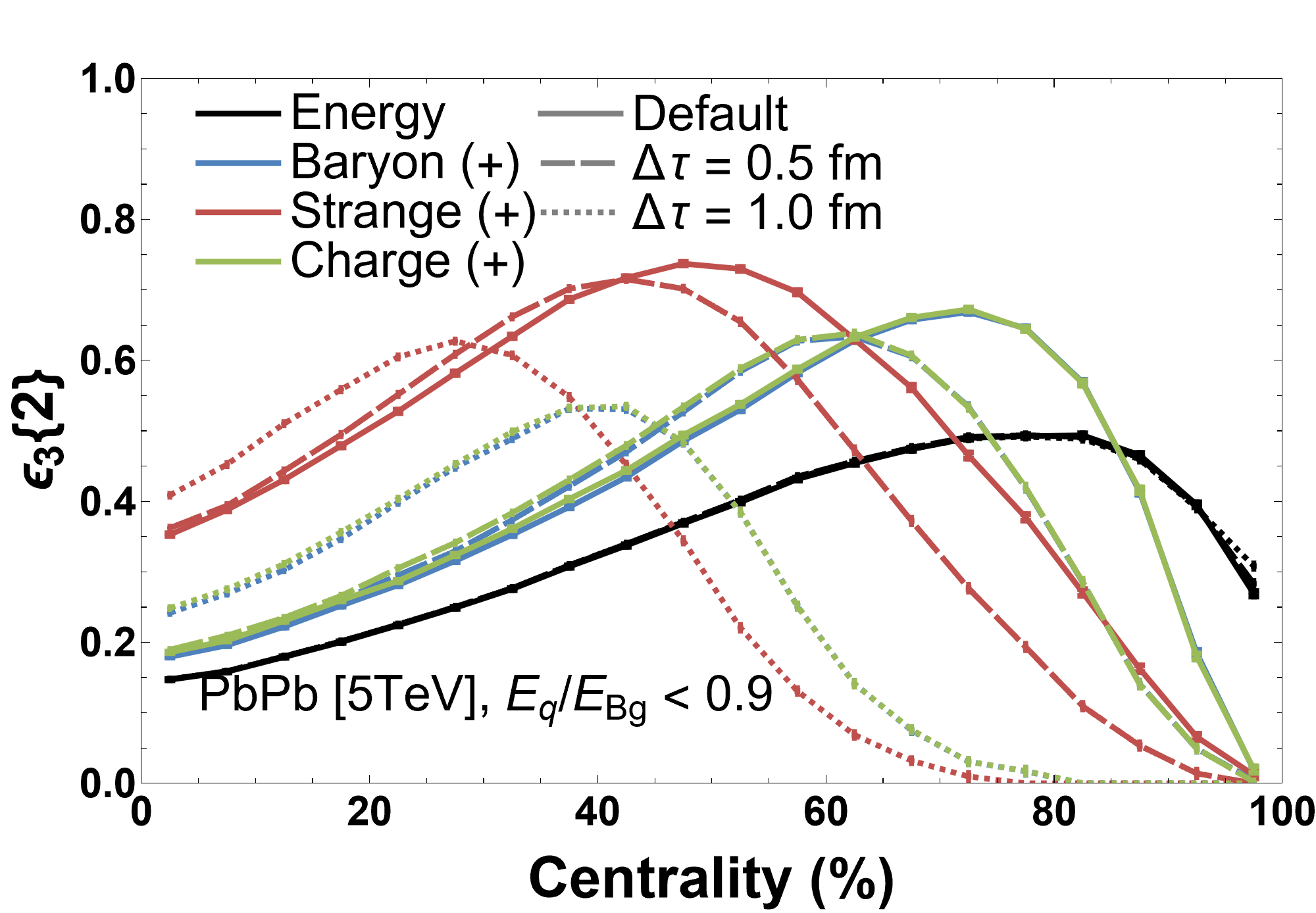}
	\end{centering}
	\caption{Comparison of $\varepsilon_n \{2\}$ across energy and BSQ distributions for different Green's function evolution times using the perturbation cutoff. }
	\label{f:EccsGreens}
\end{figure}

Finally putting all the pieces together, Fig.~\ref{f:EccsGreens} shows the Green's functions evolution with the perturbative cutoff for different evolution times, $\Delta\tau=0.5$ fm and $\Delta\tau=1$ fm, for both elliptical (left) and triangular eccentricities (right). For an evolution time of $\Delta\tau=0.5$ fm/c, we consistently see a  shift in the peak of all BSQ charge eccentricities toward the left, reflecting an increase in the dominance of number effects on the geometry as supported by the rarity of quark producing events.  However, for $\Delta\tau=0.5$ fm/c most central collisions do not appear to be strongly affected by the expansion. For an evolution of $1.0$ fm/c, there is a much greater suppression from the perturbation correction and this significantly affects all centrality classes, such that the most central collisions see enhanced eccentricities but peripheral collisions are suppressed. The shift in the peak towards smaller centrality classes (combined with a suppressed eccentricity in peripheral collisions) indicates that the model starts to break down the further in time the evolution is pushed.  Thus, there appears to be a small window in which we can apply the Green's function expansion and still obtain a reasonable number of quark/anti-quark pairs (after applying the perturbative cutoffs).  Generally, we find very similar qualitative behaviors in both $\varepsilon_2$ and $\varepsilon_3$.  However,  $\varepsilon_3$ is much more sensitive to BSQ densities and has the most significant difference between the energy eccentricities vs the BSQ eccentricities. Therefore, high-order harmonics will likely provide the best observable when comparison to experimental data.

%
%
%
%

\begin{figure}
    \begin{centering}
	\includegraphics[width=0.45 \textwidth]{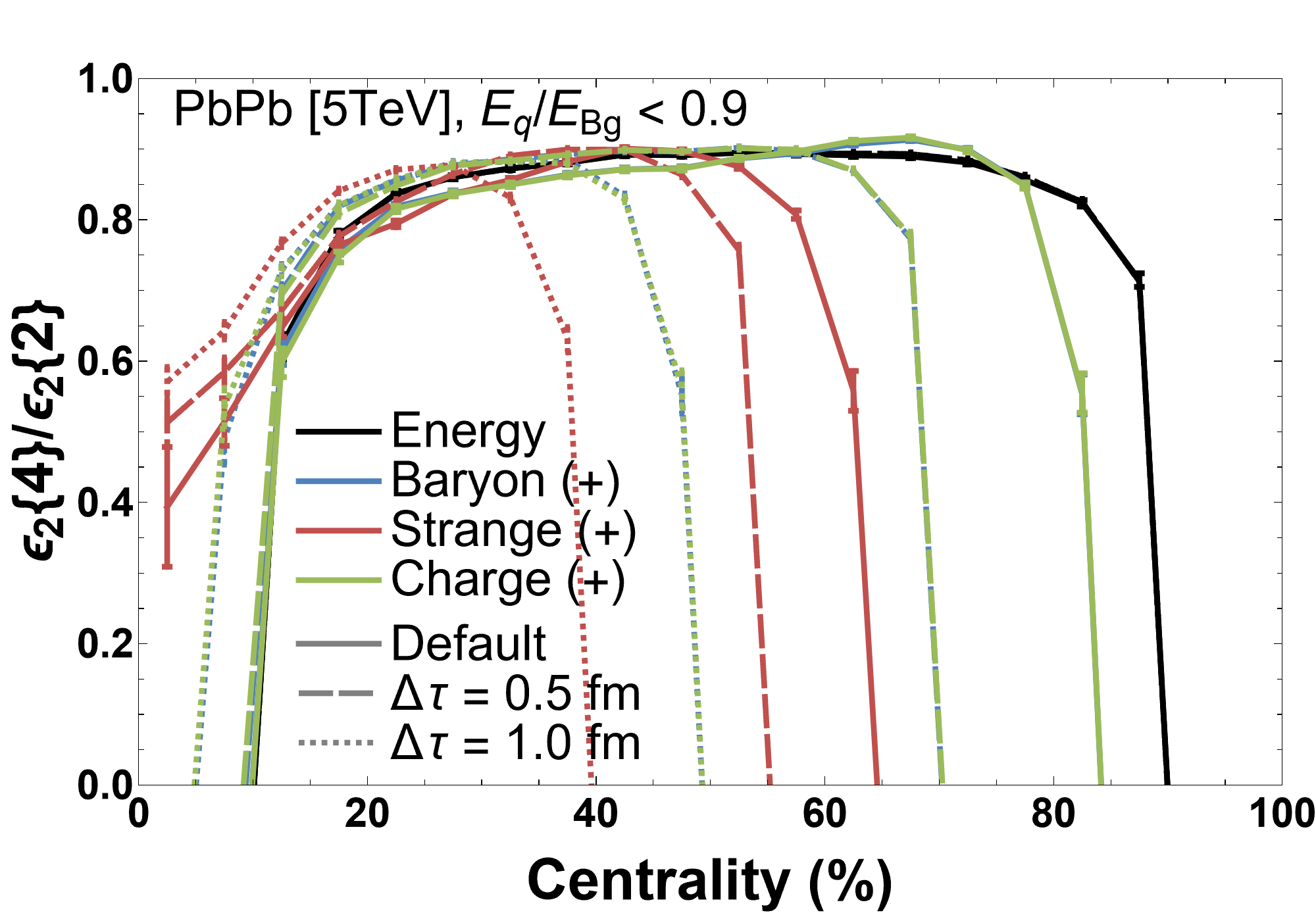} \,
	\includegraphics[width=0.45 \textwidth]{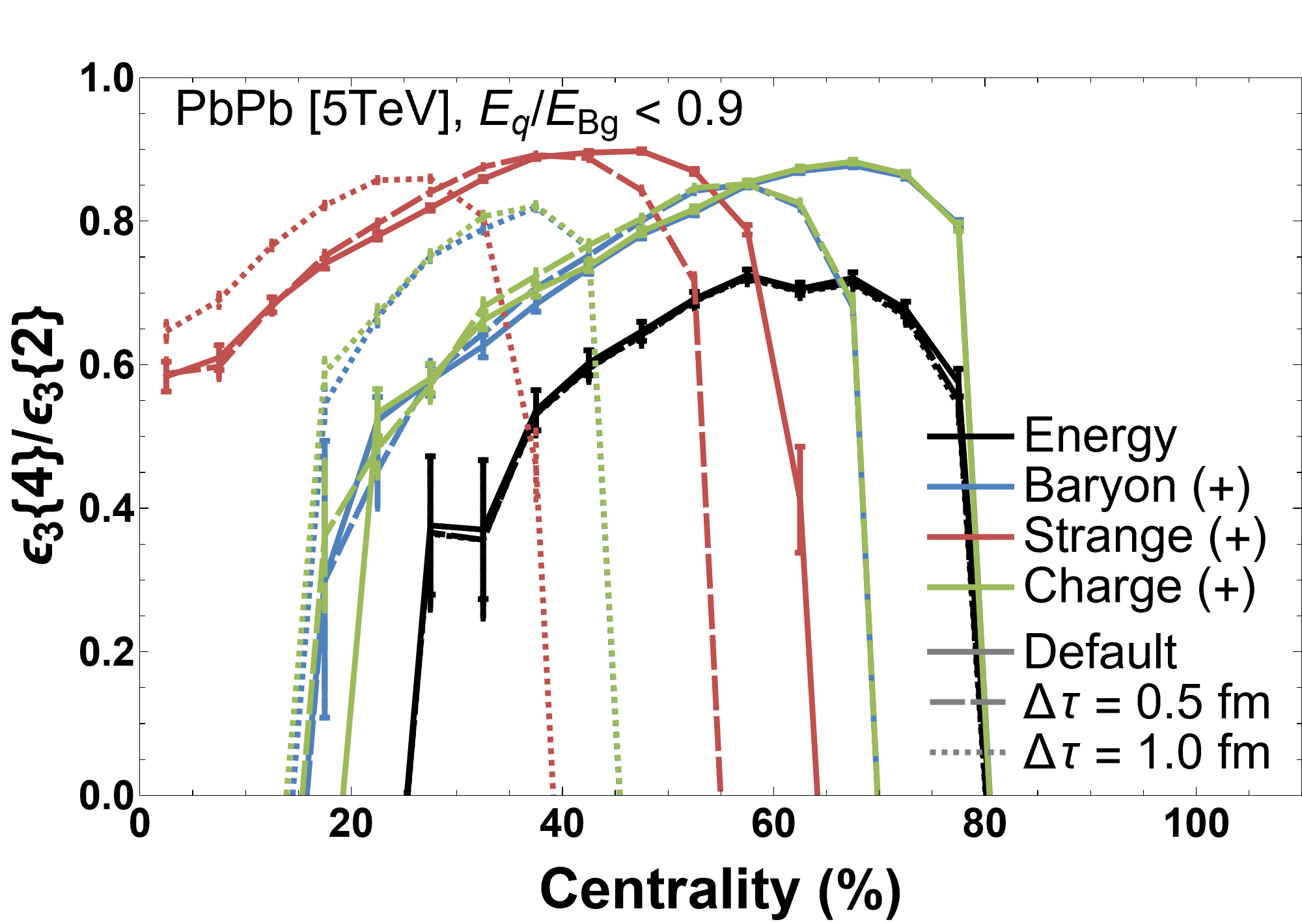}
	\end{centering}
	\caption{Comparison of $\varepsilon_n \{4\} / \varepsilon_n \{2\}$ across energy and BSQ distributions for different Green's function evolution times using the perturbation cutoff.}
	\label{f:GreensEccentricityRatios}
\end{figure}

One of the most important quantities for direct comparisons of initial state models to experimental data  is $\varepsilon_n \{4\} / \varepsilon_n \{2\}$ because medium effects cancel in the most central collisions (especially for $n=3$ \cite{Carzon:2020xwp}). 
The eccentricity ratios, $\varepsilon_n \{4\} / \varepsilon_n \{2\}$, are shown in Fig.~\ref{f:GreensEccentricityRatios} for $n=2$ (left) and $n=3$ (right). The ratio $\varepsilon_n \{4\} / \varepsilon_n \{2\}$ measure the fluctuations of geometry with values close to $1$ indicating few fluctuations wheres small values indicate a large amount of fluctuations.  Comparing elliptical and triangular flow, we find quite different results.  For elliptical flow, for more centrality to mid-central collisions the fluctuations appear to be nearly identical to the energy density fluctuations  (although ultra central collisions have some small differences).  However, for peripheral collisions where the perturbative cutoff plays a strong role, then we see there is always a centrality wherein large deviations are seen compared to the energy density distribution. Electric charge and baryon density fluctuations, for default ICCING, are nearly identical to the energy density fluctuations.  However, the longer you have a Green's function evolution, then you see deviations at lower and lower centralities (i.e. for $\Delta \tau=1$ fm, the deviation occurs at $\sim 40\%$ centrality).  Naturally for strangeness this effect is larger because one is dealing with a smaller number of quark/anti-quark pairs.

The effect for triangular flow is quite different.  Generally, we find that the application of ICCING leads to an overall decrease in the triangular flow fluctuations, regardless of the BSQ charge. Additionally, the Green's function evolution appears to enhance that effect further for $\Delta \tau$. In contrast, for elliptical flow we did not see this effect and the fluctuations were the same (at least within some centrality classes) before and after applying ICCING.  That being said, we do find that the effect of certain centralities being strongly affected by the perturbative cutoff showing up in triangular flow as well.  These are features that could eventually be looked for in experimental data, if  measurements of $v_n \{4\} / v_n \{2\}$ are made with identified particles.

%
%
%
%

\section{Conclusion \& Outlook \label{sec:Conclusion}}

We extended the method to compute non-equilibrium Green's functions of the energy-momentum tensor developed in \cite{Kamata:2020mka} by adding conserved charges and computing the corresponding Green’s functions for the charge current for perturbations around vanishing background charge densities. Using the ICCING model that initializes conserved charges through $g\rightarrow q \bar{q}$ splittings, we successfully coupled these Green's function to ICCING allowing for a pre-equilibrium phase with conserved charges. The inclusion of this pre-equilibrium evolution is a non-trivial addition to the ICCING algorithm and the successful implementation demonstrates the flexibility of the algorithm.

In order to quantify the system's response to initial perturbations in terms of Green's functions, it is necessary to consider the background evolution (see App.~\ref{app:BackgroundEvolutionInConformalSystems}) which is used in the energy evolution of the system. We find that the systems dynamics can be quantified in terms of the moments $\dE{m}{l}$, $\dN{m}{l}$ (Eq. (\ref{eq:PerturbedEandNMoments})). Furthermore the Green's functions can be obtained directly from these moments, which makes this method a powerful tool to obtain the response functions. When comparing the energy and charge Green’s function we see distinct differences in their behaviors. In the energy case we find the propagation of sound waves, where in free-streaming and even at early times they propagate with almost the speed of light, while at later times this shifts towards the speed of sound.
In contrast, for the evolution of conserved charges we find a transition from free-streaming propagation in the beginning to a diffusive behavior at late time as the system continues to thermalize.

To understand the effect the Green's functions have on initial state charge geometries, we compare between the default version of ICCING and ICCING with the Green's functions, supplemented by an approximation which simplifies the spatial structure of the Green's functions to simple Gaussian smearing. We see that for $\varepsilon_n \{2\}$ there is no difference when including the complicated structure of the Green's functions to the Gaussian smearing, although that structure becomes important for observables sensitive to 'microscopic' differences. A mismatch between the evolution of the background, described as a local process, and the charge perturbations, described as a non-local process, leads to the possibility of sites with negative energy. This issue is fixed by suppressing quark/anti-quark production that would violate some perturbative condition. This pertubative corrective measure significantly suppresses quark/anti-quark production in peripheral events but less so in central to mid-central. The primary difference then between default ICCING and ICCING with the Green's functions arises from a combination of smearing effects that occur during an expansion in time and the suppression of non-perturbative quark-anti-quark pairs. This leads to large eccentricities in central collisions but nearly vanishing eccentricities in peripheral collisions.

This work constitutes the first step toward including charge evolution in \kompost and illustrates the effect pre-equilibrium evolution has on conserved charge densities. An implementation of this method in \kompost would solve the mismatch between the background and perturbation evolutions which are local and non-local, respectively. Exploring the effect of the pre-equilibrium evolution of conserved charges on the hydrodynamic evolution of the system and on final state observables would be beyond the scope of this paper but is an interesting open question that will be explored in a future work. It would also be interesting to compute these Green's functions for charges in QCD kinetic theory and compare them to the approximation introduced in this paper. Another possible direction is extending these Green's functions around a non-vanishing background which would be useful when looking at systems that contain baryon stopping.

Last but not least, the assumption of conformal symmetry can be loosened and exploring the effect of breaking conformality in the pre-equilibrium stage is an interesting topic that deserved further detailed studies.

\section*{Acknowledgements}
P.P. and S.S acknowledge support by the Deutsche
Forschungsgemeinschaft (DFG, German Research Foundation) through the CRC-TR 211 ’Strong-interaction
matter under extreme conditions’– project number
315477589 – TRR 211. J.N.H. and P.C. acknowledge the support from the US-DOE Nuclear Science Grants No. DE-SC0020633 and DE-SC0023861 and
the support from the Illinois Campus Cluster, a computing resource that is operated by the Illinois Campus Cluster Program (ICCP) in conjunction with the National
Center for Supercomputing Applications (NCSA), and
which is supported by funds from the University of Illinois at Urbana-Champaign. M.S. is supported by a start-up grant from New Mexico State University. M. M. was supported in part by the US Department of Energy Grant No. DE-FG02-03ER41260 and BEST (Beam Energy Scan Theory) DOE Topical Collaboration. The authors also acknowledge computing time provided by the Paderborn Center for Parallel Computing (PC2) and the National Energy Research Scientific Computing Center, a DOE Office of Science User Facility supported by the Office of Science of the U.S. Department of Energy under Contract No. DE-AC02-05CH11231.

\clearpage

\appendix

\section{Background evolution}\label{app:BackgroundEvolution}

\subsection{Evolution Equations for the spherical harmonic moments}\label{app:EvolutionForMoments}
In order to solve Eq. (\ref{eq:BEBackgroundRTA}), we will adopt the ideas of \cite{Grad}, where, instead of finding solutions for the distribution functions, one studies the moments of the distribution function. \\
For the distribution functions $f_{\BG}$ and $f_{a,\BG}$ we will consider the following moments
\begin{subequations}\label{eq:EandNMomentsApp}
\begin{align}
\E{m}{l}(\tau) &= \tau^{1/3} \imsr p^\tau \Ylm\pqty{\phi_\pt,\theta_\pt} f_{\BG}\pqty{\tau,p_T,\vqty{p_\eta}}\, ,\label{eq:EMomentsApp} \\
\N{m}{l}(\tau) &= \imsr \tens{Y}{m}{l} \pqty{\phi_\pt,\theta_\pt} f_{a,\BG}\pqty{\tau, p_T, \vqty{p_\eta}}\, .\label{eq:NMomentsApp}
\end{align}
\end{subequations}
In Eq. (\ref{eq:EandNMomentsApp}) the angles are defined by $\tan\phi_\pt=p^1/p^2$ and $\cos\theta_\pt=p_\eta/(\tau p^\tau)$, while $\Ylm$ are the spherical harmonics given by
\begin{subequations}
\begin{align}
\Ylm(\phi,\theta) = \tens{y}{m}{l} \tens{P}{m}{l}(\cos\theta) e^{im\phi}
\end{align}
with
\begin{align}
\tens{y}{m}{l} = \sqrt{\frac{\pqty{2l+1}\pqty{l-m}!}{4\pi\pqty{l+m}!}}
\end{align}
and
\begin{align}
\tens{P}{m}{l}(x) = \frac{\pqty{-1}^m}{2^l l!} \pqty{1-x^2}^{m/2}\dv[l+m]{}{x}\pqty{x^2-1}^l\, .
\end{align}
\end{subequations}
being the associated Legendre polynomials.

Based on the explicit form of the spherical harmonics one can find the non-vanishing components of the background energy-momentum tensor by low order moments as well as the tracelessness condition
\begin{subequations}
\begin{align}
e(\tau) &= \frac{\sqrt{4\pi}}{\tau^{4/3}} \E{0}{0}(\tau)\, , \label{eq:EnergyAsMoment}\\
P_T(\tau) &= \frac{\sqrt{4\pi}}{\tau^{4/3}} \bqty{\frac{1}{3} \E{0}{0}(\tau) - \sqrt{\frac{1}{45}} \E{0}{2}(\tau)}\, , \\
P_L(\tau) &= \frac{\sqrt{4\pi}}{\tau^{4/3}} \bqty{\frac{1}{3} \E{0}{0}(\tau) + \sqrt{\frac{4}{45}} \E{0}{2}(\tau)}\, .
\end{align}
\end{subequations}
Furthermore, by plugging in the equilibrium distribution function, we find that
\begin{align}
\E{m}{l}\big|_\teq (\tau) = \frac{\tau^{4/3}}{\sqrt{4\pi}}e(\tau)\tens{\delta}{}{l0}\tens{\delta}{m0}{}\, ,
\end{align}
representing the rotational symmetry of the equilibrium.
In the same way we are able to reconstruct the components of the charge current via low-order moments. The net particle number of specie $a$ is given by
\begin{align}\label{eq:ChargeAsMoment}
n_a(\tau) = \frac{\sqrt{4\pi}}{\tau} \N{0}{0}(\tau) \, .
\end{align}
The chemical potential can be extracted by inverting the Landau conditions Eq. (\ref{eq:LandauMatchingT}) and Eq. (\ref{eq:LandauMatchingN}).

Applying $\tau\partial_\tau$ to the definitions of the moments and using identities for Legendre Polynomials (see App. \ref{app:IdentitiesSphericalHarmonics}) leads directly to the equations of motion, which are given as
\begin{subequations}
\begin{align}
\tau\del_\tau \E{m}{l}(\tau) &= \tens{b}{m}{l,-2} \E{m}{l-2}(\tau) + \tens{b}{m}{l,0} \E{m}{l}(\tau) + \tens{b}{m}{l,+2} \E{m}{l+2}(\tau) - \frac{\tau}{\tau_R} \bqty{\E{m}{l}(\tau) - \left. \E{m}{l}(\tau)\right|_\text{eq}}\, ,\label{eq:BackgroundEvolutionE} \\
\tau\del_\tau \N{m}{l}(\tau) &= \tens{B}{m}{l,-2} \N{m}{l-2}(\tau) + \tens{B}{m}{l,0} \N{m}{l}(\tau) + \tens{B}{m}{l,+2} \N{m}{l+2}(\tau) - \frac{\tau}{\tau_R} \bqty{\N{m}{l}(\tau) - \left. \N{m}{l}(\tau)\right|_\text{eq}} \, . \label{eq:BackgroundEvolutionN}
\end{align}
\end{subequations}
The appearing coefficients $\tens{b}{m}{l}$ and $\tens{B}{m}{l}$ are given by
\begin{subequations}\label{eq:bCoefficientsApp}
\begin{align}
\tens{b}{m}{l,-2} &= \frac{\pqty{2+l}\pqty{l+m-1}\pqty{l+m}}{\pqty{1-4l^2}} \sqrt{\frac{\pqty{2l+1}\pqty{l-m-1}\pqty{l-m}}{\pqty{2l-3}\pqty{l+m-1}\pqty{l+m}}}\, , \\
\tens{b}{m}{l,0} &= -\frac{5}{3} \frac{l\pqty{l+1}-3m^2}{4l\pqty{l+1}-3}\, , \\
\tens{b}{m}{l,+2} &= \frac{\pqty{l-1}}{\pqty{2l+3}} \sqrt{\frac{\pqty{l-m+1}\pqty{l-m+2}\pqty{l+m+1}\pqty{l+m+2}}{\pqty{2l+1}\pqty{2l+5}}}\, .
\end{align}
\end{subequations}
resp.
\begin{subequations}\label{eq:BCoefficientsApp}
\begin{align}
\tens{B}{m}{l,-2} &= \frac{\pqty{1+l}\pqty{l+m-1}\pqty{l+m}}{\pqty{1-4l^2}} \sqrt{\frac{\pqty{2l+1}\pqty{l-m-1}\pqty{l-m}}{\pqty{2l-3}\pqty{l+m-1}\pqty{l+m}}}\, , \\
\tens{B}{m}{l,0} &= - \frac{l\pqty{l+1}-3m^2}{4l\pqty{l+1}-3}\, ,\\
\tens{B}{m}{l,+2} &= \frac{l}{\pqty{2l+3}} \sqrt{\frac{\pqty{l-m+1}\pqty{l-m+2}\pqty{l+m+1}\pqty{l+m+2}}{\pqty{2l+1}\pqty{2l+5}}}\, .
\end{align}
\end{subequations}

\subsection{Initial Conditions}\label{app:InitialConditionsBackground}
In order to solve the equations of motion for $\E{m}{l}$ and $\N{m}{l}$ we need to specify the initial conditions for the moments. For early time dynamics at $\tau\ll\tau_R$ the system cannot maintain considerable longitudinal momenta. Therefore the initial distribution is naturally of the form that the transverse momentum is much larger than the longitudinal one. Taking also into account previous results \cite{Blaizot:2019scw,Blaizot:2017ucy,Blaizot:2017lht,Behtash:2018moe,Behtash:2019qtk,Strickland:2018ayk,Dash:2020zqx} one sees that the case of a (longitudinal) support in form of a Dirac delta function corresponds to a non-equilibrium attractor of the kinetic equations, i.e. that different initial conditions will approach the same curve for later times. We therefore choose
\begin{subequations}\label{eq:BackgroundInitialConditions}
\begin{align}f_{\BG}\pqty{\tau_0,p_T,\vqty{p_\eta}} &= (2\pi)^3 \delta(p_\eta) \bqty{\frac{1}{\nu_g}\frac{\dd{N_{0,g}}}{\dd{\eta}\dd[2]{\pt}\dd[2]{\xt}} + \frac{1}{\nu_q}\sum_a\pqty{\frac{\dd{N_{0,q_a}}}{\dd{\eta}\dd[2]{\pt}\dd[2]{\xt}} + \frac{\dd{N_{0,\overline{q}_a}}}{\dd{\eta}\dd[2]{\pt}\dd[2]{\xt}}}} \nonumber \\
&\equiv( 2\pi)^3 \delta(p_\eta) \frac{\dd{\tilde{N}_{0}}}{\dd{\eta}\dd[2]{\pt}\dd[2]{\xt}}\, ,\label{eq:BackgroundInitialConditionsE} \\
f_{a,\BG}\pqty{\tau_0,p_T,\vqty{p_\eta}} &= (2\pi)^3 \delta(p_\eta) \frac{1}{\nu_q} \bqty{\frac{\dd{N_{0,q_a}}}{\dd{\eta}\dd[2]{\pt}\dd[2]{\xt}} - \frac{\dd{N_{0,\overline{q}_a}}}{\dd{\eta}\dd[2]{\pt}\dd[2]{\xt}}} \equiv (2\pi)^3 \delta(p_\eta) \frac{\dd{\tilde{N}_{0,a}}}{\dd{\eta}\dd[2]{\pt}\dd[2]{\xt}}\, ,\label{eq:BackgroundInitialConditionsN}
\end{align}
\end{subequations}
where we choose the normalization such that the initial energy and charge densities are kept constant
\begin{subequations}
\begin{align}
\frac{\dd{E_{0}}}{\dd{\eta}\dd[2]{\xt}} &= \frac{\dd{E_{0,g}}}{\dd{\eta}\dd[2]{\xt}} + \sum_a\pqty{\frac{\dd{E_{0,q_a}}}{\dd{\eta}\dd[2]{\xt}} + \frac{\dd{E_{0,\overline{q}_a}}}{\dd{\eta}\dd[2]{\xt}}} =  \lim_{\tau_0\rightarrow 0}\tau_0e(\tau_0) = (e\tau)_0 = \textsl{const}\, ,\label{eq:BackgroundInitialEnergyEApp} \\
\frac{\dd{N_{0,a}}}{\dd{\eta}\dd[2]{\xt}} &= \frac{\dd{N_{0,q_a}}}{\dd{\eta}\dd[2]{\xt}} - \frac{\dd{N_{0,\overline{q}_a}}}{\dd{\eta}\dd[2]{\xt}} =  \lim_{\tau_0\rightarrow 0}\tau_0n_a(\tau_0) = (n_a\tau)_0 = \textsl{const} \, .\label{eq:BackgroundInitialEnergyN}
\end{align}
\end{subequations}
On the level of the moments the initial conditions are given by
\begin{subequations}\label{eq:InitialConditionsBackgroundMoments}
\begin{align}
\E{m}{l}(\tau_0) &= \tau_0^{1/3} \, (e\tau)_0 \, \tens{y}{m}{l} \tens{P}{m}{l}(0) \tens{\delta}{m0}{} \, , \\
\N{m}{l}(\tau_0) &= (n_a\tau)_0 \, \tens{y}{m}{l} \tens{P}{m}{l}(0) \tens{\delta}{m0}{}\, .
\end{align}
\end{subequations}

\subsection{Relation of Intensive and Extensive Quantities}\label{app:ConnectingIntensiveAndExtensiveQuantities}
In contrast to the cases in \cite{Kamata:2020mka}, we additionally need to invert
\begin{subequations}\label{eq:EandNasPolylogs}
\begin{align}
e &= T^4 \bqty{\frac{\nu_g\pi^2}{30} - \frac{3\nu_q}{\pi^2} \sum\nolimits _a \bqty{\Li_4\pqty{-z^{-1}_a} + \Li_4\pqty{-z_a}}}\, , \\
n_{a} &= \nu_q \frac{T^3}{\pi^2} \bqty{\Li_3\pqty{-z^{-1}_a} - \Li_3\pqty{-z_a}} = \frac{\nu_q}{6\pi^2} \bqty{\pi^2 T^2 \mu_a + \mu_a^3}\, ,
\end{align}
\end{subequations}
with $z_a\equiv\exp(\mu_a/T)$ to determine the temperature $T$ and the chemical potentials $\mu_a$ as a function of energy density $e$ and number density $n_a$. This is done numerically for each time step. The relations are given by
\begin{subequations}\label{eq:IntensivePerturbationsFiniteDensity}
\begin{align}
\delta T &= - T \frac{\chi_u\chi_d\chi_s\delta e - 3n_u\chi_d\chi_s\delta n_u - 3n_d\chi_u\chi_s\delta n_d - 3n_s\chi_u\chi_d\delta n_s}{9n_u^2\chi_d\chi_s + 9n_d^2\chi_u\chi_s + 9n_s^2\chi_u\chi_d - 4e\chi_u\chi_d\chi_s}\, , \\
\delta \mu_u &= \frac{\bqty{9\pqty{n_d^2\chi_s+n_s^2\chi_d}+\pqty{3n_u\mu_u-4e}\chi_d\chi_s}\delta n_u}{9n_u^2\chi_d\chi_s + 9n_d^2\chi_u\chi_s + 9n_s^2\chi_u\chi_d - 4e\chi_u\chi_d\chi_s} +\frac{ - 3\alpha_u n_d\chi_s\delta n_d - 3\alpha_u n_s\chi_d\delta n_s + \alpha_u\chi_d\chi_s\delta e}{9n_u^2\chi_d\chi_s + 9n_d^2\chi_u\chi_s + 9n_s^2\chi_u\chi_d - 4e\chi_u\chi_d\chi_s}\, , \\
\delta \mu_d &= \frac{\bqty{9n_s^2\chi_u + \pqty{9n_u^2 + \pqty{3n_d\mu_d - 4e}\chi_u}\chi_s}\delta n_d}{9n_u^2\chi_d\chi_s + 9n_d^2\chi_u\chi_s + 9n_s^2\chi_u\chi_d - 4e\chi_u\chi_d\chi_s} +\frac{-3\alpha_d n_u\chi_s\delta n_u - 3\alpha_d n_s\chi_u\delta n_s + \alpha_d\chi_u\chi_s\delta e}{9n_u^2\chi_d\chi_s + 9n_d^2\chi_u\chi_s + 9n_s^2\chi_u\chi_d - 4e\chi_u\chi_d\chi_s}\, , \\
\delta \mu_s &= \frac{\bqty{9n_d^2\chi_u + \pqty{9n_u^2 + \pqty{3n_s\mu_s - 4e}\chi_u}\chi_d}\delta n_s}{9n_u^2\chi_d\chi_s + 9n_d^2\chi_u\chi_s + 9n_s^2\chi_u\chi_d - 4e\chi_u\chi_d\chi_s} +\frac{-3\alpha_s n_u\chi_d\delta n_u - 3\alpha_s n_d\chi_u\delta n_d + \alpha_s\chi_u\chi_d\delta e}{9n_u^2\chi_d\chi_s + 9n_d^2\chi_u\chi_s + 9n_s^2\chi_u\chi_d - 4e\chi_u\chi_d\chi_s}\, .
\end{align}
\end{subequations}
Here $\chi_a$ is the susceptibility, which is given by
\begin{align}
\chi_a \equiv \frac{\nu_q}{6} \pqty{\frac{3\mu^2_a}{\pi^2} + T^2}
\end{align}
For the special case of perturbations around $n_a=0$ the susceptibilities reduce to $\chi_a=\frac{\nu_q}{6}T^2$ which yields then
\begin{subequations}\label{eq:IntensivePerturbationsZeroDensity}
\begin{align}
\delta T &= T \frac{\delta e}{4e}\, , \\
\delta \mu_a &= \frac{6}{\nu_q} \frac{\delta n_a}{T^2}\, .
\end{align}
\end{subequations}

\subsection{Background Evolution in Conformal Systems\label{app:BackgroundEvolutionInConformalSystems}}
In this section we will consider the evolution of a conformal system. In conformal systems $\tau_R$ is proportional to the inverse temperature such that \cite{Denicol:2018wdp}
\begin{align}
\tau_R T(\tau) = 5\frac{\tilde{\eta}T}{e+P}=\textsl{const}\, ,
\end{align}
where $\tilde{\eta}$ is the shear viscosity, $e$  the energy density, $P$ the pressure and $T$ the effective temperature of the system.

At this point we introduce the dimensionless time variable $x=\tau/\tau_R$, as this produces a natural time scale for the evolution of the system. Due to the change of variable we need to transform also the appearing derivatives according to
\begin{align}
\tau\del_\tau &= \tau \pdv{x}{\tau} \del_x = \tau \bqty{\frac{1}{\tau_R} - \frac{\tau}{\tau_R^2}\pqty{\del_\tau\tau_R}}\del_x = \bqty{1-\frac{1}{\tau_R}\tau\del_\tau\tau_R}x\del_x \equiv a(x)x\del_x\, ,
\end{align}
where we will call
\begin{align}
a(x)\equiv 1-x\del_\tau\tau_R = 1-\frac{1}{\tau_R}\tau\del_\tau\tau_R
\end{align}
the scale factor. As $\tau_R$ is not constant, the scale factor will take a complicated form. However, it can be related to the moments again as we find
\begin{align}
a(x)&= 1 + \frac{\tau\del_\tau T}{T} = 1 - \frac{\chi_u\chi_d\chi_s\tilde{a}(x) e + 3n_u^2\chi_d\chi_s + 3n_d^2\chi_u\chi_s + 3n_s^2\chi_u\chi_d}{9n_u^2\chi_d\chi_s + 9n_d^2\chi_u\chi_s + 9n_s^2\chi_u\chi_d - 4e\chi_u\chi_d\chi_s}\, ,
\end{align}
where
\begin{align}
\tilde{a}(x) = -\frac{4}{3} + \tens{b}{0}{0,0}\E{0}{0} + \tens{b}{0}{0,+2} \frac{\E{0}{2}(x)}{\E{0}{0}(x)} \, .
\end{align}
The appearing quantities can be expressed in terms of the low order moments (see Eq. (\ref{eq:EnergyAsMoment}), (\ref{eq:ChargeAsMoment})). We note at this point that for $n_a=0$ the scale factor reduces to
\begin{align}
a(x)&= 1 + \frac{\tau\del_\tau T}{T} = 1 - \frac{\chi_u\chi_d\chi_s\tilde{a}(x) e}{- 4e\chi_u\chi_d\chi_s} = \frac{2}{3} + \frac{1}{4}\pqty{\tens{b}{0}{0,0}\E{0}{0} + \tens{b}{0}{0,+2} \frac{\E{0}{2}(x)}{\E{0}{0}(x)}}\, ,
\end{align}
which is the form used in \cite{Kamata:2020mka}.

Using the change of variables the equations of motion can be written in terms of $x$ as
\begin{subequations}
\begin{align}
a(x)x\del_x \E{m}{l} &= \tens{b}{m}{l,-2} \E{m}{l-2} + \tens{b}{m}{l,0} \E{m}{l} + \tens{b}{m}{l,+2} \E{m}{l+2} - x \bqty{\E{m}{l} - \left. \E{m}{l}\right|_\text{eq}}\, , \\
a(x)x\del_x \N{m}{l} &= \tens{B}{m}{l,-2} \N{m}{l-2} + \tens{B}{m}{l,0} \N{m}{l} + \tens{B}{m}{l,+2} \N{m}{l+2} - x \bqty{\N{m}{l} - \left. \N{m}{l}\right|_\text{eq}}\, .
\end{align}
\end{subequations}
For our analysis we are varying the initial charge number densities in order to see its impact on the evolution of the background. Nevertheless, we keep the ratios between the three species the same, namely
\begin{align}
\begin{aligned}
\frac{n_{u,\BG}}{n_{d,\BG}} = \frac{8}{7}\, , \\
n_{s,\BG}=0.0\, ,
\end{aligned}
\end{align}
as these ratios correspond to typical values in heavy-ion collisions.
\begin{figure}[ht]
\begin{center}
\begin{minipage}{0.6\textwidth}
\includegraphics[width=\textwidth]{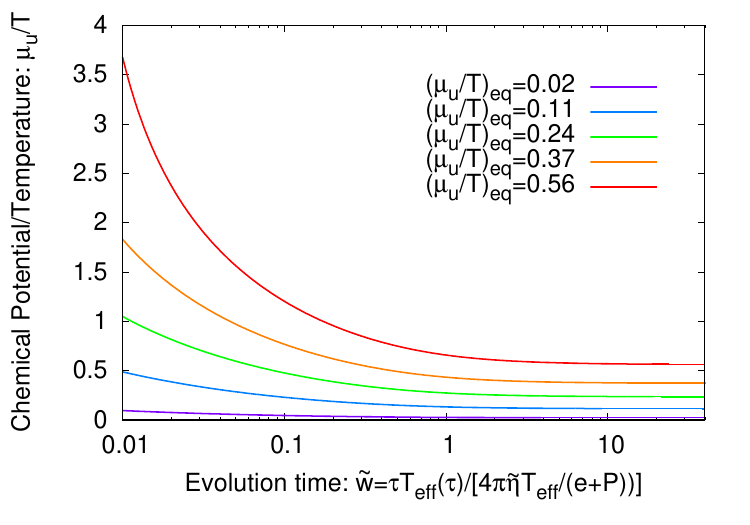}
\end{minipage}
\begin{minipage}{0.49\textwidth}
\includegraphics[width=\textwidth]{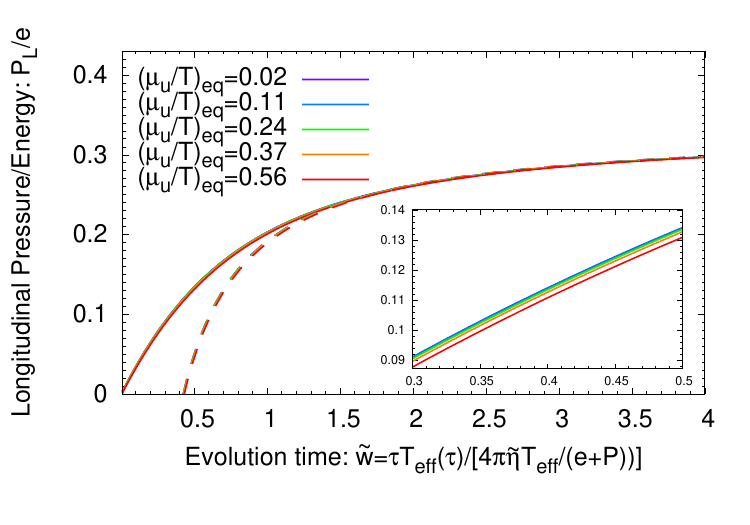}
\end{minipage}
\begin{minipage}{0.49\textwidth}
\includegraphics[width=\textwidth]{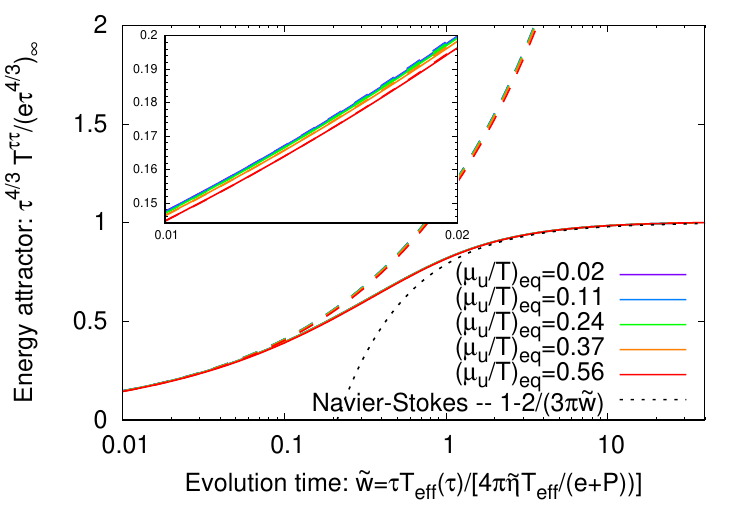}
\end{minipage}
\caption{Background evolution for conformal systems. \textsl{Top}: $\mu_u/T$ ratio for different initial values for $n_u$, \textsl{Bottom left}: Longitudinal pressure over energy for different values of $\pqty{\mu_u/T}_\teq$; The coloured dashed curves in the $P_L/e$ plot correspond to the hydrodynamical behavior at later times, $P_L/e=\frac{1}{3} - \frac{4}{9\pi\tilde{w}}$ for $\tilde{w}\rightarrow\infty$, \textsl{Bottom right}: Energy attractor for different values of $\pqty{\mu_u/T}_\teq$. The coloured dashed curves in the $\pqty{\tau^{4/3}e}/\pqty{\tau^{4/3}e}_\infty$ plot correspond to the free streaming behavior of the energy attractors at early times, $\pqty{\tau^{4/3}e}/\pqty{\tau^{4/3}e}_\infty=\frac{1}{C_\infty}\tilde{w}^{4/9}$ for $\tilde{w}\ll 1$. More details on how to fit the dashed curves in the two plots are given in the text. Inset plots are given in order to show that there are deviations between the curves. On the two axes of the inset plots are the same quantities plotted as for the larger plot, but the labels are omitted for better readability. \label{fig:BackgroundPlotsForConformalRTA}}
\end{center}
\end{figure}
\newline
\noindent
Our results for a conformal system can be seen in Fig. \ref{fig:BackgroundPlotsForConformalRTA}. In conformal systems without conserved charges it was found that the evolution is controlled by the dimensionless time variable $\tilde{w}=\tau T(\tau)/(4\pi\tilde{\eta}/s)$ \cite{Giacalone:2019ldn}. We will generalise this to systems with conserved charges and choose to present the different quantities as functions of the dimensionless time variable
\begin{align}
\tilde{w} = \frac{\tau T(\tau)}{4\pi} \frac{(e+P)}{\tilde{\eta}T} = \frac{5}{4\pi} x\, ,
\end{align}
where $T(\tau)=\bqty{\frac{30}{\nu_\text{eff}\pi^2}e(\tau)}^{1/4}$ 
and $\nu_\text{eff}=\nu_g+ 3 \cdot \frac{7}{4}\nu_q$.

At the top of Fig. \ref{fig:BackgroundPlotsForConformalRTA} we show the different curves we obtain for the ratio $\mu_u/T$. We see that at late times, when hydrodynamics is applicable, the ratio becomes constant according to
\begin{align}
\pqty{\frac{\mu_a}{T}}_\text{eq} = \frac{6}{\nu_q} \frac{n_a}{T} = \frac{6}{\nu_q}\pqty{\frac{\nu_\text{eff}\pi^2}{30}}^\frac{3}{4} \pqty{\frac{n_a}{e^\frac{3}{4}}}_\text{eq} = \textsl{const}\, .
\end{align}
In the bottom left panel we show the ratio of longitudinal pressure and energy, $P_L/e$. We see 
that the ratio is essentially zero at early times as the longitudinal pressure needs to build up first and that at late times a smooth transition to the hydrodynamical behavior 
\begin{align}
\pqty{\frac{P_L}{e}}_{\text{vHydro}} = \frac{1}{3} - \frac{4}{9\pi\tilde{w}}\label{eq:PlOverEHYdro}
\end{align}
is provided around time $\tilde{w}\sim 1.5$. In the corresponding figure this behavior is indicated by the coloured dashed curves. Note that the validity of the hydrodynamic limit Eq. (\ref{eq:PlOverEHYdro}) is guaranteed for small values of $\pqty{\mu_u/T}_\teq$, while for larger values of $\pqty{\mu_u/T}_\teq$ this is a priori not clear and needs further studies. We also see the effect of the chemical potential. As one can see in the inset plot, the $P_L/e$-ratio increases slower for for increasing $\pqty{\mu_u/T}$-ratio. Nevertheless the differences are very small, which can be explained by the assumption that we choose the same relaxation scale for all particles. It is expected to improve the results if one assumes different time scales for gluons and for quarks like following the approach of \cite{Bhadury:2020ngq}.
Regarding the bottom right panel of Fig. \ref{fig:BackgroundPlotsForConformalRTA} we show the results for the energy attractor
\begin{align}
\pqty{\tau^{4/3}e}/\pqty{\tau^{4/3}e}_\infty\, ,
\end{align}
where
\begin{align}\pqty{\tau^{4/3}e}_\infty = \lim_{\tau\rightarrow\infty} \tau^{4/3}e(\tau) = \textsl{const}
\end{align}
describes the asymptotic energy density scaled with $\tau^{4/3}$. It is convenient to consider $\tau^{4/3}e$ as this becomes constant at late times as ideal hydrodynamics predicts. The value of the constant can be obtained by the numerical solution of the equations of motion and depends on the chemical potential which is considered as the energy evolution couples to the charge number via the scale factor. In the figure we also show the free-streaming behavior, which we can parametrise according to \cite{Kamata:2020mka}
\begin{align}
\frac{\pqty{\tau^{4/3}e}}{\pqty{\tau^{4/3}e}_\infty} = C_\infty^{-1} \tilde{w}^\frac{4}{9}
\end{align}
at early times (corresponds to the dashed coloured curves) and the hydrodynamical behavior
\begin{align}
\frac{\pqty{\tau^{4/3}e}}{\pqty{\tau^{4/3}e}_\infty} = 1 - \frac{2}{3\pi\tilde{w}}
\end{align}
at late times (corresponding to the black dashed curve) \cite{Kamata:2020mka}. We emphasise that in the case of a conformal system we also observe a smooth transition from the early time free-streaming regime to the late time viscous hydrodynamical regime, which starts to describe the evolution around times $\tilde{w}\sim 1.5$.

Regarding the free-streaming behavior we fit the energy attractor at early times for the curves corresponding to different chemical potentials using
\begin{align}\label{eq:FreeStreamingEnergyAttractorRTAConformal}
\frac{\pqty{\tau^{4/3}e}}{\pqty{\tau^{4/3}e}_\infty} = C_\infty^{-1} \tilde{w}^\frac{4}{9}
\end{align}
to extract the values of $C_\infty$. The results can be seen in Tab. (\ref{tab:ParametersBackground}). We emphasise that the role of $C_\infty$ is as follows. At late times the system can be described by viscous hydrodynamics. However, the approach to this regime depends on the theory, such that different theories approach viscous hydrodynamics differently. This difference in the approach to the late time behavior results in a mismatch of the ratios of initial energy density to the final energy density (see \cite{Kamata:2020mka} for a comparison of \kompost QCD kinetic theory to results obtained in conformal relaxation time approximation without conserved charges), where $C_\infty$ is used to express the late time energy density in terms of the initial energy density, such that we find Eq. (\ref{eq:FreeStreamingEnergyAttractorRTAConformal}) at early times. In Yang-Mills kinetic theory one finds $C_\infty\approx 0.9$ \cite{Kurkela:2018vqr,Kurkela:2018wud,Giacalone:2019ldn}. Looking at Tab. (\ref{tab:ParametersBackground}) we see that in conformal relaxation time approximation with conserved charges the value of $C_\infty$ increases as we increase the initial charge number density, however it will stay below the value in \cite{Kurkela:2018vqr,Kurkela:2018wud,Giacalone:2019ldn}.

\begin{table}[h!]
\centering
\caption{Values of $C_\infty$ obtained by numerical fits.\label{tab:ParametersBackground}}
\begin{tabular}{c||c|c|c|c|c}
$\pqty{\mu_u/T}_\teq$ & 0.02 & 0.11 & 0.24 & 0.37 & 0.56 \\
\hline
\hline
$C_\infty$ & 0.87643 & 0.87712 & 0.87927 & 0.88374 & 0.89349 \\
\end{tabular}
\end{table}

\section{Perturbations Around Bjorken Flow\label{app:Perturbations}}

\subsection{Linearised equations of motion and Landau matching}\label{app:LinEoM}
By linearising the kinetic equations around the boost invariant and homogeneous background one finds an evolution equation for the perturbation of the distribution functions $\delta f$ and $\delta f_a$
\begin{subequations}\label{eq:GeneralEoMForPerturbationApp}
\begin{align}
\begin{aligned}
\bqty{p^\tau \del_\tau + p^{i}\del_i - \frac{p_\eta}{\tau^2}\del_\eta} \delta f(x,p) &= - \frac{p^\tau}{\tau_R} \delta f\pqty{x,p} + \frac{p_\mu\delta u^\mu(x)}{\tau_R} \bqty{\pqty{f_{\text{eq}} - f\pqty{x,p}} + \frac{p^\tau}{T(\tau)} f^{\scriptscriptstyle (1,0)}_{\text{eq}}} \\
& \qquad -\frac{p^\tau}{\tau_R} \frac{\delta T(x)}{T(\tau)} \bqty{\frac{T(\tau)}{\tau_R} \pdv{\tau_R}{T} \pqty{f_{\text{eq}} - f\pqty{x,p}} + \frac{p^\tau}{T(\tau)} f^{\scriptscriptstyle (1,0)}_{\text{eq}}} \\
& \qquad +\frac{p^\tau}{\tau_R} \sum_a \delta \mu_{a}(x)  \bqty{f^{\scriptscriptstyle (0,1)}_{q_a,\text{eq}} + \overline{f}^{\scriptscriptstyle (0,1)}_{q_a,\text{eq}}}
\end{aligned}
\end{align}
and
\begin{align}
\begin{aligned}
\bqty{p^\tau \del_\tau + p^{i}\del_i - \frac{p_\eta}{\tau^2}\del_\eta} \delta f_{a}(x,p) &= - \frac{p^\tau}{\tau_R} \delta f_{a}\pqty{x,p} + \frac{p_\mu\delta u^\mu(x)}{\tau_R} \bqty{\pqty{f_{a,\text{eq}} - f_{a}\pqty{x,p}} + \frac{p^\tau}{T(\tau)} f^{\scriptscriptstyle (1,0)}_{a,\text{eq}}} \\
& \qquad -\frac{p^\tau}{\tau_R} \frac{\delta T(x)}{T(\tau)} \bqty{\frac{T(\tau)}{\tau_R} \pdv{\tau_R}{T} \pqty{f_{a,\text{eq}} - f_{a}\pqty{x,p}} + \frac{p^\tau}{T(\tau)} f^{\scriptscriptstyle (1,0)}_{a,\text{eq}}} \\
& \qquad +\frac{p^\tau}{\tau_R}\delta \mu_a(x)  f^{\scriptscriptstyle (0,1)}_{a,\text{eq}}\, ,
\end{aligned}
\end{align}
\end{subequations}
The perturbations of the rest-frame velocity, $\delta u^\mu(x)$, the temperature, $\delta T(x)$, and the chemical potential, $\delta \mu_{a}(x)$, are obtained by the linearised Landau-matching.

As the velocity $u^\mu$ is normalised to $u_\mu u^\mu=+1$, we immediately find that $u_\mu\delta u^\mu =0$,
from which
\begin{align}
\delta u^\tau = 0\,
\end{align}
directly follows. The perturbed energy-momentum tensor and the perturbed charge current are given by
\begin{subequations}
\begin{align}
\tens{\delta T}{\mu\nu}{} &= \int \frac{\dd[4]{p}}{(2\pi)^4} \frac{2\pi}{\sqrt{-g(x)}}\delta(p^2)2\theta(p^0) p^\mu p^\nu \delta f(x,p)\, , \\
\tens{\delta N}{\mu}{\! a} &= \int \frac{\dd[4]{p}}{(2\pi)^4} \frac{2\pi}{\sqrt{-g(x)}}\delta(p^2)2\theta(p^0) p^\mu \delta f_{a}(x,p)\, .
\end{align}
\end{subequations}
Using this, the perturbed eigenvalue problem for the energy-momentum tensor reads
\begin{subequations}
\begin{align}
\pqty{u_\mu + \delta u_\mu} \pqty{\tensor*{T}{*^{\mu\nu}} + \tensor*{\delta T}{*^{\mu\nu}}} &= \pqty{e + \delta e} \pqty{u^\nu + \delta u^\nu} \, ,\label{eq:PerturbedLandauMatching}
\end{align}
\end{subequations}
while the one for the charge current is given by
\begin{align}
\pqty{u_\mu+\delta u_\mu}\pqty{\tens{N}{\mu}{a} + \tens{\delta N}{\mu}{a}} = n_a +\delta n_a \, .
\end{align}
By using the leading order solutions we can deduce the different components, namely
\begin{align}\label{eq:LinearisedLandauMatching}
\begin{aligned}
\delta e &= \tensor*{\delta T}{*^{\tau\tau}} \, , \\
\delta u^\tau = 0 \quad , \quad \delta u^{i} &= \frac{\tensor*{\delta T}{*^{\tau i}}}{e + P_T} \quad , \quad \delta u^\eta = \frac{\tensor*{\delta T}{*^{\tau\eta}}}{e + P_L} \, , \\
\delta n_a &= \tens{\delta N}{\tau}{a} \, .
\end{aligned}
\end{align}

\subsection{Evolution Equations for the Perturbed Moments}\label{app:EoMPerturbedMoments}
The perturbation of the distribution functions are expanded in terms of spherical harmonics according to
\begin{subequations}
\begin{align}
\dE{m}{l}(\tau) &= \tau^{1/3} \imsr p^\tau \Ylm\pqty{\phi_{\pt\kt},\theta_\pt} \delta f_\kt\pqty{\tau,\pt,\vqty{p_\eta}}\, , \label{eq:EPerturbedMomentsApp} \\
\dN{m}{l}(\tau) &= \imsr \Ylm\pqty{\phi_{\pt\kt},\theta_\pt} \delta f_{a,\kt}\pqty{\tau,\pt,\vqty{p_\eta}}\, , \label{eq:NPerturbedMomentsApp}
\end{align}
\end{subequations}
Similar to the background one can obtain the components of $\tens{\delta T}{\mu\nu}{\kt}$ as combinations of low order moments \cite{Kamata:2020mka}
\begin{subequations}\label{eq:dTObtainedFromMoments}
\begin{alignat}{4}
\tau^{4/3}\tens{\delta T}{\tau\tau}{\kt}& &&= &&\sqrt{4\pi}\dE{0}{0} \, , \\
\tens{\delta}{ij}{}\frac{i\kt^{i}}{\abs{\kt}}\tau^{4/3}\tens{\delta T}{\tau j}{\kt}& &&= -i &&\sqrt{\frac{2\pi}{3}}\qty\Big(\dE{+1}{1} - \dE{-1}{1}) \, , \\
\tens{\epsilon}{ij}{}\frac{i\kt^{i}}{\abs{\kt}}\tau^{4/3}\tens{\delta T}{\tau j}{\kt}& &&= -&&\sqrt{\frac{2\pi}{3}}\qty\Big(\dE{+1}{1} + \dE{-1}{1}) \, , \\
\tau^{4/3}(-\tau)\tens{\delta T}{\tau\eta}{\kt}& &&= &&\sqrt{\frac{4\pi}{3}}\dE{0}{1} \, , \\
\tens{\delta}{ij}{}\tau^{4/3}\tens{\delta T}{ij}{\kt}& &&= &&\sqrt{\frac{16\pi}{9}}\dE{0}{0} - \sqrt{\frac{16\pi}{45}}\dE{0}{2} \, , \\
\frac{\kt^{i}\kt^{j}}{\kt^2}\tau^{4/3}\tens{\delta T}{ij}{\kt}& &&= &&\sqrt{\frac{4\pi}{9}}\dE{0}{0} - \sqrt{\frac{4\pi}{45}}\dE{0}{2} + \sqrt{\frac{2\pi}{15}}\qty\Big(\dE{+2}{2} + \dE{-2}{2}) \, , \\
\tens{\epsilon}{lj}{}\frac{\kt^{i}\kt^{l}}{\kt^2}\tau^{4/3}\tens{\delta T}{ij}{\kt}& &&= -i &&\sqrt{\frac{2\pi}{15}}\qty\Big(\dE{+2}{2} - \dE{-2}{2}) \, , \\
\tens{\delta}{ij}{}\frac{i\kt^{i}}{\abs{\kt}}\tau^{4/3}(-\tau)\tens{\delta T}{\eta j}{\kt}& &&= -i &&\sqrt{\frac{2\pi}{15}}\qty\Big(\dE{+1}{2} - \dE{-1}{2}) \, , \\
\tens{\epsilon}{ij}{}\frac{i\kt^{i}}{\abs{\kt}}\tau^{4/3}(-\tau)\tens{\delta T}{\eta j}{\kt}& &&= -&&\sqrt{\frac{2\pi}{15}}\qty\Big(\dE{+1}{2} + \dE{-1}{2}) \, , \\
\tau^{4/3}\tau^2\tens{\delta T}{\eta\eta}{\kt}& &&= &&\sqrt{\frac{16\pi}{45}}\dE{0}{2} + \sqrt{\frac{4\pi}{9}}\dE{0}{0} \, .
\end{alignat}
\end{subequations}
It is also possible to obtain the components of $\tens{\delta N}{\mu}{\! a,\hspace{0.8pt} \kt}$, which are given by
\begin{subequations}\label{eq:dNObtainedFromMoments}
\begin{alignat}{4}
\tau\tens{\delta N}{\tau}{\! a,\hspace{0.8pt} \kt}& &&= &&\sqrt{4\pi}\dN{0}{0} \, , \\
\tens{\delta}{ij}{}\frac{i\kt^{i}}{\abs{\kt}}\tau\tens{\delta N}{j}{\! a,\hspace{0.8pt} \kt}& &&= -i&&\sqrt{\frac{2\pi}{3}}\qty\Big(\dN{+1}{1} - \dN{-1}{1}) \, , \\
\tens{\epsilon}{ij}{}\frac{i\kt^{i}}{\abs{\kt}}\tau\tens{\delta N}{j}{\! a,\hspace{0.8pt} \kt}& &&= -&&\sqrt{\frac{2\pi}{3}}\qty\Big(\dN{+1}{1} + \dN{-1}{1}) \, , \\
\tau(-\tau)\tens{\delta N}{\eta}{\! a,\hspace{0.8pt} \kt}& &&= &&\sqrt{\frac{4\pi}{3}}\dN{0}{1} \, .
\end{alignat}
\end{subequations}
Note that we decomposed transverse components parallel and perpendicular to the wave vector $\kt$. \\
In order to shorten the notation in the following we define
\begin{subequations}
\begin{align}
\tens{(\Delta E)}{m}{l} &\equiv \tens{(E_{\text{eq}} - E + E^{\scriptscriptstyle (1,0)}_{\text{eq}})}{m}{l} \, , \\
\tens{(\Delta N_{\! a})}{m}{l} &\equiv \tens{(N_{\! a,\text{eq}} - N_{\! a} + N^{\scriptscriptstyle (1,0)}_{\! a,\text{eq}})}{m}{l} \, .
\end{align}
\end{subequations}
By direct application of the time derivative to the moments we can find their evolution equations to be
\begin{align}\label{eq:EoMPerturbedE}
\begin{aligned}
\tau \del_\tau &\dE{m}{l} \\
&= \tens{b}{m}{l,-2} \dE{m}{l-2} + \tens{b}{m}{l,0} \dE{m}{l} + \tens{b}{m}{l,+2} \dE{m}{l+2} \\
& \qquad - \frac{i\abs{\kt}\tau}{2} \qty\Big[\tens{u}{m}{l,-} \dE{m+1}{l-1} + \tens{u}{m}{l,+} \dE{m+1}{l+1} + \tens{d}{m}{l,-} \dE{m-1}{l-1} + \tens{d}{m}{l,+} \dE{m-1}{l+1}] \\
& \qquad -\frac{\tau}{\tau_R} \bqty{\dE{m}{l} + \frac{\delta T_\mathbf{k}}{T} \tens{(E^{\scriptscriptstyle (1,0)}_{\text{eq}})}{m}{l} - \sum\nolimits_a \delta\mu_{a,\hspace{0.8pt} \mathbf{k}} \qty\Big[\tens{(E^{\scriptscriptstyle (0,1)}_{\! q_a,\text{eq}} + \overline{E}^{\scriptscriptstyle (0,1)}_{\! q_a,\text{eq}})}{m}{l}] } \\
& \qquad -\frac{\tau}{\tau_R} \frac{\delta T_\mathbf{k}}{T} \frac{T(\tau)}{\tau_R} \pdv{\tau_R}{T} \tens{(E_{\text{eq}} - E)}{m}{l} \\
& \qquad -\frac{\tau}{\tau_R} \frac{\delta u^\parallel_\mathbf{k}}{2} \Big[ \tens{u}{m}{l,-} \tens{(\Delta E)}{m+1}{l-1} + \tens{u}{m}{l,+} \tens{(\Delta E)}{m+1}{l+1}  + \tens{d}{m}{l,-} \tens{(\Delta E)}{m-1}{l-1} + \tens{d}{m}{l,+} \tens{(\Delta E)}{m-1}{l+1} \Big] \\
& \qquad -\frac{\tau}{\tau_R} \frac{\delta u^\perp_\mathbf{k}}{2i} \Big[ \tens{u}{m}{l,-} \tens{(\Delta E)}{m+1}{l-1} + \tens{u}{m}{l,+} \tens{(\Delta E)}{m+1}{l+1} - \tens{d}{m}{l,-} \tens{(\Delta E)}{m-1}{l-1} - \tens{d}{m}{l,+} \tens{(\Delta E)}{m-1}{l+1} \Big]\, ,
\end{aligned}
\end{align}
and
\begin{align}\label{eq:EoMPerturbedN}
\begin{aligned}
\tau\del_\tau &\dN{m}{l} \\
&= \tens{B}{m}{l,-2} \dN{m}{l-2} + \tens{B}{m}{l,0} \dN{m}{l} + \tens{B}{m}{l,+2} \dN{m}{l+2} \\
& \qquad - \frac{i\abs{\kt}\tau}{2} \qty\Big[\tens{u}{m}{l,-} \dN{m+1}{l-1} + \tens{u}{m}{l,+} \dN{m+1}{l+1} + \tens{d}{m}{l,-} \dN{m-1}{l-1} + \tens{d}{m}{l,+} \dN{m-1}{l+1}] \\
& \qquad -\frac{\tau}{\tau_R} \bqty{\dN{m}{l} + \frac{\delta T_\mathbf{k}}{T} \tens{(N^{\scriptscriptstyle (1,0)}_{\! a,\text{eq}})}{m}{l} - \delta\mu_{a,\hspace{0.8pt} \mathbf{k}} \tens{(N^{\scriptscriptstyle (0,1)}_{\! a,\text{eq}})}{m}{l} } \\
& \qquad -\frac{\tau}{\tau_R} \frac{\delta T_\mathbf{k}}{T} \frac{T(\tau)}{\tau_R} \pdv{\tau_R}{T} \tens{(N_{\! a,\text{eq}} - N_{\! a})}{m}{l} \\
& \qquad -\frac{\tau}{\tau_R} \frac{\delta u^\parallel_\mathbf{k}}{2} \Big[ \tens{u}{m}{l,-} \tens{(\Delta N_{\! a})}{m+1}{l-1} + \tens{u}{m}{l,+} \tens{(\Delta N_{\! a})}{m+1}{l+1} + \tens{d}{m}{l,-} \tens{(\Delta N_{\! a})}{m-1}{l-1} + \tens{d}{m}{l,+} \tens{(\Delta N_{\! a})}{m-1}{l+1} \Big] \\
& \qquad -\frac{\tau}{\tau_R} \frac{\delta u^\perp_\mathbf{k}}{2i} \Big[ \tens{u}{m}{l,-} \tens{(\Delta N_{\! a})}{m+1}{l-1} + \tens{u}{m}{l,+} \tens{(\Delta N_{\! a})}{m+1}{l+1} - \tens{d}{m}{l,-} \tens{(\Delta N_{\! a})}{m-1}{l-1} - \tens{d}{m}{l,+} \tens{(\Delta N_{\! a})}{m-1}{l+1} \Big]\, .
\end{aligned}
\end{align}
where we used App. \ref{app:IdentitiesSphericalHarmonics} in order to express the angle relations in terms of moments. The coefficients $\tens{u}{m}{l,\pm}$ and $\tens{d}{m}{l,\pm}$ are given by
\begin{subequations}\label{eq:upmAnddpm}
\begin{alignat}{12}
u_{l,-}^{m}& &&= &&+\sqrt{\frac{(l-m)(l-m-1)}{4l^2-1}} \quad &&, \quad &&u_{l,+}^{m}&&= &&-\sqrt{\frac{(l+m+1)(l+m+2)}{3+4l(l+2)}} \, , \\
d_{l,-}^{m}& &&= &&-\sqrt{\frac{(l+m)(l+m-1)}{4l^2-1}} \quad &&, \quad &&d_{l,+}^{m}&&= &&+\sqrt{\frac{(l-m+1)(l-m+2)}{3+4l(l+2)}} \, .
\end{alignat}
\end{subequations}
while the $\tens{b}{m}{l}$ and $\tens{B}{m}{l}$ are the same as in the background evolution equations (see Eq. (\ref{eq:bCoefficientsApp}),(\ref{eq:BCoefficientsApp})). \\
The derivatives of the moments are defined to be
\begin{subequations}
\begin{align}
\tens{(E^{\scriptscriptstyle (1,0)}_{\text{eq}})}{m}{l}(\tau) &= \tau^{1/3}\imsr p^\tau\Ylm\pqty{\phi_{\pt\kt},\theta_\pt} \frac{p^\tau}{T(\tau)} f^{\scriptscriptstyle (1,0)}_{\text{eq}}\pqty{\frac{p^\tau}{T(\tau)},\mu(\tau)} \, , \\
\tens{(E^{\scriptscriptstyle (0,1)}_{\text{eq}})}{m}{l}(\tau) &= \tau^{1/3}\imsr p^\tau\Ylm\pqty{\phi_{\pt\kt},\theta_\pt} f^{\scriptscriptstyle (0,1)}_{\text{eq}}\pqty{\frac{p^\tau}{T(\tau)},\mu(\tau)} \, , \\
\tens{(N^{\scriptscriptstyle (1,0)}_{\! a,\text{eq}})}{m}{l}(\tau) &= \imsr \tens{Y}{m}{l}\pqty{\phi_{\pt\kt},\theta_\pt} \frac{p^\tau}{T(\tau)} f^{\scriptscriptstyle (1,0)}_{a,\text{eq}}\pqty{\frac{p^\tau}{T(\tau)},\mu(\tau)} \, , \\
\tens{(N^{\scriptscriptstyle (0,1)}_{\! a,\text{eq}})}{m}{l}(\tau) &= \imsr \tens{Y}{m}{l}\pqty{\phi_{\pt\kt},\theta_\pt} f^{\scriptscriptstyle (0,1)}_{a,\text{eq}}\pqty{\frac{p^\tau}{T(\tau)},\mu(\tau)} \,  .
\end{align}
\end{subequations}
A straightforward computation of the derivatives shows that we can relate them to
\begin{subequations}
\begin{align}
\tens{(E^{\scriptscriptstyle (1,0)}_{\text{eq}})}{m}{l}(\tau) &= -4\tens{\pqty{E_\teq}}{m}{l}(\tau)\, , \\
\tens{(E^{\scriptscriptstyle (0,1)}_{\text{eq}})}{m}{l}(\tau) &= 3\tau^{1/3}\sum_a \tens{\pqty{N_{a,\teq}}}{m}{l}(\tau)\, , \\
\tens{(N^{\scriptscriptstyle (1,0)}_{\! a,\text{eq}})}{m}{l}(\tau) &= -3\tens{\pqty{N_{\! a,\teq}}}{m}{l}(\tau) \, , \\
\tens{(N^{\scriptscriptstyle (0,1)}_{\! a,\text{eq}})}{m}{l}(\tau) &= \frac{\tau}{\sqrt{4\pi}}\chi_a\tens{\delta}{}{l0}\tens{\delta}{m0}{} \, .
\end{align}
\end{subequations}
However, as we are interested in perturbations around vanishing background density, the equations will simplify since
\begin{align}
\tens{\pqty{N_{a,\teq}}}{m}{l}(\tau) = 0\,
\end{align}
for $\mu(x)=0$. Nevertheless, the susceptibilities
\begin{align}
\chi_a = \frac{\nu_q}{6} \pqty{\frac{3\mu^2_a}{\pi^2} + T^2}
\end{align}
are non-zero for zero density but reduce to
\begin{align}
\chi_a\pqty{\mu_a=0} = \frac{\nu_q}{6} T^2\, .
\end{align}
We can also relate the perturbations of the intensive quantities to the perturbation of the extensive quantities for $n_a=0$ according to Eq. (\ref{eq:IntensivePerturbationsZeroDensity})
\begin{subequations}
\begin{align}
\frac{\delta T_\kt}{T} &= \frac{\delta e_\kt}{4e}\, , \\
\delta \mu_{a,\hspace{0.8pt} \kt} &= \frac{6}{\nu_q} \frac{\delta n_{a.\hspace{0.8pt} \kt}}{T^2}\, .
\end{align}
\end{subequations}
Therefore we can replace $\delta T_\kt$ and $\delta\mu_{a,\hspace{0.8pt} \kt}$ with $\delta e_\kt$ and $\delta n_{a,\hspace{0.8pt} \kt}$, which is useful since we can express these quantities by low order moments again
\begin{subequations}\label{eq:PerturbationsWrittenAsMomentsApp}
\begin{align}
\tau^{4/3} \delta e_\kt(\tau) &= \sqrt{4\pi} \dE{0}{0}(\tau) \, , \\
\tau^{4/3} (e+P_T) \delta u^\parallel_\kt(\tau) &= -\sqrt{\frac{2\pi}{3}} \qty\Big(\dE{+1}{1}(\tau) - \dE{-1}{1}(\tau)) \, , \\
\tau^{4/3} (e+P_T) \delta u^\perp_\kt(\tau) &= i\sqrt{\frac{2\pi}{3}} \qty\Big(\dE{+1}{1}(\tau) + \dE{-1}{1}(\tau)) \, , \\
\tau \delta n_{a,\hspace{0.8pt} \kt} &= \sqrt{4\pi}\dN{0}{0}\, ,
\end{align}
\end{subequations}
This results in a closed set of equations since all appearing perturbations can be written as linear combinations of moments.

At the level of the equations of motion we see that the dependence on the direction of the transverse wave-vector $\kt$ has disappeared and the equations only depend on $\vkt$. This is due to the decomposition of the distribution functions into spherical harmonics and represents the azimuthal rotation symmetry of the background in the transverse plane.

The equations of motion above (Eq. (\ref{eq:EoMPerturbedE}) and Eq. (\ref{eq:EoMPerturbedN})) are considered at a fixed value of the wave-number $\abs{\kt}$. However, it is more convenient to rewrite the equation of motion in a mode where we consider it for a fixed value of the propagation phase
\begin{align}
\kappa = \vkt (\tau-\tau_0)\, .
\end{align}
By making this change of variable from $\vkt$ to $\kappa$, we also need to rewrite the time derivative according to
\begin{align}
\tau\del_\tau\big|_\kt = \tau\del_\tau\big|_{\vkt(\tau-\tau_0)} + \frac{\tau}{\tau-\tau_0}\vkt(\tau-\tau_0)\del_{\vkt(\tau-\tau_0)}\big|_\tau\, .
\end{align}
This means, that we find an additional term resulting from the change of variables.

Furthermore, for conformal systems it is convenient to work again with the dimensionless variable $x=\tau/\tau_R$. Following the same procedure as for the background, we need to transform the derivative by making use of the scale factor (see App. \ref{app:BackgroundEvolutionInConformalSystems}). By introducing $s(\tau)=(\tau-\tau_0)/\tau$ with
\begin{align}\label{eq:DEforS}
a(x)x\del_x s(x) = 1 - s(x)
\end{align}
we the finally find the equations we are using to compute the Green's functions
\begin{subequations}\label{eq:EoMPerturbedEandNinKappaMode}
\begin{align}\label{eq:EoMPerturbedEinKappaMode}
\begin{aligned}
\Big[ s(x) &a(x)x\del_x + \kappa\del_\kappa \Big] \dE{m}{l} \\
&= s(x) \Big[ \tens{b}{m}{l,-2} \dE{m}{l-2} + \tens{b}{m}{l,0} \dE{m}{l} + \tens{b}{m}{l,+2} \dE{m}{l+2} \Big] \\
& \phantom{=} - \frac{i\kappa}{2} \qty\Big[\tens{u}{m}{l,-} \dE{m+1}{l-1} + \tens{u}{m}{l,+} \dE{m+1}{l+1} + \tens{d}{m}{l,-} \dE{m-1}{l-1} + \tens{d}{m}{l,+} \dE{m-1}{l+1}] \\
& \phantom{=} -xs(x) \bqty{\dE{m}{l} + \frac{\delta T_\mathbf{k}}{T} \tens{(E^{\scriptscriptstyle (1,0)}_{\text{eq}})}{m}{l}} - xs(x) \frac{\delta T_\mathbf{k}}{T} \frac{T(\tau)}{\tau_R} \pdv{\tau_R}{T} \tens{(E_{\text{eq}} - E)}{m}{l} \\
& \phantom{=} -xs(x) \frac{\delta u^\parallel_\mathbf{k}}{2} \Big[ \tens{u}{m}{l,-} \tens{(\Delta E)}{m+1}{l-1} + \tens{u}{m}{l,+} \tens{(\Delta E)}{m+1}{l+1} + \tens{d}{m}{l,-} \tens{(\Delta E)}{m-1}{l-1} + \tens{d}{m}{l,+} \tens{(\Delta E)}{m-1}{l+1} \Big] \\
& \phantom{=} -xs(x) \frac{\delta u^\perp_\mathbf{k}}{2i} \Big[ \tens{u}{m}{l,-} \tens{(\Delta E)}{m+1}{l-1} + \tens{u}{m}{l,+} \tens{(\Delta E)}{m+1}{l+1} - \tens{d}{m}{l,-} \tens{(\Delta E)}{m-1}{l-1} - \tens{d}{m}{l,+} \tens{(\Delta E)}{m-1}{l+1} \Big]\, ,
\end{aligned}
\end{align}
and
\begin{align}
\begin{aligned}
\Big[ s(x) &a(x)x\del_x + \kappa\del_\kappa \Big] \dN{m}{l} \\
&= s(x) \Big[ \tens{B}{m}{l,-2} \dN{m}{l-2} + \tens{B}{m}{l,0} \dN{m}{l} + \tens{B}{m}{l,+2} \dN{m}{l+2} \Big] \\
& \phantom{=} - \frac{i\kappa}{2} \qty\Big[\tens{u}{m}{l,-} \dN{m+1}{l-1} + \tens{u}{m}{l,+} \dN{m+1}{l+1} + \tens{d}{m}{l,-} \dN{m-1}{l-1} + \tens{d}{m}{l,+} \dN{m-1}{l+1}] \\
& \phantom{=} -xs(x) \bqty{\dN{m}{l} - \delta\mu_{a,\hspace{0.8pt} \mathbf{k}} \tens{(N^{\scriptscriptstyle (0,1)}_{\! a,\text{eq}})}{m}{l} } - xs(x) \frac{\delta T_\mathbf{k}}{T} \frac{T(\tau)}{\tau_R} \pdv{\tau_R}{T} \tens{(N_{\! a,\text{eq}} - N_{\! a})}{m}{l} \\
& \phantom{=} -xs(x) \frac{\delta u^\parallel_\mathbf{k}}{2} \Big[ \tens{u}{m}{l,-} \tens{(\Delta N_{\! a})}{m+1}{l-1} + \tens{u}{m}{l,+} \tens{(\Delta N_{\! a})}{m+1}{l+1} + \tens{d}{m}{l,-} \tens{(\Delta N_{\! a})}{m-1}{l+1} \Big] \\
& \phantom{=} -xs(x) \frac{\delta u^\perp_\mathbf{k}}{2i} \Big[ \tens{u}{m}{l,-} \tens{(\Delta N_{\! a})}{m+1}{l-1} + \tens{u}{m}{l,+} \tens{(\Delta N_{\! a})}{m-1}{l-1} - \tens{d}{m}{l,+} \tens{(\Delta N_{\! a})}{m-1}{l+1} \Big]\, ,
\end{aligned}
\end{align}
\end{subequations}
where
\begin{subequations}
\begin{align}
\tens{(\Delta E)}{m}{l} &\equiv \tens{(E_{\text{eq}} - E + E^{\scriptscriptstyle (1,0)}_{\text{eq}})}{m}{l} \, , \\
\tens{(\Delta N_{\! a})}{m}{l} &\equiv \tens{(N_{\! a,\text{eq}} - N_{\! a})}{m}{l} \, .
\end{align}
\end{subequations}
Note that $N^{\scriptscriptstyle (1,0)}_{\! a,\text{eq}}=0$ since it is proportional to $\tens{\pqty{N_{a,\teq}}}{m}{l}$, which is zero for vanishing background density.

\subsection{Initial Energy Perturbations}
So far we considered the evolution of linearised perturbations on top of a background modeled as a Bjorken flow. In order to describe the early time dynamics of heavy-ion collisions we need suitable initial conditions for the perturbations to solve the equations of motion. Here we will consider initial energy and charge perturbations. \\
We follow the idea of \cite{Kurkela:2018vqr}, which means initial energy perturbations will be associated with an infinitesimal change of the energy scale of the background distribution. For the initial distribution function of the perturbations we will find therefore
\begin{align}
\delta f_\kt \pqty{\tau_0,\pt,\vqty{p_\eta}} = -\pqty{\frac{\vqty{\pt}}{3} \del_\vqty{\pt} \tens{f}{(0)}{\BG}} e^{-i\kt\cdot\frac{\pt}{\vqty{\pt}}\tau_0}\, .\label{eq:DeltaFkEnergyPerturbation}
\end{align}
The factor $e^{-i\kt\cdot\frac{\pt}{\vqty{\pt}}\tau_0}$ takes into account the free-streaming behavior for times $\tau<\tau_0\ll\tau_R$, while $\tens{f}{(0)}{\BG}$ is given by Eq. (\ref{eq:BackgroundInitialConditionsE}). We can insert Eq. (\ref{eq:DeltaFkEnergyPerturbation}) into the definition of $\dE{m}{l}$ to translate the initial condition to the moments according to
\begin{align}\label{eq:DeltaEForEnergyPerturbation}
\begin{aligned}
\dE{m}{l}(\tau_0) &= \tau_0^{1/3} (-i)^m J_m\pqty{\vkt\tau_0} \tens{y}{m}{l} \tens{P}{m}{l}\pqty{0} (e\tau)_0
\end{aligned}
\end{align}
with $(e\tau)_0$ being the asymptotic energy density of the background Eq. (\ref{eq:BackgroundInitialEnergyEApp}) and $J_m\pqty{x}$ being the Bessel function of the first kind of order $m$. In agreement with \cite{Kurkela:2018vqr} we find for the energy and velocity perturbations
\begin{subequations}
\begin{alignat}{4}
\frac{\delta e_\kt(\tau_0)}{e}& &&= &&J_0\pqty{\vkt\tau_0}\, , \\
\frac{e+P_T}{e}\delta u_\kt^\parallel (\tau_0)& &&= -i &&J_1\pqty{\vkt\tau_0}\, , \\
\frac{e+P_T}{e}\delta u_\kt^\perp (\tau_0)& &&= 0\, . &&
\end{alignat}
\end{subequations}

\subsection{Initial Charge Perturbations}
The natural choice for the moments of conserved charges is an initial perturbation in terms of the number of quarks respectively the number of anti-quarks. Therefore we choose initial perturbations of the form
\begin{align}
\delta f_{a,\kt} \pqty{\tau_0,\pt,\vqty{p_\eta}} &= \delta f_{q_a,\kt} \pqty{\tau_0,\pt,\vqty{p_\eta}} - \delta \overline{f}_{q_a,\kt} \pqty{\tau_0,\pt,\vqty{p_\eta}} \nonumber \\
&= \pqty{1 + \frac{1}{2} \alpha_a - 1 + \frac{1}{2}\alpha_a}\tens{f}{(0)}{a,\BG} e^{-i\kt\cdot\frac{\pt}{\vqty{\pt}}\tau_0} \nonumber \\
&= \alpha_a \tens{f}{(0)}{a,\BG} e^{-i\kt\cdot\frac{\pt}{\vqty{\pt}}\tau_0}\, .
\end{align}
In this particular case we choose
\begin{align}
\alpha_a = \frac{\delta n_a(\tau_0)}{n_a(\tau_0)}\, .
\end{align}
Translated to the level of the charge moments the initial conditions are given by
\begin{align}
\begin{aligned}
\dN{m}{l}(\tau_0) &= (-i)^m J_m\pqty{\vkt\tau_0} \tens{y}{m}{l} \tens{P}{m}{l}\pqty{0} \alpha_a \pqty{n_a\tau}_0\, ,
\end{aligned}
\end{align}
where $\pqty{n_a\tau}_0$ is given by Eq. (\ref{eq:BackgroundInitialEnergyN}).
For the perturbation $\delta n_a$ we find
\begin{align}
\frac{\delta n_{a,\kt} (\tau_0)}{n_a} = \alpha_a J_0(\vkt \tau_0)\, .
\end{align}

\subsection{Numerics} \label{app:Numerics}
The procedure for finding the Green's functions numerically is more or less the same as for the background. However we will consider perturbations around zero densities, i.e. we set the initial values for $n_a$ to zero. At a given $l_\text{max}$ we truncate the equations of motions for the moments.

Regarding $l_\text{max}$ our numerical studies have shown that we need take a relatively high value of $l$ in order to find convergence for the charge moments. 
To check this, it is convenient to consider free-streaming. This is due to the fact that we are able to compute analytically the behavior of the response functions in free-streaming. Based on this we can use free-streaming in order to check if the code runs correctly (at least without perturbation-terms in the equations of motion).

\begin{figure}[h!]
\includegraphics[width=\textwidth]{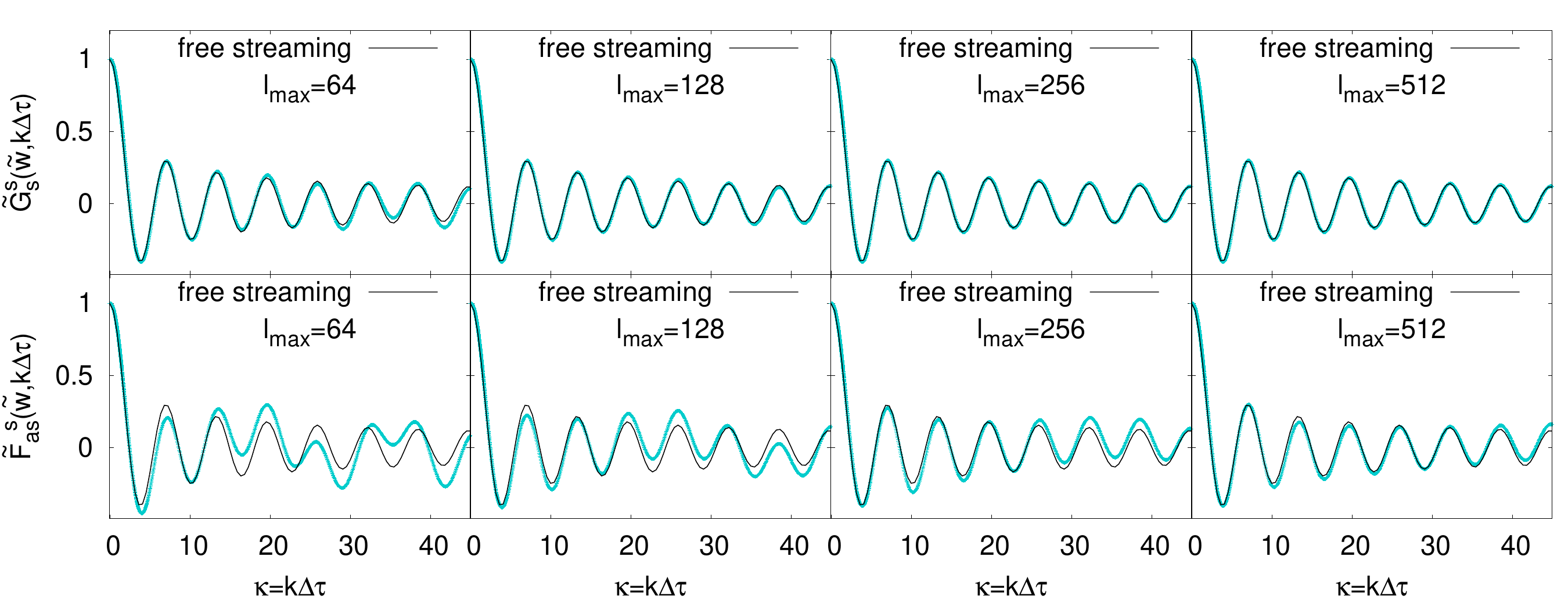}
\caption{\textsl{Top row}: $\G{s}{s}$, \textsl{bottom row}: $\F{s}{s}$. \textsl{From left to right}: Corresponding Green's functions for $l_\text{max}=64$,$128$, $256$, $512$. The black curve corresponds to analytic free-streaming solutions while the blue curve corresponds to our data. Clearly, for $\G{s}{s}$ we find convergence to the free-streaming very fast (no significant improvement for $l_\text{max}>128$). For $\F{s}{s}$ we see acceptable convergence only for $l_\text{max}\geq512$, which justifies that we choose $l_\text{max}=512$ for our numerics.\label{fig:ConvergencePlots}}
\end{figure}

It turns out that we find convergence towards the free-streaming behavior for the energy moments a lot faster than for the charge moments in terms of $l_\text{max}$. The results of this studies can be seen in Fig. \ref{fig:ConvergencePlots}. This justifies our choice of $l_\text{max}=512$.

\section{Non-Equilibrium Green's Functions of Energy-Momentum Tensor and Current of Conserved Charges}\label{app:NonEqGreensFunctions}

\subsection{Green's Functions of the Energy-Momentum Tensor}\label{app:GreensFunctionsTApp}
We follow the construction of the response functions according to \cite{Kurkela:2018vqr,Kurkela:2018wud} and express $\tens{\delta T}{\mu\nu}{\kt}(\tau)$ as
\begin{align}\label{eq:ResponseFunctionEquationForTApp}
\frac{\tens{\delta T}{\mu\nu}{\kt}(\tau)}{e(\tau)} = \frac{1}{2} \G{\mu\nu}{\alpha\beta}\pqty{\kt,\tau,\tau_0} \frac{\tens{\delta T}{\alpha\beta}{\kt}(\tau_0)}{e(\tau_0)}\, .
\end{align}
We decompose the several response functions into a basis of scalars (\textsl{s}), vectors (\textsl{v}) and tensors (\textsl{t}). For initial energy perturbations we thus have
\begin{subequations}
\begin{align}
\G{\tau\tau}{\tau\tau}\pqty{\kt,\tau} &= \G{s}{s}\pqty{\kappa,x}\, ,\\
\G{\tau i}{\tau\tau}\pqty{\kt,\tau} &= -i\frac{\kt^{i}}{\vkt}\G{v}{s}\pqty{\kappa,x}\, ,\\
\G{ij}{\tau\tau}\pqty{\kt,\tau} &= \tens{\delta}{ij}{} \G{t,\delta}{s}\pqty{\kappa,x} + \frac{\kt^{i}\kt^{j}}{\vkt^2} \G{t,k}{s}\pqty{\kappa,x}\, .
\end{align}
\end{subequations}
\noindent Since the normalization of the linearized perturbation is 
arbitrary, we adopt the convention
\begin{align}
\frac{\delta e(\tau_0)}{e(\tau_0)} = 1
\end{align}
such that we can express the decomposed response functions in terms of $\tens{\delta T}{\mu\nu}{\kt}(\tau)$ (see \cite{Kurkela:2018vqr}) according to
\begin{subequations}
\begin{align}
\G{s}{s}\pqty{\kappa,x} &= \frac{\tens{\delta T}{\tau\tau}{\kt}(x)}{e(x)} = \frac{\delta e_\kappa(x)}{e(x)}\, ,\\
\G{v}{s}\pqty{\kappa,x} &= i\tens{\delta}{}{ij}\frac{\kt_i}{\vkt} \frac{\tens{\delta T}{\tau j}{\kappa}(x)}{e(x)}\, ,\\
\G{t,\delta}{s}\pqty{\kappa,x} &= \bqty{\tens{\delta}{}{ij} - \frac{\kt_{i}\kt_{j}}{\vkt^2}} \frac{\tens{\delta T}{ij}{\kappa}(x)}{e(x)}\, ,\\
\G{t,k}{s}\pqty{\kappa,x} &= \bqty{2\frac{\kt_{i}\kt_{j}}{\vkt^2} - \tens{\delta}{}{ij}} \frac{\tens{\delta T}{ij}{\kappa}(x)}{e(x)}\, .
\end{align}
\end{subequations}

\subsection{Green's Functions of the Current of Conserved Charges}\label{app:GreensFunctionsNApp}
The Green's functions corresponding to the conserved charges are defined by
\begin{align}
\tau\tens{\delta N}{\mu}{\kt}(\tau) = \F{\mu}{\alpha} \pqty{\kt,\tau,\tau_0} \, \tau_0\tens{\delta N}{\alpha}{\kt}(\tau_0)\, .
\end{align}
Note that we dropped the flavor indices on $\F{\mu}{\alpha}$ as we consider perturbations around vanishing background densities. In such a setting the Green's functions decouple in terms of the flavor. Following the same argumentation $\tens{\delta N}{\mu}{\kt}$ does not depend on the flavor anymore neither. \\
Like before, we will decompose $\F{\mu}{\alpha}$ also in a scalar-vector-tensor basis according to
\begin{subequations}
\begin{align}
\F{\tau}{\tau}\pqty{\kt,\tau} &= \F{s}{s}\pqty{\kappa,x}\, ,\\
\F{i}{\tau}\pqty{\kt,\tau} &= -i\frac{\kt^{i}}{\vkt}\F{v}{s}\pqty{\kappa,x}\, .
\end{align}
\end{subequations}
Adapting the normalization
\begin{align}
\tau_0\tens{\delta N}{\tau}{\kt}(\tau_0) = 1
\end{align}
we find
\begin{subequations}
\begin{align}
\F{s}{s}\pqty{\kappa,x} &= \tau\tens{\delta N}{\tau}{\kappa}(x)\, ,\\
\F{v}{s}\pqty{\kappa,x} &= i \frac{\kt_{i}}{\vkt} \tau\tens{\delta N}{i}{\kappa}(x)\, .
\end{align}
\end{subequations}

\subsection{Numerical Results for the Non-Equilibrium Green's Functions of the Energy-Mo\-men\-tum Tensor}\label{app:GreensFunctionsTResults}
\begin{figure}[t!]
\begin{center}
\begin{minipage}{0.51\textwidth}
\includegraphics[width=\textwidth]{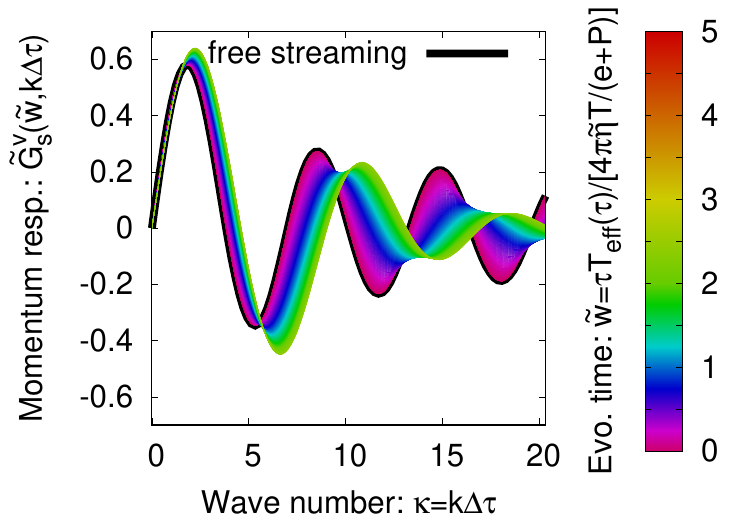}
\end{minipage}
\begin{minipage}{0.49\textwidth}
\includegraphics[width=\textwidth]{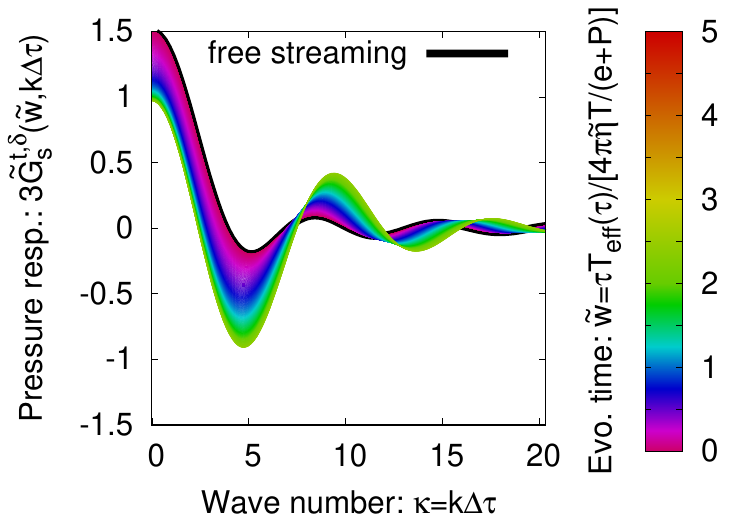}
\end{minipage}
\begin{minipage}{0.49\textwidth}
\includegraphics[width=\textwidth]{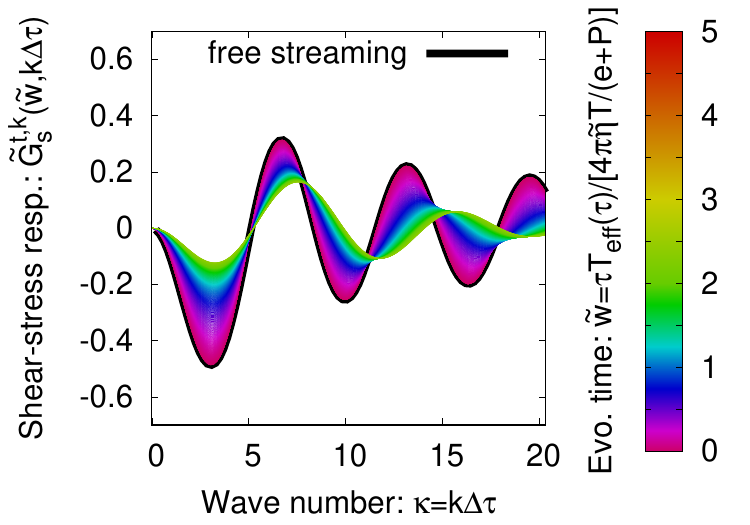}
\end{minipage}
\caption{Evolution of the energy-momentum Green's functions in response to initial energy perturbations in the constant $\kappa$-mode. The different panels correspond to different response functions; different curves in each panel corresponds to different times $\tilde{w}$.\label{fig:ResponseFunctionsToInitialEnergyPerturbationsKappaMode}}
\end{center}
\end{figure}
\begin{figure}[t!]
\begin{center}
\begin{minipage}{0.49\textwidth}
\includegraphics[width=\textwidth]{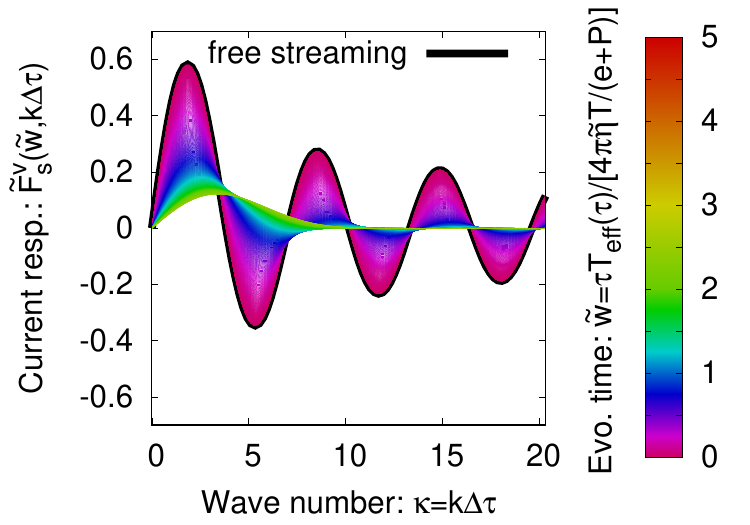}
\end{minipage}
\caption{Evolution of the charge Green's functions in response to initial charge perturbations in the constant $\kappa$-mode. The different panels correspond to different response functions; different curves in each panel corresponds to different times $\tilde{w}$.\label{fig:ResponseFunctionsToInitialChargePerturbationsKappaMode}}
\end{center}
\end{figure}

The results for $\G{s}{s}$ an $\F{s}{s}$ are presented in the main text in Sec. \ref{sec:Theory_NonEquilibriumGF}. In Fig. \ref{fig:ResponseFunctionsToInitialEnergyPerturbationsKappaMode} and Fig. \ref{fig:ResponseFunctionsToInitialChargePerturbationsKappaMode} we will show the results for the other Green's functions.
In addition to the points discussed in the main part we can very clearly see the isotropy at later times in the figure for the pressure response $\G{t,\delta}{s}$. After scaling the response function, we see that at early times the longitudinal pressure is zero while at times when the system can be described by hydrodynamics ($\tilde{w}\geq 1$), the longitudinal pressure is established and we find the effect of isotropy as the response function approaches one at zero propagation phase indicating $e=3P$ in the hydrodynamic limit.

\noindent

\subsection{Green's Functions of the Energy-Momentum Tensor in Coordinate Space}\label{app:EMGreensFunctionsInCoordinateSpace}
\begin{figure}[t!]
\begin{center}
\begin{minipage}{0.51\textwidth}
\includegraphics[width=\textwidth]{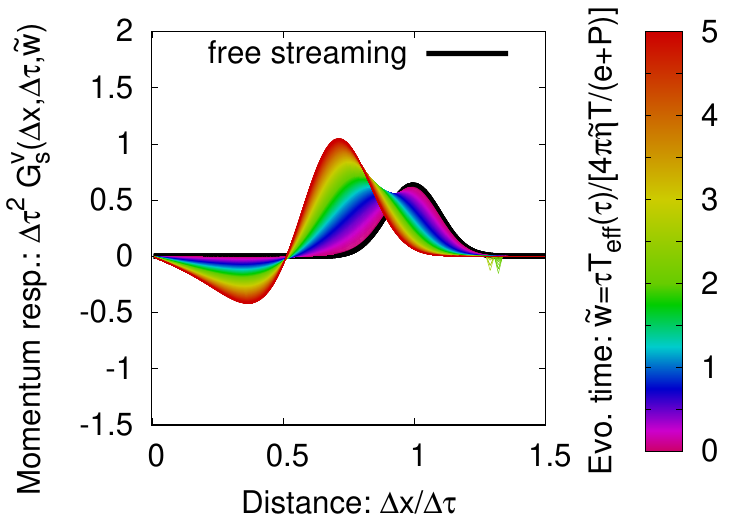}
\end{minipage}
\begin{minipage}{0.49\textwidth}
\includegraphics[width=\textwidth]{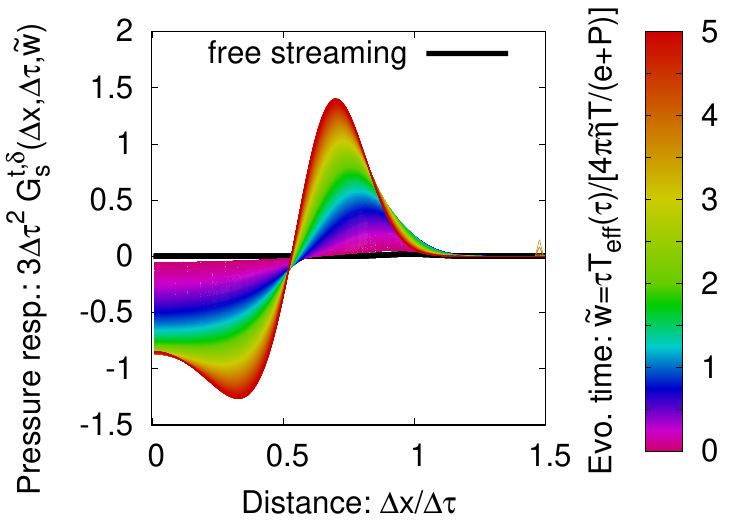}
\end{minipage}
\begin{minipage}{0.49\textwidth}
\includegraphics[width=\textwidth]{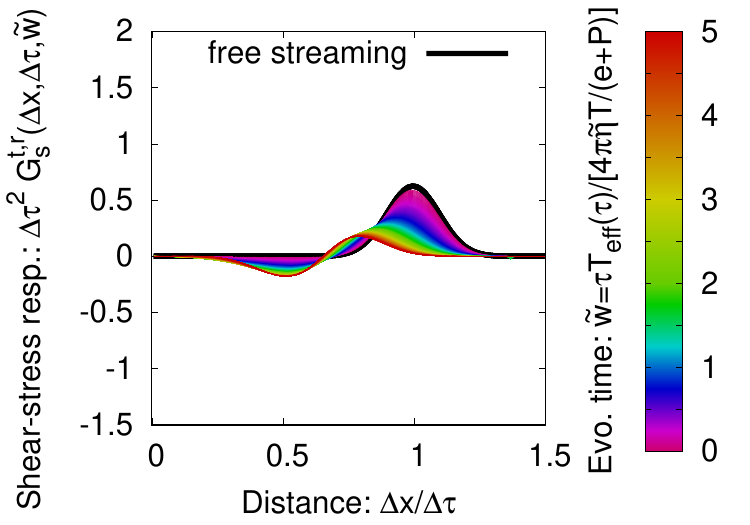}
\end{minipage}
\caption{Evolution of the energy Green's functions in response to initial energy perturbations in coordinate space. The different panels correspond to different response functions; different curves in each panel corresponds to different times $\tilde{w}$.\label{fig:ResponseFunctionsToInitialEnergyPerturbationsCoordinateSpace}}
\end{center}
\end{figure}
Similar to the decomposition in Fourier space, we can decompose the Green's functions in coordinate space as well into a basis of scalars, vectors and tensors, such that we find
\begin{subequations}
\begin{align}
\GCS{\tau\tau}{\tau\tau}\pqty{\rt,\tau} &= \GCS{s}{s}\pqty{\vrt,\tau}\, ,\\
\GCS{\tau i}{\tau\tau}\pqty{\rt,\tau} &= \frac{\rt^{i}}{\vrt}\GCS{v}{s}\pqty{\vrt,\tau}\, ,\\
\GCS{ij}{\tau\tau}\pqty{\rt,\tau} &= \tens{\delta}{ij}{} \GCS{t,\delta}{s}\pqty{\vrt,\tau} + \frac{\rt^{i}\rt^{j}}{\vrt^2} \GCS{t,r}{s}\pqty{\vrt,\tau}\, .
\end{align}
\end{subequations}
The relation to their counterparts in Fourier space is given by the following Fourier-Hankel transforms
\begin{subequations}
\begin{align}
\GCS{s}{s}\pqty{\vrt,\tau} &= \frac{1}{2\pi} \int \dd{\vkt} \vkt J_0\pqty{ \vkt \vrt } \G{s}{s} \pqty{\vkt,\tau} \, , \\
\GCS{v}{s}\pqty{\vrt,\tau} &= \frac{1}{2\pi} \int \dd{\vkt} \vkt J_1\pqty{ \vkt \vrt } \G{v}{s} \pqty{\vkt,\tau} \, , \\
\GCS{t,\delta}{s}\pqty{\vrt,\tau} &= \frac{1}{2\pi} \int \dd{\vkt} \vkt \bqty{ J_0\pqty{ \vkt \vrt } \G{t,\delta}{s} \pqty{\vkt,\tau} + \frac{J_1\pqty{ \vkt \vrt }}{\vkt \vrt} \G{t,k}{s} \pqty{\vkt,\tau}} \, , \\
\GCS{t,r}{s}\pqty{\vrt,\tau} &= \frac{-1}{2\pi} \int \dd{\vkt} \vkt J_2\pqty{ \vkt \vrt } \G{t,k}{s} \pqty{\vkt,\tau} \, .
\end{align}
\end{subequations}
The Green's function $\GCS{s}{s}$ in coordinate space was already shown in Sec. \ref{sec:GreensFunctionsInCoordinateSpace}. Although not relevant for our current study, for the sake of completeness we present the other Green's functions, corresponding to different components of the energy-momentum tensor, in Fig. \ref{fig:ResponseFunctionsToInitialEnergyPerturbationsCoordinateSpace}.

\subsection{Green's Functions of the Current of Conserved Charges in Coordinate Space}\label{app:ChargeGreensFunctionsInCoordinateSpace}
\begin{figure}[t!]
\begin{center}
\begin{minipage}{0.49\textwidth}
\includegraphics[width=\textwidth]{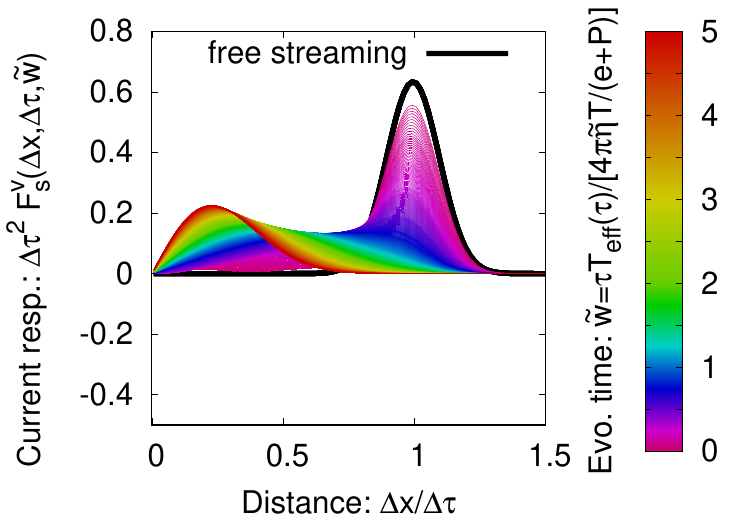}
\end{minipage}
\caption{Evolution of the charge Green's functions in response to initial charge perturbations in coordinate space. The different panels correspond to different response functions; different curves in each panel corresponds to different times $\tilde{w}$.\label{fig:ResponseFunctionsToInitialChargePerturbationsCoordinateSpace}}
\end{center}
\end{figure}
\noindent
For the charge Green's functions the decomposition in coordinate space is given by
\begin{subequations}
\begin{align}
\FCS{\tau}{\tau}\pqty{\rt,\tau} &= \FCS{s}{s}\pqty{\vrt,\tau}\, ,\\
\FCS{i}{\tau}\pqty{\rt,\tau} &= \frac{\rt^{i}}{\vrt}\FCS{v}{s}\pqty{\vrt,\tau}\, .
\end{align}
\end{subequations}
The relation to their counterparts in Fourier space is given by the Fourier-Hankel transforms
\begin{subequations}
\begin{align}
\FCS{s}{s}\pqty{\vrt,\tau} &= \frac{1}{2\pi} \int \dd{\vkt} \vkt J_0\pqty{ \vkt \vrt } \F{s}{s} \pqty{\vkt,\tau} \, , \\
\FCS{v}{s}\pqty{\vrt,\tau} &= \frac{1}{2\pi} \int \dd{\vkt} \vkt J_1\pqty{ \vkt \vrt } \F{v}{s} \pqty{\vkt,\tau} \, .
\end{align}
\end{subequations}
Again, the coordinate space Green's function for charge response, $\FCS{s}{s}$, is already shown in Sec. \ref{sec:GreensFunctionsInCoordinateSpace}, while the Green's function associated with the current response is shown in Fig. \ref{fig:ResponseFunctionsToInitialChargePerturbationsCoordinateSpace}.

\section{Identities For Spherical Harmonics and Associated Legendre Polynomials}\label{app:IdentitiesSphericalHarmonics}
\noindent
While deriving the equations of motion for $\E{m}{l}$ and $\N{m}{l}$ we used several identities for the associated Legendre polynomials and numerical coefficients. First we will list the appearing coefficients
\begin{subequations}
\begin{align}
\begin{alignedat}{16}
\tens{\Delta}{m}{l,-}& &&= &&+\frac{\pqty{l+1}\pqty{l+m}}{2l+1} &&\quad &&, \quad &&\tens{\xi}{m}{l,-}&& &&= &&\frac{l+m}{2l+1}\, ,\\
\tens{\Delta}{m}{l,+}& &&= &&-\frac{l\pqty{l-m+1}}{2l+1} &&\quad &&, \quad &&\tens{\xi}{m}{l,+}&& &&= &&\frac{l-m+1}{2l+1}\, , \\
\\
\tens{a}{m}{l,-2}& &&= &&-\tens{\xi}{(2),m}{l,-2} - \tens{\Delta}{m}{l,-}\tens{\xi}{m}{l-1,-} &&\quad &&, \quad &&\tens{b}{m}{l,-2}&& &&= &&\tens{a}{m}{l,-2}  \frac{\tens{y}{m}{l}}{\tens{y}{m}{l-2}}\, , \\
\tens{a}{m}{l,0}& &&= &&\frac{1}{3}  - \tens{\xi}{(2),m}{l,0} - \tens{\Delta}{m}{l,-}\tens{\xi}{m}{l-1,+} - \tens{\Delta}{m}{l,+}\tens{\xi}{m}{l+1,-} &&\quad &&, \quad &&\tens{b}{m}{l,0}&& &&= &&\tens{a}{m}{l,0}\, , \\
\tens{a}{m}{l,+2}& &&= &&-\tens{\xi}{(2),m}{l,+2} - \tens{\Delta}{m}{l,+}\tens{\xi}{m}{l+1,+} &&\quad &&, \quad &&\tens{b}{m}{l,+2}&& &&= &&\tens{a}{m}{l,+2} \frac{\tens{y}{m}{l}}{\tens{y}{m}{l+2}}\, , \\
 \\
\tens{A}{m}{l,-2}& &&= &&-\tens{\Delta}{m}{l,-}\tens{\xi}{m}{l-1,-} &&\quad &&, \quad &&\tens{B}{m}{l,-2}&& &&= &&\tens{A}{m}{l,-2} \frac{\tens{y}{m}{l}}{\tens{y}{m}{l-2}}\, , \\
\tens{A}{m}{l,0}& &&= &&-\tens{\Delta}{m}{l,-} \tens{\xi}{m}{l-1,+} - \tens{\Delta}{m}{l,+} \tens{\xi}{m}{l+1,-} &&\quad &&, \quad &&\tens{B}{m}{l,0}&& &&= &&\tens{A}{m}{l,0}\, , \\
\tens{A}{m}{l,+2}& &&= &&-\tens{\Delta}{m}{l,+} \tens{\xi}{m}{l+1,+} &&\quad &&, \quad &&\tens{B}{m}{l,+2}&& &&= &&\tens{A}{m}{l,+2} \frac{\tens{y}{m}{l}}{\tens{y}{m}{l+2}}\, , \\
 \\
\tens{u}{m}{l,-}& &&= &&+\frac{\tens{y}{m}{l}}{\pqty{2l+1}\tens{y}{m+1}{l-1}} &&\quad &&, \quad &&\tens{d}{m}{l,-}&& &&= &&+\frac{\tens{y}{m}{l}}{\pqty{2l+1}\tens{y}{m-1}{l-1}\tens{\sigma}{m}{l}\tens{\sigma}{-m+1}{l-1}}\, , \\
\tens{u}{m}{l,+}& &&= &&-\frac{\tens{y}{m}{l}}{\pqty{2l+1}\tens{y}{m+1}{l+1}} &&\quad &&, \quad &&\tens{d}{m}{l,+}&& &&= &&-\frac{\tens{y}{m}{l}}{\pqty{2l+1}\tens{y}{m-1}{l+1}\tens{\sigma}{m}{l}\tens{\sigma}{-m+1}{l+1}}\, . \\
\end{alignedat}
\end{align}
and
\begin{align}
\begin{aligned}
\tens{\xi}{(2),m}{l,-2} &= \tens{\xi}{m}{l,-}\tens{\xi}{m}{l-1,-}\, , \\
\tens{\xi}{(2),m}{l,0}  &= \tens{\xi}{m}{l,-}\tens{\xi}{m}{l-1,+} + \tens{\xi}{m}{l,+}\tens{\xi}{m}{l+1,-}\, , \\
\tens{\xi}{(2),m}{l,+2} &= \tens{\xi}{m}{l,+}\tens{\xi}{m}{l+1,+}\, , \\
\\
\tens{\sigma}{m}{l} &= (-1)^m \frac{\pqty{l-m}!}{\pqty{l+m}!}\, .
\end{aligned}
\end{align}
\end{subequations}
These coefficients are now used to formulate the following identities in a compact way. For the background we use
\begin{align}
\pqty{1-x^2}\dv{}{x}\tens{P}{m}{l}(x) = \tens{\Delta}{m}{l,-}\tens{P}{m}{l-1}(x) + \tens{\Delta}{m}{l,+}\tens{P}{m}{l+1}(x)
\end{align}
and
\begin{align}\label{eq:RelationXTimesP}
x\tens{P}{m}{l}(x) = \tens{\xi}{m}{l,-}\tens{P}{m}{l-1}(x) + \tens{\xi}{m}{l,+}\tens{P}{m}{l+1}(x)
\end{align}
together with
\begin{align}
x^2\tens{P}{m}{l}(x) = \tens{\xi}{(2),m}{l,-2}\tens{P}{m}{l-2}(x) + \tens{\xi}{(2),m}{l,0}\tens{P}{m}{l}(x) + \tens{\xi}{(2),m}{l,+2}\tens{P}{m}{l+2}(x)\, .
\end{align}
Combining these identities we find
\begin{align}
\bqty{\pqty{\frac{1}{3} - x^2} - x\pqty{1-x^2}\dv{}{x}}\tens{P}{m}{l}(x) = \tens{a}{m}{l,-2}\tens{P}{m}{l-2}(x) + \tens{a}{m}{l,0}\tens{P}{m}{l}(x) + \tens{a}{m}{l,+2}\tens{P}{m}{l+2}(x)
\end{align}
respectively
\begin{align}
- x\pqty{1-x^2}\dv{}{x}\tens{P}{m}{l}(x) = \tens{A}{m}{l,-2}\tens{P}{m}{l-2}(x) + \tens{A}{m}{l,0}\tens{P}{m}{l}(x) + \tens{A}{m}{l,+2}\tens{P}{m}{l+2}(x) \, .
\end{align}
\newline
\noindent
Furthermore we make use of
\begin{align}
\tens{P}{-m}{l}(x) = \tens{\sigma}{m}{l}\tens{P}{m}{l}(x)
\end{align}
and
\begin{align}
\sqrt{1-x^2}\tens{P}{m}{l}(x) = \frac{1}{2l+1} \bqty{\tens{P}{m+l}{l-1}(x) - \tens{P}{m+1}{l+1}(x)}
\end{align}
in order to find
\begin{subequations}
\begin{align}
\sin(\theta)e^{+i\phi}\Ylm(\phi,\theta) &= \tens{u}{m}{l,-}\tens{Y}{m+1}{l-1}(\phi,\theta) + \tens{u}{m}{l,+}\tens{Y}{m+1}{l+1}(\phi,\theta)\, , \label{eq:SinEpTimesY}\\
\sin(\theta)e^{-i\phi}\Ylm(\phi,\theta) &= \tens{d}{m}{l,-}\tens{Y}{m-1}{l-1}(\phi,\theta) + \tens{d}{m}{l,+}\tens{Y}{m-1}{l+1}(\phi,\theta)\, , \label{eq:SinEmTimesY}
\end{align}
\end{subequations}
which are used to compute the additional  terms in the equation of motion for the perturbed moments.

%
\section{Eccentricities, Cumulants, and Anisotropic Flow}
\label{app:Eccentricities}
%

%
\subsection{Standard Initial State Eccentricities}
%

Quantifying the geometry of the initial state is done using the standard definition of the complex eccentricity vector $\bm{\mathcal{E}_n}$ given as
\begin{align} \label{e:ecc1}
\bm{\mathcal{E}_n} \equiv \varepsilon_n \, e^{i n \psi_n} \equiv -
\frac{\int r dr d\phi \, r^n e^{i n \phi} \, f(r, \phi)}
{\int r dr d\phi \, r^n \, f(r, \phi)} ,
\end{align}
where $f(r,\phi)$ is an initial state distribution like the entropy or energy density which specifies the initial state. The magnitude of the eccentricity is $\varepsilon_n$ and $\psi_n$ is the complex (event-plane) angle. We can express this quantity in terms of the complex position vector $\bm{r} \equiv x + i y$ through $r^n e^{i n \phi} = \bm{r}^n$:
\begin{align} \label{e:ecc2}
\bm{\mathcal{E}}_n \equiv -
\frac{\int d^2 \bm{r} \, \bm{r}^n \, f(\bm{r})}
{\int d^2 \bm{r} \, |\bm{r}|^n \, f(\bm{r})} \: ,
\end{align}
where boldface is used to denote the complex vector.  Usually these definitions are specified as applying only in the center of mass frame. This can be expressed in terms of a general coordinate system:
\begin{align} \label{e:ecc3}
\bm{\mathcal{E}}_n \equiv -
\frac{\int d^2 \bm{r} \, (\bm{r} - \bm{r}_{CMS})^n \, f(\bm{r})}
{\int d^2 \bm{r} \, |\bm{r} - \bm{r}_{CMS}|^n \, f(\bm{r})}
\end{align}
with the center-of-mass vector
\begin{align}
\bm{r}_{CMS} \equiv
\frac{\int d^2 r \, \bm{r} \, f(\bm{r})}
{\int d^2 r \, f(\bm{r})}
=
\frac{1}{f_{tot}} \, \int d^2 r \, \bm{r} \, f(\bm{r}) .
\end{align}
A consequence of this definition is that the directed eccentricity $\bm{\mathcal{E}}_1$ vanishes identically.

This method for describing the initial state is well suited when the quantity $f(\bm{r})$ being described is positive definite like the energy or entropy density.  However if $f(\bm{r}) = \rho(\bm{r})$ is a charge density, particularly when total net charge is zero, it becomes impossible to define a corresponding frame such that $\bm{\mathcal{E}_1} = 0$ when the total charge vanishes.  Instead, there is always a nonzero $\bm{\mathcal{E}_1}$  proportional to the dipole moment.  Due to this inability to construct a center-of-charge frame and ensure that $\bm{\mathcal{E}_1}$ vanishes for a conserved charge with $q_{tot} = 0$, the usual definitions \eqref{e:ecc1} or \eqref{e:ecc2} must be modified.

For a conserved charge density $\rho_{\mathcal{X}}(\bm{r})$ (here we consider baryon number $B$, strangeness $S$, or electric charge $Q$ for $\mathcal{X}$), regions of positive charge with $\rho_{\mathcal{X}}(\bm{r}) > 0$ and negative charge with $\rho_{\mathcal{X}}(\bm{r}) < 0$ will be treated separately by decomposing
\begin{align}
\rho_{\mathcal{X}} \equiv \rho^{(\mathcal{X}^+)} \: \theta(\rho_{\mathcal{X}}) +
\rho^{(\mathcal{X}^-)} \: \theta(-\rho_{\mathcal{X}}) ,
\end{align}
where the position argument $\bm{r}$ is suppressed for brevity. Then the eccentricities corresponding to the positive and negative charge densities are
\begin{align}
\varepsilon_n^{(\mathcal{X}^{\pm})} &\equiv
\left|
\frac{
    \int d^2 \bm{r} \, \left( \bm{r} - \bm{r}_{CMS} \right)^n \,
    \rho^{(\mathcal{X}^{\pm})}(\bm{r})
    }{
    \int d^2 \bm{r} \, \left| \bm{r} - \bm{r}_{CMS} \right|^n \,
    \rho^{(\mathcal{X}^{\pm})}(\bm{r})
    }
\right|.
\end{align}
A consequence of these choices is that when there is no charge density, then the eccentricity is zero.

%
\subsection{Cumulants}
%

To quantify the initial state geometry, we will use the cumulants for the initial eccentricities $\varepsilon_n$:
\begin{subequations} \label{e:Ecumdefs}
\begin{align}
\varepsilon_n \{2\} &= \sqrt{ \left\langle \varepsilon_n^2 \right\rangle }
\\ \notag \\
\varepsilon_n \{4\} &= \sqrt[4]{
2 \left\langle \varepsilon_n^2 \right\rangle^2 -
\left\langle \varepsilon_n^4 \right\rangle
}
\notag \\ &=
\varepsilon_n \{2\} \,
\sqrt[4]{
1  - \frac{\mathrm{Var}(\varepsilon_n^2)}{\langle \varepsilon_n^2 \rangle^2}
} .
\end{align}
\end{subequations}

These have been shown to be good predictors of the final state flow harmonics, $v_n$, due to a linear scaling relationship \cite{Niemi:2012aj}.

\bibliography{bibliography}

\end{document}